\documentclass{amsart}
\usepackage{fullpage}
\usepackage{graphicx}
\input pictex.sty
\newtheorem{thm}{Theorem}[section]

\newtheorem{lem}[thm]{Lemma}

\theoremstyle{definition}

\newtheorem{prob}[equation]{Problem}
\theoremstyle{remark}
\newtheorem{rem}[thm]{Remark}
\numberwithin{equation}{section}
\newenvironment{Proof}{\proof}{\endproof}

\newcommand{\set}[1]{\left\{#1\right\}}
\newcommand{\Real}{\mathbb R}

\newcommand{\Cbb}{\mathbb{C}}
\newcommand{\cb}{\mathbb{C}}
\newcommand{\Ocal}{\mathcal{O}}
\newcommand{\del}{\delta}
\newcommand{\lam}{\lambda}
\newcommand{\gam}{\gamma}
\newcommand{\al}{\alpha}

\def\pmtwo#1#2#3#4{\left( \begin{array}{cc}#1&#2\\#3&#4\end{array}\right)}

\def\zz{\left(-\zeta \right)}

\begin{document}

\title[Asymptotics of RMT partition function]{Asymptotics of the partition function for random matrices
via Riemann-Hilbert techniques, and applications to graphical enumeration}%
\author{N. M. Ercolani}%
\address{Dept. of Math, Univ. of Arizona}%
\email{ercolani@math.arizona.edu}%
\author{K. D. T-R McLaughlin}%
\address{Dept. of Math., Univ. of Arizona and Dept. of Math.,
Univ. of North Carolina} \email{mcl@amath.unc.edu}

\thanks{K. D. T-R McLaughlin was supported in part by NSF
grant no. 9970328 and no. 0200749, and N. M. Ercolani was supported in part by
NSF grant no. 0073087.  The authors gratefully acknowledge the faculty
and staff of the Ecole Normale Superieur, Paris, France, for their
kind hospitality.}%

\begin{abstract}
We study the partition function from random matrix theory using a well
known connection to orthogonal polynomials, and a recently developed
Riemann-Hilbert approach to the computation of detailed asymptotics for
these orthogonal polynomials.  We obtain the first proof of a complete
large $N$ expansion for the partition function, for a general class of
probability measures on matrices, originally conjectured by Bessis,
Itzykson, and Zuber.  We prove that the coefficients in the asymptotic
expansion are analytic functions of parameters in the original
probability measure, and that they are generating functions for the
enumeration of labelled maps according to genus and valence.  Central to
the analysis is a large $N$ expansion for the mean density of
eigenvalues, uniformly valid on the entire real axis.
\end{abstract}
\maketitle
\tableofcontents

\section{Motivation and Background}

In this paper we will consider asymptotics for the following
family of integrals:
\begin{eqnarray}
\label{I.001}
& & Z_{N}(t_{1},t_{2}, \ldots, t_{\nu}) =\\
\nonumber & & \int_{\mathbb{R}} \cdots \int_{\mathbb{R}} \exp{ \left\{
-N^{2}\left[\frac{1}{N} \sum_{j=1}^{N} V(\lambda_{j}; \ t_{1}, \ldots,
t_{\nu})  -
\frac{1}{N^{2}} \sum_{j\neq \ell} \log{| \lambda_{j} -
\lambda_{\ell} | } \right] \right\} } d^{N} \lambda, \\
& & V(\lambda; \ t_{1}, \ldots, t_{\nu} ) = V_{{\bf t}}(\lambda) =
V(\lambda) = \frac{1}{2}
\lambda^{2} + \sum_{k=1}^{\nu} t_{k}\lambda^{k}.
\end{eqnarray}
where the parameters $\{t_{1},\ldots, t_{\nu}\}$
are assumed to be such that the integral converges.  For example,
one may suppose that $\nu$ is even, and $t_{\nu}>0$.  We will use 
the following set for allowable ${\bf{t}} = (t_{1}, \ldots, t_{\nu})$.
For any given $T>0$ and $\gamma > 0$, define
\begin{equation*}
\mathbb{T}(T,\gamma) = \{ {\bf{t}} \in \mathbb{R}^{\nu}: \
|{\bf{t}} | \le T, \ t_{\nu} > \gamma \sum_{j=1}^{\nu-1} |t_{j}|\}.
\end{equation*}
There are three main results in this paper, Theorems \ref{I.002thm},
\ref{GeomAsymp}, and \ref{1PTASS}.  The first of these is the
following theorem. 
\begin{thm}\label{I.002thm}
There is $T>0$ and $\gamma > 0$ so that for $\bf{t} \in \mathbb{T}(T,\gamma)$,
one has the $N \to \infty$ asymptotic expansion 
\begin{eqnarray}
\label{I.002}
\ \ \ \log{ \left(\frac{Z_{N}(\bf{t})}{Z_{N}(\bf{0})} \right)} = N^{2}
e_{0}({\bf{t}}) + e_{1}({\bf{t}}) +
\frac{1}{N^{2}} e_{2}({\bf{t}}) + \cdots.
\end{eqnarray}
The meaning of this expansion is:  if you keep terms up to order
$N^{-2k}$, the error term is bounded by $C N^{-2k-2}$, where the
constant $C$ is independent of $\bf{t}$ for all $\bf{t} \in
\mathbb{T}(T,\gamma)$.  For each $j$, the function
$e_{j}({\bf{t}})$ is
an analytic function of the (complex) vector ${\bf{t}}$, in a
neighborhood of ${\bf{0}}$. Moreover, the 
asymptotic expansion of derivatives of $\log{ \left( Z_{N} \right)}$
may be calculated via term-by-term differentiation of the above series.
\end{thm}
{\bf Remark:}  In the statement of the Theorem, ${\bf{t}} \in
\mathbb{T}(T,\gamma)$.  This is not the largest domain where the
asymptotic expansion holds true.  What is really required is the
existence of a path $\Gamma$ in $\mathbb{R}^{\nu}$ connecting ${\bf{t}}$
to ${\bf{0}}$ such that for all ${\bf{t}} \in \Gamma$, the associated
equilibrium measure (see Section \ref{Sec:2}) is supported on a single
interval, with strict variational inequality off the support, strict
positivity on the interval of the support, and square-root vanishing at
the endpoints.  A global characterization of a maximal domain where the
expansion holds true would lead us too far from the main focus of this paper,
and we will not pursue this here.

The integral $Z_{N}(t_{1},\ldots, t_{\nu}) = Z_{N}({\bf t}) $ appears in
a number of areas of mathematics and mathematical physics.  We will explain
some of the motivations for studying the asymptotics of $Z_{N}({\bf t})$
in the following subsections.  In what follows it will be useful to
define the ratio: 
\begin{eqnarray}
\hat{Z}_{N}({\bf{t}}) = Z_{N}({\bf{t}})/Z_{N}({\bf{0}}).
\end{eqnarray}

\subsection{Motivation:  statistical mechanics of a log-gas.}  The
integral (\ref{I.001}) has the natural interpretation as the
partition of function for a statistical mechanical system of
particles on the line, with logarithmic interaction potential, in
the presence of an external field whose potential is $V_{{\bf
t}}(\lambda)$.  The asymptotic behavior for
$N \to \infty$ may then be interpreted as the limiting behavior of
the statistical mechanical system in the low temperature and many
particle asymptotic limit.

With $N$ fixed, the integral $Z_{N}(t_{1},\ldots, t_{\nu})$, and
indeed all relevant statistical observables, may be expressed in
terms of families (parameterized by $\{ t_{1},\ldots, t_{\nu}\}$,
and $N$) of associated orthogonal polynomials (see Section \ref{SSLT} below). By
itself, this observation is not terribly useful.  However, it
turns out that these families of orthogonal polynomials are in a
certain sense completely integrable. Indeed, all limiting
asymptotic questions about the orthogonal polynomials (and hence
concerning the statistical mechanical system) are in principle
{\it explicitly computable} (see Section \ref{Sec:2} below).

\subsection{Motivation: random matrix theory}The integral
(\ref{I.001}) is also of fundamental importance in the theory of
random matrices.  As is well known (see, for example, the review
text \cite{Mehta}), in the theory of random Hermitian matrices
from the so-called "Unitary ensemble", one considers the measure
on $N \times N$ Hermitian matrices given by
\begin{eqnarray}
\label{I.003}
 d\mu_{\bf t} = \frac{1}{{Z}_{N}} \exp{\left\{-N \mbox{ Tr }
\left[ V_{{\bf t}} (M) \right]
\right\} } dM,
\end{eqnarray}
where $dM$ is Lebesgue measure on the matrix entries, i.e.
\begin{eqnarray}
\nonumber
 dM = \prod_{j<k} d M_{jk}^{\mbox{R}} dM_{jk}^{\mbox{I}}
\prod_{j=1}^{N} dM_{jj},
\end{eqnarray}
where $M_{jk}^{\mbox{R}}$ denotes the real part of the matrix
entry $M_{jk}$, and $M_{jk}^{\mbox{I}}$ denotes the imaginary
component of the matrix entry $M_{jk}$. It is a basic fact that
the measure (\ref{I.003}) induces a probability measure on the
eigenvalues, with density
\begin{eqnarray}
\label{I.004} \hspace{0.3in}
\frac{1}{Z_{N}} \exp{ \left\{
-N^{2}\left[\frac{1}{N} \sum_{j=1}^{N} V_{{\bf t}}
(\lambda_{j})  +
\frac{1}{N^{2}} \sum_{j\neq \ell} \log{| \lambda_{j} -
\lambda_{\ell} | } \right] \right\} } d^{N} \lambda,
\end{eqnarray}
where $Z_{N} = Z_{N}({\bf t})$ appearing in (\ref{I.004}) is
precisely the partition function defined in (\ref{I.001}).  Thus
asymptotics such as those contained in (\ref{I.002}) yield
asymptotic information concerning the statistics of the
eigenvalues of these random matrices.  For example, by
differentiating $\log{ Z_{N} }$, one obtains
\begin{eqnarray}
\label{I.004a}
 \frac{\partial}{\partial t_{\ell}} \log{ Z_{N}} = -N
\mathbb{E} \left(  \mbox{ Tr } M^{\ell} \right),
\end{eqnarray}
where $\mathbb{E}$ denotes the expectation with respect to the probability
measure $d\mu_{\bf t}$.  
Now if one has established (\ref{I.002}), then in conjunction with
(\ref{I.004a}), one learns the following:
\begin{eqnarray}
\label{I.005}
 & &\lim_{N \to \infty} \mathbb{E} \left(
\frac{1}{N}\mbox{ Tr } M^{\ell} \right) = \frac{\partial}{\partial
t_{\ell} }
e_{0}(t_{1}, \ldots, t_{\nu}).
\end{eqnarray}
Similar calculations show that (\ref{I.002}) implies
\begin{eqnarray}
\label{I.006}
 & & \lim_{N \to \infty} \left\{\mathbb{E} \left(
\mbox{ Tr }\left( M^{n} \right) \cdot \mbox{ Tr } \left( M^{m}
\right) \right) - \mathbb{E} \left( \mbox{ Tr } M^{m} \right)
\cdot \mathbb{E} \left( \mbox{ Tr } M^{n} \right) \right\}\\
\nonumber
 & & \hspace{0.3in}= \frac{\partial^{2} }{\partial t_{m} \partial t_{n} }
e_{0}(t_{1}, \ldots, t_{\nu}).
\end{eqnarray}
Observe that (\ref{I.005}) and (\ref{I.006}) together imply that
the fundamental random variables $\left\{ \frac{1}{N}\mbox{Tr
}M^{\ell}\right\}_{\ell=1}^{\infty}$ are asymptotically
uncorrelated.

\subsection{Motivation: Graphical Enumeration}

\subsubsection{The Viewpoint of Gaussian Expectations} \label{Gaussview}
One way to try to think about the partition function $\hat{Z}_{N}$
in (\ref{I.003}) is as a Gaussian expectation of the {\it interaction
term}:
$\exp{\left\{-N \mbox{ Tr }\left[ V_{{\bf t}} (M) 
  -\frac{1}{2} M^2\right]\right\}}$.
This is the viewpoint taken in much of the physical literature on random
matrix theory, for example \cite{BIZ}.

To better explain this we briefly review some facts about Gaussian measures:

\noindent A measure $\mu$ on $\mathbb{R}^n$ is called {\it Gaussian} if its
characteristic function, $\phi({\bf k}):=\int_{\mathbb{R}^n}$ $ e^{i ({\bf k},{\bf x})} d\mu$, has the form
\begin{equation}
\phi({\bf k})= \exp\{i({\bf \rho},{\bf k}) -\frac{1}{2} ({\bf Qk},{\bf k})\},
\end{equation}
where ${\bf \rho}$ is the mean and ${\bf Q}$ is called the
covariance of $\mu$.
For mean zero and nonsingular covariance,
\begin{equation}
\label{dmu11}
d\mu = \exp\{-\frac{1}{2} ({\bf Ax},{\bf x})\}{\bf dx}
\end{equation}
where ${\bf A} = {\bf Q}^{-1}$. We consider $d \mu$ of the form (\ref{dmu11}).

The salient feature of Gaussian measures, and the one which is fundamental
for applications to graphical enumeration, is that expectations of
general polynomial functions can be reduced to quadratic expectations. The
explicit recipe for this, generally referred to as the Wick formula, states
that for linear functions $\ell_i$ on $\mathbb{R}^n$,

\begin{equation}
\langle \ell_1 \ell_2 \dots \ell_{2k}\rangle =
\sum \langle \ell_{r_1}\ell_{s_1} \rangle \langle \ell_{r_2}\ell_{s_2} \rangle
\dots \langle \ell_{r_k}\ell_{s_k} \rangle
\end{equation}
where $\langle \cdot \rangle$ denotes expectation with respect to $d\mu$
and the sum is taken over the $(2k-1)!!$ ($= (2k-1)(2k-3)$ $\dots 1$)
Wick couplings of $1,2,\dots,2k$. A {\it Wick coupling} is a partition of
$1,2,\dots,2k$ into couples $(r_i,s_i)$ such that $r_1 < r_2 \dots < r_k$
and $s_i > r_i$. The quadratic expectations are completely
determined by $\langle x_i y_j \rangle = q_{ij}$ where $q_{ij}$ is the
$ij^{th}$ entry of the covariance matrix $Q$. Expectations of odd polynomials
vanish.

The proof of the Wick formula follows from a comparison of the Taylor
coefficients of the characteristic function $\phi({\bf k})$ and its logarithm.
We refer the reader to \cite[P. 9]{Sim74} for details.

The random matrix measure defining the Gaussian Unitary Ensemble (GUE)
and given by

\begin{equation}
    \label{I.011W}
d\mu = 2^{-\frac{N}{2}}\pi^{-\frac{N^2}{2}}
\exp{\left\{-\frac{1}{2}\mbox{Tr}M^2\right\}} dM
\end{equation}
is manifestly a Gaussian measure. It is straightforward to work out the
covariance matrix from which one may conclude that
\begin{eqnarray}
    \label{I.012W}
    \langle m_{ij} m_{ji}\rangle = 1, \mbox{ and }\ \langle m_{ij} m_{kl}\rangle = 0
    \mbox{ for } \ (i,j) \ne (k,l).
    \end{eqnarray}

\subsubsection{Enumerating Maps}

There are by now a number of striking examples which illustrate the
power of random matrix methods for calculating explicit solutions
of combinatorial problems related to graphical enumeration. One of these
is the problem of enumerating maps.

A {\it map}  is a graph which is embedded into a Riemann surface so that

\begin{enumerate}
\item the (images of the) edges do not intersect;
\item dissecting the surface along the edges decomposes it into a union
of open cells; these cells are called the {\it faces} of the map.
\end{enumerate}

Given this definition, one can give a precise combinatorial description of
a class of maps in terms of edge identifications between a collection
of faces. For instance one can pose the question of how many maps can be
constructed from a single face, having $2k$ edges around its boundary, by
identifying its edges in pairs. It is straightforward to see that this
number is $(2k-1)!!$. A more subtle question is to ask how many of these
maps lie on a surface of genus g. It turns out that this counting problem
has a direct and natural reinterpretation in terms of the combinatorics
of Wick couplings. Let $\varepsilon_g(k)$ denote the number of one face
maps with $2k$ edges on a Riemann surface of genus $g$. A generating
function for these numbers can be directly expressed in terms of a random
matrix moment \cite{HarZag}:

\begin{equation} \label{oneface}
\langle \mbox{Tr} M^{2k} \rangle =
N^{k+1} \sum_{g=0}^{[k/2]}\varepsilon_g(k)\left( \frac{1}{N^2}\right)^g
\end{equation}

There are extensions of these kinds of calculations to more general classes
of maps along with remarkable applications to the calculation of geometric
invariants of moduli spaces of Riemann surfaces. We refer the reader to
\cite{HarrMorr} for a good general description of these results.

\subsubsection{Diagrammatic Expansions}
\label{DiagExp}

We return now to considering the evaluation of the partition function for
the {\it deformed unitary ensemble}  $\hat{Z}_{N}({\bf t})$. Here we
will summarize some beautiful work relating the asymptotic expansion
(\ref{I.002}) to some problems in enumerative geometry.  As mentioned
above, in \cite{BIZ} the authors asserted the existence of the expansion
(\ref{I.002}), and presented a very elegant consistency argument, which
is described in subsection \ref{history}.  The components of
${\bf t}$ are viewed as deformation parameters; when ${\bf t}=0$ one has the
partition function of the original Gaussian Unitary Ensemble.

To fix ideas, will will restrict our attention to the case considered in
\cite{BIZ}.  So in this subsection, we will set all parameters
$t_{j}=0$, except for $t = t_{4}$.  (While in \cite{BIZ} the authors
considered more general deformations, the authors gave a more detailed
discussion for this case.)

Consider the Taylor series expansion corresponding to all 
parameters $t_j = 0$ except for $t=t_4$,
\begin{equation}
\exp\left\{-\frac{t}{N}\mbox{Tr} M^4\right\} = 
\sum_{n\geq 
0}\frac{1}{n!}\left(\frac{-t}{N}\right)^{n}\left(TrM^4\right)^{n}.
\end{equation}
This is clearly globally convergent for all ${t}$. The difficulty arises
when one considers the Gaussian expectation (using (\ref{I.011W})) 
of this exponential function.
Commuting the expectation integrals with the sum on the right hand side, \cite{BIZ} 
produced the {\it formal} Taylor series expansion of the partition
function around ${t=0}$:

\begin{eqnarray} \label{Formal1}
\hat{Z}_{N}(t) &=& \left\langle\exp\left\{-\frac{t}{N}
\mbox{Tr} M^4\right\}\right\rangle = 
\frac{1}{Z_{N}({\bf 0})} Z_{N}(0,0,0,t_{4}=t,0,\ldots,0)
\\ \nonumber
&\mbox{``}=\mbox{''}&  \ {\sum_{n\geq
0}\frac{1}{n!}\left(\frac{-t}{N}\right)^{n} 
\langle(TrM^4)^n\rangle}.
\end{eqnarray}

The quotation marks indicate where one has proceeded formally.  Indeed
it is manifest that this is not a convergent series expansion
since the integral corresponding to $\langle\exp\{-(t/N)\mbox{Tr}
M^4\}\rangle$ converges only for $t$ with positive real part.  Properly
speaking, this is really shorthand for a series of identities relating
{\it {derivatives}} of $\hat{Z}_{N}$ evaluated at $t=0$ to expectations
of powers of $\mbox{ Tr }M^{4}$.

Again, use of the Wick calculus allows one to replace the matrix moments
appearing in the above formal series by a generating function for
enumerating a certain class of {\it labelled} 4-valent maps (or disjoint 
unions of maps) which we will refer to as diagrams. A {\it four valent
diagram} consists of 
\begin{enumerate}
\item n 4-valent vertices;
\item a labelling of the vertices by the numbers 1,2,...,n;
\item a labelling of the edges incident to vertex $\sigma$ (for $\sigma
= 1, \ldots, n$) by letters
$i_{\sigma},j_{\sigma},k_{\sigma},\ell_{\sigma}$ (This
alphabetic order corresponds to the cyclic order of the edges
around the vertex);
\item a partitioning of the labelled edges into pairs.
\end{enumerate}
(Once the number $n$ is specified, the set of all $n$-vertex, four valent
diagrams is in one-to-one correspondence with the set of all pairings.)
Connecting the edges according to the pairing yields a
graph, together with a cyclic ordering of the edges around each vertex.
For definiteness, we will adhere to the rule that this cyclic
ordering corresponds to a clockwise
orientation of the edges around a vertex.  

If the underlying graph is connected, the cyclic ordering determines a
map associated to the diagram. In that case we will call the diagram a
g-map if the map is on a Riemann surface of genus g.  (Note that a {\it
g-map} as defined here has more structure than a {\it map} as defined in
the previous subsection since a g-map carries a labelling of its
vertices and edges.) If the graph is not connected, then the edge labelling
associates a map to each connected component, and the conglomerate of
maps will be called a g-diagram.

The use of the Wick calculus yields the following different
representation of (\ref{Formal1}), which clearly demonstrates that the
partition function is a generating function for the enumeration of {\it
diagrams}:
\begin{equation} \label{genfcn}
\hat{Z}_{N}(t) \mbox{``}= \mbox{''} \sum_{n\geq 0}\frac{1}{n!}\left(\frac{-t}{N}\right)^{n}\sum_g
\#\{\mbox{4-valent, n-vertex, g-diagrams}\}N^{2-2g+n} .
\end{equation}

Exploiting the relationship between the terms of a Taylor series and those
of its logarithm one can write down an equally formal representation for the
the {\bf logarithm} of the partition function which can be regarded as a
generating function for connected n-vertex diagrams:

\begin{equation}\label{logconn}
\log{ \hat{Z}_{N}(t) } \mbox{``}=\mbox{''} 
\sum_{n\geq 0}\frac{1}{n!}\left(\frac{-t}{N}\right)^{n}\sum_{g \geq 0}
\#\{\mbox{4-valent, n-vertex g-maps}\}N^{2-2g+n}.
\end{equation}

Finally one can make another leap beyond rigor and resum the terms of the
previous formal series, as was done in \cite{BIZ}, to formally
order the series by the genus of the surface to which the diagram maps.
This yields what is referred to in the physics literature as the genus
expansion,

\begin{equation}
\label{genusexp4V}
\log\hat{Z}_{N}(t) \mbox{``}= \mbox{''} \ {\sum_g E_g(t) N^{2-2g}}, 
\end{equation}
where $E_g(t) =  \sum_{n\geq 1}\frac{1}{n!}(-t)^n \kappa_g(n)$ is a
formal series (possibly convergent) in which each of the coefficients
$\kappa_g(n)$ is the number of connected maps of genus g
with n vertices (all 4-valent).

The authors of \cite{BIZ} fully appreciated that this intuitive ``deduction''
was completely formal and in fact suggested an approach for providing a
rigorous derivation. This approach was never fully pursued (we briefly
review its history in Section \ref{history}).

We have taken a different approach and our proof of Theorem
\ref{I.002thm} provides a rigorous validation of the form of the
representation (\ref{genusexp4V}).  When restricted to the four-valent
case, our result may be summarized as follows.
\begin{itemize}
\item[A.] There is a positive number $T_{4}$ such that for $0 \le t \le
T_{4}$, the partition function $\hat{Z}_{N}(t)$ possesses the 
following asymptotic expansion.
\begin{eqnarray}
\label{4VPartExp}
N^{-2} \log{ \hat{Z}_{N}(t)} = e_{0}(t) + N^{-2} e_{1}(t) + N^{-4}
e_{2}(t) + \cdots.
\end{eqnarray}
The meaning of this expansion is:  if you keep terms up to order
$N^{-2k}$, the error term is bounded by $C N^{-2k-2}$.  The coefficients
$e_{g}(t)$ possess analytic continuations to a neighborhood of $t=0$.
\item[B.]  The coefficients $e_{g}(t)$ are related to the counting of
4-valent g-maps via the following formula:
$$
(-1)^n \frac{\partial^n}{\partial t^n} e_g(0)
=
\#  \{\mbox{4-valent, n-vertex, g-maps}\}. 
$$
\end{itemize}

For the case of more general deformations, and the full asymptotic
expansion (\ref{I.002}), we have the following theorem concerning the
coefficients $e_g(t_1 \dots t_\nu)$.

\begin{thm}
\label{GeomAsymp}
The coefficients in the asymptotic
expansion (\ref{I.002}) satisfy the following relations.  Let $g$ be
a nonnegative integer.  Then
\begin{equation} \label{genusexpA}
e_g(t_1 \dots t_\nu)
=  \sum_{n_j\geq 1}\frac{1}{n_1!\dots n_\nu!}
(-t_1)^{n_1}\dots (-t_\nu)^{n_\nu}\kappa_g(n_1,\dots,n_\nu)
\end{equation}
in which each of the coefficients $\kappa_g(n_1, \dots, n_\nu)$ is the
number of g-maps with $n_j$  j-valent vertices for $j=1,\dots,\nu$.
\end{thm}

\subsection{Motivation:  The Connection to Orthogonal Polynomials}
\label{SSLT}

As mentioned above, the integral (\ref{I.001}) is intimately
connected to the theory of orthogonal polynomials.  Consider the measure
  \begin{eqnarray}
\label{OPMeas}
  w_{N}(x) dx := \exp{\left[   -N V_{{\bf t}}(x)\right] } dx.
  \end{eqnarray}
\bigskip

Let us define $\{p_{j}(x; \ N, \ {\bf t})  \}_{j=0}^{\infty}$ to be the
sequence of polynomials
orthogonal with respect to the measure $w_{N}(x) dx$.  That is,
$\{p_{j}(x; \  N, \ {\bf t})\}_{j=0}^{\infty}$
satisfies
  \begin{eqnarray}
\int_{-\infty}^{\infty}  p_{j} p_{k} w_{N} dx  = \left\{
\begin{array}{cc}
0 & j \neq k \\
1 & j = k \\
\end{array} \right.,
\end{eqnarray}
and $p_{j}(x; \ N, \ {\bf t}) = \gamma_{j}^{(N)}
x^{j} + \cdots $, $\gamma_{k}^{(N)}> 0$. (The leading coefficient
$\gamma_{k}^{(n)}$ is of course dependent on the parameters
$t_{1}, \ldots, t_{\nu}$; however, we suppress this dependence for
notational convenience.)  The fact of the matter is that
$Z_{N}({\bf t})$ may also be defined via
\begin{eqnarray}
\label{I.010}
 Z_{N}({\bf t}) = N! \
\prod_{\ell=0}^{N-1} \left( \gamma_{\ell}^{(N)}\right)^{-2}.
\end{eqnarray}

\bigskip

$Z_{N}$ is also defined via
\begin{eqnarray}
Z_{N}(t_{1}, \ldots, t_{\nu}) = N! \ \left|
\begin{array}{cccc}
c_{0} & c_{1} &\cdots & c_{N-1} \\
c_{1} & c_{2} & \cdots & c_{N} \\
\vdots & \vdots &\ddots & \vdots \\
\vdots & \vdots & \ddots &\vdots \\
c_{N-1}& c_{N} & \cdots & c_{2 N - 2} \\
\end{array} \right|,
\end{eqnarray}
where $c_{j} = \int_{\mathbb{R}} x^{j} w_{N}(x) dx$ are the
moments of the measure $w_{N}(x) dx$, and the determinant above is
called a Hankel determinant (see, for example, Szeg\"{o}'s classic
text \cite{Szego}).
The asymptotic expansion (\ref{I.002}) constitutes a version of
the strong Szeg\"{o} limit theorem for Hankel determinants. The
strong Szeg\"{o} limit theorem concerns the asymptotic behavior of
Toeplitz determinants associated to a given measure on the
interval $(0, 2 \pi)$.  We refer the interested reader to
\cite{Szego} for more information.

\bigskip

It is well known in both the approximation theory literature and
the random matrix theory literature (see, for example,
\cite{Mehta},\cite{SaffTotik} and the references therein, or
\cite{DKM} and references) that $Z_{N}({\bf t})$
satisfies the following leading order asymptotic behavior:
\begin{eqnarray}
& & \nonumber \lim_{N\to\infty} \frac{1}{N^{2}} \log{ Z_{N}(t_{1},
\ldots t_{\nu})} \\
\nonumber & & = \sup_{\mu \in {\mathbb{A}} } \left\{- \int
V_{{\bf t}} (\lambda)
 d \mu(\lambda) + \int \int \log{ \left| \lambda
- \eta \right| } d\mu(\lambda) d\mu(\eta) \right\},
\end{eqnarray}
where $\mathbb{A}$ is the set of all positive Borel measures on
the real axis, with unit mass.  The variational problem posed
above is referred to in the approximation theory literature as the
problem of determining the equilibrium measure for logarithmic
potentials in the presence of an external field (see
\cite{SaffTotik}).  It is well known that the supremum is achieved
at a unique measure $\mu^{*}$ (see, for example,
\cite{SaffTotik}), and that for real analytic external fields, the
equilibrium measure $\mu^{*}$ is supported on finitely many
disjoint intervals, and on the interior of each interval, the
equilibrium measure has analytic density \cite{DKM}.

\subsection{Brief discussion of history} \label{history}

In \cite{BIZ}, Bessis, Itzykson and Zuber considered asymptotics for the
integral (\ref{I.001}), but for the special case in which $t_{j} \equiv
0$ for $j \neq 4$.  That is, they considered the integral
\begin{eqnarray}
& &Z_{N}^{(\mbox{\tiny{BIZ}})}(t) =\\
\nonumber & & \int \cdots \int \exp{ \left\{
-N^{2}\left[\frac{1}{N} \sum_{j=1}^{N} \left( \frac{1}{2}
\lambda_{j}^{2} +t \lambda_{j}^{4} \right)  - \frac{1}{N^{2}}
\sum_{j\neq \ell} \log{| \lambda_{j} - \lambda_{\ell} | } \right]
\right\} } d^{N} \lambda.
\end{eqnarray}
They computed asymptotics for this integral by using
(\ref{I.010}), together with some reasonable assumptions on the
asymptotics of ratios of the leading coefficients
$\gamma_{j}^{(n)}$. The basic idea is to re-write (\ref{I.010}) in
yet another form, namely
\begin{eqnarray}
\label{I.012}
 & & \frac{1}{N^{2}} \log {
Z_{N}^{(\mbox{\tiny{BIZ}})}(t)} =
\\& &
\nonumber
 \frac{1}{N^{2}} \log{(N!) } - \frac{2}{N}\log{
\left(\gamma_{0}^{(N)} \right)} + \sum_{j=1}^{N-1} \frac{2 ( N-j
)}{N^{2}} \log{ \left( b_{j-1}^{(N)}\right)},
\end{eqnarray}
where $b_{j}^{(N)}$ denotes the recurrence coefficient associated
to the orthogonal polynomial sequence, $x p_{j}(x) = b_{j}
p_{j+1}(x) + a_{j}p_{j}(x) + b_{j-1} p_{j-1}(x)$.  It easy easy to
see that the recurrence coefficients are related directly to the
leading coefficients $\gamma_{j}^{(N)}$, via $b_{j} =
{\gamma_{j}^{(N)}} / {\gamma_{j+1}^{(N)}}$.

The authors then made the following asymptotic assumptions
concerning the recurrence coefficients $b_{j}^{(N)}$:
\begin{eqnarray}
\label{I.013}
 & &
 \hspace{0.1in}
\left( b_{j}^{(N)} \right)^{2} = r_{0}\left(\frac{j}{N},
 t \right) + \frac{1}{N^{2}} r_{2} \left( \frac{j}{N}, t \right) +
 \cdots + \frac{1}{N^{2k}} r_{2k}\left( \frac{j}{N}, t \right) +
 \cdots \ ,
\end{eqnarray}
where the functions $r_{2 \ell}(x,t)$ are assumed to be infinitely
differentiable functions of the variable $x$.  With these
assumptions in hand, Bessis, Itzykson, and Zuber then substitute
(\ref{I.013}) into (\ref{I.012}), and apply the Euler-Maclaurin
summation formula to replace each summation by a series of
integrals.

Following that, there is an argument that {\it{if an asymptotic
description such as (\ref{I.013}) holds, then one may deduce the
form of the functions $r_{2\ell}$}}.  This is achieved by studying
a nonlinear recursion relation satisfied by the recurrence
coefficients $b_{j}^{(N)}$.  This in turn leads to a prediction
for the asymptotics of $Z_{N}^{(\mbox{\tiny{BIZ}})}$.  The authors
compute explicitly the first three terms, and present a conjecture
for all terms.  Their conjecture is that for $j \ge 2$,
$e_{j}^{(\mbox{\tiny{BIZ}})}(t)$ is a rational function of $t$,
with one pole in the finite plane, of order $5(j-1)$, located at
$t = -1/48$.

Quite separately, a rigorous analysis of the recurrence
coefficients for polynomials orthogonal with respect to weights of
the form $e^{-x^{2m}}dx$ \cite{Magnus} was carried out by
analyzing the very same nonlinear recursion relation.  This
yielded a proof of the leading order information contained in
(\ref{I.013}), but valid only if $j \to \infty$.

On the other hand, in \cite{BIZ}, the authors use (\ref{I.013}) as
a uniform expansion, valid even for $j=1$, as $N \to \infty$.  It
would seem plausible that such an expansion would hold for $j \to
\infty$, $N \to \infty$, such that $j/N \to x > 0$.  However, it
is entirely possible that for $j$ such that $j/N$ approaches $0$,
there is some correction to (\ref{I.013}) which may affect the
asymptotic computation of $\log{ (Z_{N}^{(\mbox{\tiny{BIZ}})})}$
at some finite algebraic order in $N$.

For this reason, it is of great interest to find some independent
way to compute the asymptotics for $\log{( Z_{N}(t_{1}, \ldots,
t_{\nu}))}$.

In a different direction, in \cite{BI}, techniques from the theory
of isomonodromic deformations, as well as techniques for the
asymptotic analysis of Riemann-Hilbert problems were applied to
the problem of determining asymptotics for the polynomials
$p_{n}(z; \ N, \ t )$ orthogonal with respect to the 1-parameter
family of measures $\exp{[-N( x^{2}/2 + t x^{4})]}$.  The results
were used to prove that the associated random matrix model obeys
the universality conjecture. That is, it was proved that the local
statistics of the eigenvalues of random matrices with the
probability measure $\exp{[ - N \mbox{ Tr } ( M^{2}/2 + t M^{4})]}
dM$ converge as $N \to \infty$ to a universal random point
process, whose correlation functions are given in terms of the
so-called sine kernel.

In addition, in \cite{DKMVZ1}-\cite{DKMVZ3}, recent techniques for
the asymptotic analysis of Riemann-Hilbert problems were applied
to the problem of determining asymptotics for a wide class of
orthogonal polynomials, which contain the orthogonal polynomials
$p_{n}(z; \ N , \ {\bf t})$ as a subset. The
results were used to establish this universality conjecture of
random matrix theory, for a family of real analytic probability
measures. As a by-product of the analysis in \cite{DKMVZ3},
asymptotics for the recurrence coefficients $b_{N-1}^{(N)}$ were
obtained.

The upshot of this is that there is emerging a new way to compute
asymptotics of $\log{(  Z_{N}({\bf t})) }$.  In both
\cite{BI} and \cite{DKMVZ1}-\cite{DKMVZ3}, an important asymptotic
technique is the Deift-Zhou steepest descent / stationary phase
method for the asymptotic analysis of Riemann-Hilbert problems.
This method was introduced in \cite{DZ1}, and further developed in
\cite{DZ2} and \cite{DVZ}.

One way to proceed is to compute the asymptotics for the
orthogonal polynomials, and then deduce the asymptotics for the
recurrence coefficients $b_{j}^{(N)}$. However, this still
requires an analysis of the recurrence coefficients which is {\it
uniform} in $j$, i.e. for $j = 0$ through $j=N-1$.  This appears
to be an onerous task, and so we shall not adopt this approach
here.  Instead, we will proceed, starting in the next subsection, by the direct analysis of yet another
representation of $\log{(Z_{N})}$ (formula (\ref{I*5})).

\subsection{A statistical mechanical formula, and orthogonal polynomials.}

In this paper we will analyze the large $N$ behavior of
$Z_{N}({\bf t})$ by direct analysis of the formula
(\ref{I.SMF}) for the logarithmic derivative of $Z_{N}$.  The
derivation of this formula proceeds as follows.  One begins by
computing the logarithmic derivative of $Z_{N}$:
\begin{eqnarray}
& & \ \frac{\partial }{\partial t_{\ell}} \log{(Z_{N})} =
\frac{1}{Z_{N}} \int \cdots \int \left(-N \sum_{j=1}^{N}
\lambda_{j}^{\ell} \right) \\
\nonumber & & \hspace{0.3in} \times \exp{ \left\{
-N^{2}\left[\frac{1}{N} \sum_{j=1}^{N} V_{{\bf t}}(\lambda_{j})
  -
\frac{1}{N^{2}} \sum_{j\neq \ell} \log{| \lambda_{j} -
\lambda_{\ell} | } \right] \right\} } d^{N} \lambda.
\end{eqnarray}
Of course, this has the immediate interpretation as an expectation
value, with respect to the probability measure (\ref{I.003}):
\begin{eqnarray}
& & \ \frac{\partial }{\partial t_{\ell}} \log{(Z_{N})} =
\mathbb{E}_{N} \left( -N \sum_{j=1}^{N} \lambda_{j}^{\ell}
\right).
\end{eqnarray}
Now it is well known (see, for example, \cite{Mehta}) that
expectation values of functions of the $\lambda_{j}$s can be
expressed in terms of the associated orthogonal polynomials.  In
particular, the mean density of the $\lambda_{j}$'s is given by
the so-called "one-point function",
\begin{eqnarray}& & \ \
\rho_{N}^{(1)} (\lambda) = \frac{1}{N} \exp{ \left[ -N V_{{\bf
t}}(\lambda) \right]
} \sum_{j=0}^{N-1} p_{j}(\lambda; \ N, {\bf t})^{2}.
\end{eqnarray}
Using this connection, one has the following remarkable formula
for the partial derivatives of $\log{ (Z_{N})}$:
\begin{eqnarray}
\label{I.SMF}
 & & \ \frac{\partial }{\partial t_{\ell}}
\log{(Z_{N})} = -N^{2} \mathbb{E} \left(\frac{1}{N}\mbox{ Tr } M^{\ell}\right)
              = -N^{2} \int_{-\infty}^{\infty} \lambda^{\ell}
\rho_{N}^{(1)}(\lambda) d \lambda
\end{eqnarray}
in which the first equality follows from (\ref{I.004a}).
The fundamental theorem of calculus readily implies
\begin{eqnarray}
\label{I*5}
Z_N({\bf t}\, ) = Z_N({\bf 0}) \exp{ \left\{ -N^{2}\int_0^{\bf t} \int_{\Real}
\rho_1^{(N)} (\lam ) \, \overset{\longrightarrow}{\nabla_{ t}}
V {d\lam} \cdot \vec{d\ell} \right\} }.
\end{eqnarray}

Since $Z_N({\bf t} =0)$ is explicitly known $(Z_N({\bf 0}) =
\cdots )$, the formula (\ref{I*5}) shows that if we have global
asymptotics for $\rho_1^{(N)}$, we can deduce an asymptotic
expansion for $Z_N({\bf t}\, )$.
We will analyze the one-point function, globally on the real axis,
and determine a uniform asymptotic representation for it.  Then we
will evaluate the integral appearing in the right hand side of
(\ref{I.SMF}), and from this obtain a uniform asymptotic expansion for
(\ref{I*5}).  This is the procedure by which we shall compute the
asymptotic expansion of $\log{ Z_{N} }$.

The following theorem is a fundamental result which, following the above
prescription, yields Theorem \ref{I.002thm}.

\begin{thm}
\label{1PTASS}
There is $T>0$ and $\gamma > 0$ so that for all ${\bf t} \in
\mathbb{T}(T,\gamma)$, the following expansion holds true:
\begin{eqnarray}
\label{I.Weakasym}
\int_{-\infty}^{\infty} f(\lambda) \rho_{N}^{(1)}(\lambda) d \lambda =
f_{0} + N^{-2} f_{1} + N^{-4} f_{2} + \cdots,
\end{eqnarray}
provided the function $f(\lambda)$ is $C^{\infty}$ smooth, and grows no faster
than a polynomial for $\lambda \to \infty$.  The coefficients $f_{j}$
depend analytically on ${\bf t}$ for ${\bf t} \in \mathbb{T}(T, \gamma)$,
and the asymptotic expansion may be differentiated term by term.  
\end{thm}
In the statement of the above Theorem, the meaning of the asymptotic
expansion is this:  if you stop after the term $N^{-2j} f_{j}$, the
error term is ${\Ocal}(N^{-(2j+2)})$.
\bigskip

\begin{rem}
We note that in some very special instances results of the above form may
be deduced from known results in the literature. For instance, taking the
Gaussian limit of (\ref{I.SMF}) one has
\begin{eqnarray*}
 & & \ \frac{\partial }{\partial t_{2k}}
\log{(Z_{N})}|_{\bf t=0}
= -N^{2} \int_{-\infty}^{\infty} \lambda^{2k}
         \rho_{N}^{(1)}(\lambda,{\bf t=0}) d \lambda
= -N^{2} \langle \frac{1}{N} N^{-k}\mbox{ Tr } M^{2k}\rangle.
\end{eqnarray*}
The last term, $-N^{-k+1} \langle\mbox{ Tr } M^{2k}\rangle$, which
by formula (\ref{oneface}) is equal to a {\it finite} series,
$-N^2 \sum_{g=0}^{[k/2]}\varepsilon_g(k)\left( \frac{1}{N^2}\right)^g$, is
a generating function for counting one-face maps with $k$ edges. This series
is manifestly even in N. In the setting of (\ref{I.SMF}) this series is
actually a generating function for counting $2k$-valent one-vertex maps
(see also (\ref{1Vert.24})-(\ref{1Vert.26})).
There is no inconsistency in this since by duality these two countings are
equivalent.
\end{rem}

\begin{rem}
Additionally, in \cite{APS}, Albeverio et al have recently considered
asymptotic expansions of the form (\ref{I.Weakasym}), for the special
function $f_{z}(\lambda) = (z - \lambda)^{-1}$, under the assumption
that the external field is even, but they allow for the equilibrium
measure to be supported on one or two intervals.  They assert an
asymptotic expansion for such integrals which is in inverse powers of
$N$; our analysis shows that there is a complete expansion, in inverse
powers of $N^{2}$.
\end{rem}

The paper is organized as follows.  In Section \ref{sec:enumer} we
present a self-contained description of the connection between random
matrix integrals and the enumeration of maps on Riemann surfaces,
according to genus and valence; we also prove Theorem \ref{GeomAsymp},
using Theorem \ref{I.002thm}.  In Section \ref{Sec:2} we summarize the
procedure developed in \cite{DKMVZ1}-\cite{DKMVZ3} to establish rigorous
uniform global asymptotics for the associated orthogonal polynomials.  In
Section \ref{sec:MDOS} we use the results of Section \ref{Sec:2} to derive
exact formulae for $\rho_{N}^{(1)}(\lambda)$,
the mean density of eigenvalues at finite $N$, in terms of the
Riemann--Hilbert procedure.  Explicit in the formulae derived is {\it a
complete asymptotic expansion} for $\rho_{N}^{(1)}$.  In Section
\ref{ProofMT}, we use the formulae derived in Section \ref{sec:MDOS}, to
establish Theorem \ref{1PTASS}.

\section{Enumeration of g-maps}
\label{sec:enumer}

In this section we will explain the connection between the evaluation of
the Gaussian moments $\langle \prod_{j=1}^\nu$ $ (TrM^{j})^{n_j} \rangle$, and 
g-diagrams and g-maps (as defined in subsection \ref{DiagExp}).
Following that we will prove Theorem \ref{GeomAsymp} which states  
that the asymptotic expansion
(\ref{I.002}) is a generating function for g-maps, enumerating them
according to the number of vertices with different valences and the
genus.

\subsection{Gaussian matrix integrals and enumeration of diagrams}

For simplicity we first discuss in detail the case of the pure moment
$(TrM^4)^n$; afterward we will indicate how this extends to general 
mixed moments.

The expression $(TrM^4)^n$ has the form
\begin{equation} \label{trcycle}
\sum  m_{i_1, j_1}m_{j_1, k_1}m_{k_1, \ell_1} m_{\ell_1, i_1}
      m_{i_2, j_2}m_{j_2, k_2} \dots
      m_{i_n, j_n}m_{j_n, k_n}m_{k_n, \ell_n} m_{\ell_n, i_n}
\end{equation}
where the sum is taken over all configurations, where by a {\it
configuration} one means an assignment to each of the $4n$ labels
appearing in the list $ \{ i_{\nu}, j_{\nu},
k_{\nu},\ell_{\nu}\}_{\nu = 1}^{n}$ a value from $\{1, \dots,
N\}$. There are thus $N^{4n}$ configurations. Each configuration
corresponds to a selection of $4n$ matrix entries.

We compute the expectation of this quantity with respect to the
original probability measure (\ref{I.011W}), interchanging integration with
summation:
\begin{eqnarray}
\label{TRExpec}
& & \hspace{0.2in}
\left< \left( \mbox{Tr } M^{4} \right)^{n} \right> \\
\nonumber
& & =
\sum_{\mbox{configs}} \left< m_{i_1, j_1}m_{j_1, k_1}m_{k_1, \ell_1} m_{\ell_1, i_1}
      m_{i_2, j_2}m_{j_2, k_2} \dots
      m_{i_n, j_n}m_{j_n, k_n}m_{k_n, \ell_n} m_{\ell_n, i_n}
      \right>
\end{eqnarray}
By the Wick calculus, the expectation of each term in the above
sum can itself be evaluated as a sum over Wick couplings as they
were described in Section \ref{Gaussview}. To appreciate this one
should first observe that each of the matrix entries appearing in
the expectation on the right hand side of (\ref{TRExpec}) is a
linear function.  Next, we introduce the following notation:
\begin{eqnarray}
& &     \left< m_{i_1, j_1}m_{j_1, k_1}m_{k_1, \ell_1} m_{\ell_1, i_1}
      m_{i_2, j_2}m_{j_2, k_2} \dots
      m_{i_n, j_n}m_{j_n, k_n}m_{k_n, \ell_n} m_{\ell_n, i_1}
      \right> \\
      \nonumber
      & & = \left< f_{1} f_{2} \cdots f_{4n} \right>,
\end{eqnarray}
where
\begin{eqnarray}
    \label{I.019W}
    & & f_{1} = m_{i_{1},j_{1}}, \ \
     f_{2} = m_{j_{1},k_{1}}, \ \
     f_{3} = m_{k_{1},\ell_{1}}, \ \
    f_{4} =  m_{\ell_{1},i_{1}},
    \end{eqnarray}
and for $\nu = 2, \ldots, n$,
\begin{eqnarray}
    & &f_{4(\nu-1)+1} = m_{i_{\nu},j_{\nu}}, \\
    & & f_{4(\nu-1)+2} = m_{j_{\nu},k_{\nu}}, \\
    & & f_{4(\nu-1)+3} = m_{k_{\nu},\ell_{\nu}}, \\
    & & f_{4(\nu-1)+4} = m_{\ell_{\nu},i_{\nu}}.
    \label{I.023W}
\end{eqnarray}
  
The definitions
(\ref{I.019W})-(\ref{I.023W}) implicitly define an invertible
mapping $(r,s)(\cdot) = (r(\cdot),s(\cdot))$ between the integers
$\{1, \ldots 4n\}$ and the matrix indices $\{ (i_{\nu},j_{\nu}) $,
$(j_{\nu},k_{\nu})$, $(k_{\nu},\ell_{\nu})$, $(\ell_{\nu},i_{\nu})
\}$ which we will describe as follows:
\begin{eqnarray}
\label{I.025map}
 & &    (r,s)(\cdot) = (r(\cdot),s(\cdot)) \ : \
    \{1, \ldots, 4n\} \to
    \{ (i_{\nu},j_{\nu}) , (j_{\nu},k_{\nu}),
(k_{\nu},\ell_{\nu}), (\ell_{\nu},i_{\nu}) \}_{\nu = 1}^{n}, \\
\nonumber & & k \mapsto (r(k),s(k)), \mbox{ so that } f_{k} =
m_{r(k),s(k)} \mbox{ (using (\ref{I.019W})-(\ref{I.023W}))}.
\end{eqnarray}

We now apply the Wick calculus:
\begin{eqnarray}
    \label{I.024W}
    & & \sum_{\mbox{configs}} \left< f_{1} \cdots f_{4n} \right> \\
    \nonumber
    & & = \sum_{\mbox{configs}} \ \sum_{\omega \in { W}_{4n}}
    \left< f_{\omega_{1,1}} f_{\omega_{1,2}} \right> \left<
    f_{\omega_{2,1}}f_{\omega_{2,2}}\right> \cdots
    \left<f_{\omega_{2n,1}} f_{\omega_{2n,2}} \right>
    \end{eqnarray}
In this setting a Wick coupling, $\omega = \omega_{i,j}$, $i = 1,
\ldots, 2n$, $j=1,2$, is a partition of $1, 2, \ldots, 4n$ into $2n$
couples $(\omega_{i,1},\omega_{i,2})$ so that $\omega_{1,1} <
\omega_{2,1} < \cdots < \omega_{2n,1}$, and $\omega_{i,2}> \omega_{i,1}$
for $i = 1, \ldots, 2n$.  We denote the set of all such Wick couplings of the
integers $1, \ldots, 4n$ by $W_{4n}$.

Now the outer summation over configurations appearing on the right hand 
side of (\ref{I.024W}) is over the different realizations of the variables 
$f_{1}, \ldots, f_{4n}$ determined by the configuration. We may
interchange orders of summation:

\begin{eqnarray}
\label{I.025W}
  & & \left< \left( \mbox{Tr }M^{4} \right)^n \right> =
  \sum_{\mbox{configs}} \left< f_{1} \cdots f_{4n} \right> \\
    \nonumber
    & & \hspace{1.0in}=  \sum_{\omega \in { W}_{4n}} \sum_{\mbox{configs}}
    \left< f_{\omega_{1,1}} f_{\omega_{1,2}} \right> \left<
    f_{\omega_{2,1}}f_{\omega_{2,2}}\right> \cdots
    \left<f_{\omega_{2n,1}} f_{\omega_{2n,2}} \right>.
    \end{eqnarray}

The individual terms
\begin{eqnarray}
    \label{I.027Term}
    \left< f_{\omega_{1,1}} f_{\omega_{1,2}} \right> \left<
f_{\omega_{2,1}}f_{\omega_{2,2}}\right> \cdots
\left<f_{\omega_{2n,1}} f_{\omega_{2n,2}} \right>
\end{eqnarray}
are either $0$ or $1$.  For each Wick coupling $\omega$, certain
configurations contribute a $1$ to the inner summation, and other
configurations do not contribute.  The question of evaluating the original
integral $\left< \left(
\mbox{Tr }M^{4} \right)^{n} \right>$ has been reduced to computing,
for each Wick ordering $\omega \in W_{4n}$, the number of
configurations providing unit contribution to the inner sum appearing
in (\ref{I.025W}).  For each different Wick coupling $\omega$, it turns out 
that this number is of the form $N^{F(\omega)}$, and accepting this for the
moment, we have the formula
\begin{eqnarray}
    \label{I.027form}
 \sum_{\mbox{configs}}\left< f_{\omega_{1,1}}
f_{\omega_{1,2}} \right> \left<
    f_{\omega_{2,1}}f_{\omega_{2,2}}\right> \cdots
    \left<f_{\omega_{2n,1}} f_{\omega_{2n,2}} \right> =
    N^{F(\omega)}.
\end{eqnarray}

To determine the contribution $N^{F(\omega)}$ for each Wick coupling,
and to explain the quantity $F(\omega)$, let us first recall the
direct relationship (\ref{I.019W})-(\ref{I.023W}) between
$\{f_{j}\}$ and the original variables of integration
$\{m_{i_{\nu},j_{\nu}},m_{j_{\nu},k_{\nu}},
m_{k_{\nu},\ell_{\nu}},m_{\ell_{\nu},i_{\nu}} \}_{\nu=1}^{n}$, and
the identities (\ref{I.012W}).  Using the mapping $(r,s)(\cdot)$
defined in (\ref{I.025map}) , the inner sum in (\ref{I.025W}),
over all configurations, is of the form
\begin{eqnarray}
    \label{I.030W}
    & & \hspace{-0.3in}
    \sum_{\mbox{configs}}
    \left< m_{r(\omega_{1,1}), s(\omega_{1,1})}
    m_{r(\omega_{1,2}),s(\omega_{1,2})} \right>
\left< m_{r(\omega_{2,1}), s(\omega_{2,1})}
    m_{r(\omega_{2,2}),s(\omega_{2,2})} \right>
\cdots \\
\nonumber
& & \hspace{1.75in} \cdots
\left< m_{r(\omega_{2n,1}), s(\omega_{2n,1})}
    m_{r(\omega_{2n,2}),s(\omega_{2n,2})} \right>.
\end{eqnarray}
Each different configuration appearing in (\ref{I.030W}) is a
different choice of the matrix indices $\{ (i_{\nu},j_{\nu}) $,
$(j_{\nu},k_{\nu})$, $(k_{\nu},\ell_{\nu})$, $(\ell_{\nu},i_{\nu})
\}$, and for each one we then ask if the contribution is $0$ or
$1$.  Using (\ref{I.012W}), the contribution will be $1$ if and
only if for each $\mu = 1, \ldots, 2n$,
\begin{eqnarray}
    & &
    \left< m_{r(\omega_{\mu,1}), s(\omega_{\mu,1})}
    m_{r(\omega_{\mu,2}),s(\omega_{\mu,2})} \right>
    =1,
    \end{eqnarray}
and this is true if and only if
\begin{eqnarray}
    \label{I.032W}
    & &
    r(\omega_{\mu,1}) = s(\omega_{\mu,2}) \mbox{ and } s(\omega_{\mu,1}) =
    r(\omega_{\mu,2}) \mbox{ for each } \mu = 1, \ldots, 2n.
\end{eqnarray}

The sum (\ref{I.030W}) is precisely the number of configurations
for which the $4n$ equalities (\ref{I.032W}) hold true, and since
$r(\omega_{\mu,1})$ is, for each $\mu$, one of
$\{i_{\nu},j_{\nu},k_{\nu}, \ell_{\nu}\}_{\nu=1}^{n}$ (and
similarly for $s(\omega_{\mu,2})$, $s(\omega_{\mu,1})$, and
$r(\omega_{\mu,2})$ ), it follows that the sum over
configurations, is now over
\begin{eqnarray}
    \nonumber
    & & i_{\nu}, j_{\nu}, k_{\nu},  \ell_{\nu} = 1, \ldots N, \\
    \nonumber
    & & \mbox{ so that (\ref{I.032W}) holds.}
    \end{eqnarray}

The calculation of (\ref{I.030W}) now proceeds as follows.  For
each Wick coupling $\omega$ we write out equalities amongst the
$4n$ indices of summation $\{i_{\nu}, j_{\nu}, k_{\nu}, \ell_{\nu}
\}_{\nu=1}^{n}$ enforced by (\ref{I.032W}) and group them in
``chains'' (for example, one might have $i_{3} = j_{5} = \ell_{7}
\cdots$). Each index of summation appears in two equalities
(because we are computing $\left< \left( \mbox{Tr }M^{4}
\right)^{n} \right>$, see (\ref{TRExpec})) and so each distinct
chain is, fact, a closed cycle.  For each closed cycle, the
original summation is reduced to a summation over a single free
parameter, ranging from $1$ to $N$.  Finally then, the summation 
(\ref{I.030W})
is $N^{F(\omega)}$ where $F(\omega)$ is precisely the number of
closed cycles determined by the Wick coupling $\omega$, and we have 
explained (\ref{I.027form}).

Summarizing, we have shown that
\begin{eqnarray}
\label{Trace.032}
\left< \left( \mbox{ Tr }M^{4} \right)^{n} \right> = \sum_{\omega
\in W_{4n}} N^{F(\omega)},
\end{eqnarray}
where $F(\omega)$ is the number of closed cycles determined by
$\omega$ via the equalities (\ref{I.032W}).  Note that the right hand side of (\ref{Trace.032}) is a finite sum, as $W_{4n}$ is a finite set.

\vskip 0.2in

 A key insight of the classic paper of \cite{BIZ} was
to realize that counting the Wick couplings could be reduced to
listing all possible diagrams (as defined in between (\ref{Formal1})
and (\ref{genfcn}) above). To see this one begins with a collection of $n$ 4-valent
vertex configurations labelled as in Figure \ref{Vertex}.  
A given coupling determines a unique way of
connecting the outgoing and incoming rays of each of the $n$ 4-valent
vertices, to form ``roads''.  The result of such
a glueing will be a diagram.  Indeed, if we interpret each road in
Figure \ref{Vertex} as an edge, and each edge is labelled by the
``outgoing'' side of the road, we clearly have a diagram.  (The
``faces'' of the diagram correspond 1-1 to the closed index cycles of
that coupling.)  Conversely, given a diagram as defined, which
includes the labellings, we can read off directly, from the way
the incident edges are paired, what the couplings must be. This
gives a bijection between diagrams and Wick couplings.

\vskip 0.1in
\font\thinlinefont=cmr5
\centerline{\beginpicture
\setcoordinatesystem units <.600000cm,.600000cm>
\unitlength=1.00000cm
\linethickness=1pt
\setplotsymbol ({\makebox(0,0)[l]{\tencirc\symbol{'160}}})
\setshadesymbol ({\thinlinefont .})
\setlinear
%
%
\linethickness= 0.500pt
\setplotsymbol ({\thinlinefont .})
\ellipticalarc axes ratio  0.635:0.635  360 degrees 
	from  8.255 13.970 center at  7.620 13.970
%
%
\linethickness= 0.500pt
\setplotsymbol ({\thinlinefont .})
\ellipticalarc axes ratio  0.635:0.635  360 degrees 
	from  2.540 13.970 center at  1.905 13.970
%
%
\linethickness= 0.500pt
\setplotsymbol ({\thinlinefont .})
\ellipticalarc axes ratio  0.635:0.635  360 degrees 
	from 13.970 13.970 center at 13.335 13.970
%
%
\linethickness= 0.500pt
\setplotsymbol ({\thinlinefont .})
%
%
\plot  0.385 13.610  0.131 13.547  0.385 13.483 /
\putrule from  0.131 13.547 to  1.433 13.547
%
%
\linethickness= 0.500pt
\setplotsymbol ({\thinlinefont .})
%
%
\plot  1.491 13.298  1.425 13.551  1.364 13.296 /
\plot  1.425 13.551  1.439 12.232 /
%
%
\linethickness= 0.500pt
\setplotsymbol ({\thinlinefont .})
\putrule from  0.127 14.383 to  1.429 14.383
%
%
\plot  1.175 14.319  1.429 14.383  1.175 14.446 /
%
%
%
\linethickness= 0.500pt
\setplotsymbol ({\thinlinefont .})
%
%
\plot  1.487 15.442  1.420 15.695  1.360 15.440 /
\plot  1.420 15.695  1.435 14.376 /
%
%
\linethickness= 0.500pt
\setplotsymbol ({\thinlinefont .})
\putrule from  2.381 14.383 to  3.683 14.383
%
%
\plot  3.429 14.319  3.683 14.383  3.429 14.446 /
%
%
%
\linethickness= 0.500pt
\setplotsymbol ({\thinlinefont .})
\plot  2.364 15.699  2.379 14.381 /
%
%
\plot  2.313 14.634  2.379 14.381  2.440 14.635 /
%
%
%
\linethickness= 0.500pt
\setplotsymbol ({\thinlinefont .})
%
%
\plot  2.629 13.610  2.375 13.547  2.629 13.483 /
\putrule from  2.375 13.547 to  3.677 13.547
%
%
\linethickness= 0.500pt
\setplotsymbol ({\thinlinefont .})
\plot  2.366 13.532  2.381 12.213 /
%
%
\plot  2.315 12.466  2.381 12.213  2.442 12.468 /
%
%
%
\linethickness= 0.500pt
\setplotsymbol ({\thinlinefont .})
\plot  8.100 15.699  8.115 14.381 /
%
%
\plot  8.049 14.634  8.115 14.381  8.176 14.635 /
%
%
%
\linethickness= 0.500pt
\setplotsymbol ({\thinlinefont .})
%
%
\plot  7.172 15.442  7.106 15.695  7.045 15.440 /
\plot  7.106 15.695  7.120 14.376 /
%
%
\linethickness= 0.500pt
\setplotsymbol ({\thinlinefont .})
\putrule from  8.117 14.383 to  9.419 14.383
%
%
\plot  9.165 14.319  9.419 14.383  9.165 14.446 /
%
%
%
\linethickness= 0.500pt
\setplotsymbol ({\thinlinefont .})
\putrule from  5.821 14.383 to  7.123 14.383
%
%
\plot  6.869 14.319  7.123 14.383  6.869 14.446 /
%
%
%
\linethickness= 0.500pt
\setplotsymbol ({\thinlinefont .})
%
%
\plot  6.111 13.610  5.857 13.547  6.111 13.483 /
\putrule from  5.857 13.547 to  7.159 13.547
%
%
\linethickness= 0.500pt
\setplotsymbol ({\thinlinefont .})
%
%
\plot  7.223 13.298  7.156 13.551  7.096 13.296 /
\plot  7.156 13.551  7.171 12.232 /
%
%
\linethickness= 0.500pt
\setplotsymbol ({\thinlinefont .})
\plot  8.090 13.532  8.105 12.213 /
%
%
\plot  8.038 12.466  8.105 12.213  8.165 12.468 /
%
%
%
\linethickness= 0.500pt
\setplotsymbol ({\thinlinefont .})
%
%
\plot  8.365 13.610  8.111 13.547  8.365 13.483 /
\putrule from  8.111 13.547 to  9.413 13.547
%
%
\linethickness= 0.500pt
\setplotsymbol ({\thinlinefont .})
\plot 13.803 15.699 13.818 14.381 /
%
%
\plot 13.751 14.634 13.818 14.381 13.878 14.635 /
%
%
%
\linethickness= 0.500pt
\setplotsymbol ({\thinlinefont .})
\putrule from 13.822 14.383 to 15.124 14.383
%
%
\plot 14.870 14.319 15.124 14.383 14.870 14.446 /
%
%
%
\linethickness= 0.500pt
\setplotsymbol ({\thinlinefont .})
%
%
\plot 14.059 13.610 13.805 13.547 14.059 13.483 /
\putrule from 13.805 13.547 to 15.107 13.547
%
%
\linethickness= 0.500pt
\setplotsymbol ({\thinlinefont .})
\plot 13.813 13.532 13.828 12.213 /
%
%
\plot 13.762 12.466 13.828 12.213 13.889 12.468 /
%
%
%
\linethickness= 0.500pt
\setplotsymbol ({\thinlinefont .})
%
%
\plot 11.815 13.610 11.561 13.547 11.815 13.483 /
\putrule from 11.561 13.547 to 12.863 13.547
%
%
\linethickness= 0.500pt
\setplotsymbol ({\thinlinefont .})
%
%
\plot 12.915 13.298 12.848 13.551 12.788 13.296 /
\plot 12.848 13.551 12.863 12.232 /
%
%
\linethickness= 0.500pt
\setplotsymbol ({\thinlinefont .})
\putrule from 11.557 14.383 to 12.859 14.383
%
%
\plot 12.605 14.319 12.859 14.383 12.605 14.446 /
%
%
%
\linethickness= 0.500pt
\setplotsymbol ({\thinlinefont .})
%
%
\plot 12.919 15.442 12.852 15.695 12.792 15.440 /
\plot 12.852 15.695 12.867 14.376 /
%
%
\put{$1$} [lB] at  1.736 13.805
%
%
\put{$i_1$} [lB] at  3.175 14.603
%
%
\put{$j_1$} [lB] at  3.313 13.001
%
%
\put{$j_1$} [lB] at  2.551 12.347
%
%
\put{$k_1$} [lB] at  0.889 12.347
%
%
\put{$k_1$} [lB] at  0.159 12.90
%
%
\put{$\ell_1$} [lB] at  0.032 14.610
%
%
\put{$\ell_1$} [lB] at  0.889 15.307
%
%
\put{$i_1$} [lB] at  2.455 15.307
%
%
\put{$\cdots$} [lB] at  9.980 13.801
%
%
\put{$2$} [lB] at  7.535 13.785
%
%
\put{$n$} [lB] at 13.250 13.785
%
%
\put{$\ell_2$} [lB] at  6.503 15.307
%
%
\put{$i_2$} [lB] at  8.20 15.307
%
%
\put{$i_2$} [lB] at  8.904 14.613
%
%
\put{$j_2$} [lB] at  8.915 13.01
%
%
\put{$j_2$} [lB] at  8.2580 12.447
%
%
\put{$k_2$} [lB] at  6.503 12.447
%
%
\put{$k_2$} [lB] at  5.694 13.0155
%
%
\put{$\ell_2$} [lB] at  5.620 14.610
%
%
\put{$\ell_n$} [lB] at 12.228 15.307
%
%
\put{$i_n$} [lB] at 13.889 15.307
%
%
\put{$\ell_n$} [lB] at 11.4 14.673
%
%
\put{$k_n$} [lB] at 11.494 13.031
%
%
\put{$k_n$} [lB] at 12.1045 12.447
%
%
\put{$j_n$} [lB] at 13.9893 12.447
%
%
\put{$i_n$} [lB] at 14.694 14.673
%
%
\put{$j_n$} [lB] at 14.694 13.031
\linethickness=0pt
\putrectangle corners at  0.032 15.725 and 15.149 12.188
\endpicture}
\vskip 0.2in
\centerline{Figure \refstepcounter{equation}\theequation\label{Vertex}.}

The full  count of non-vanishing contributions is then given by summing over diagrams with
a weighting, $N^{F(\omega)}$, that counts the number of configurations that 
give rise to a nonvanishing contribution in the coupling, $\omega$, 
associated to this diagram. Also the exponent of the weighting, $F(\omega)$,
which was defined to be the number of closed index cycles in the coupling 
is now also seen to be the number of faces, $F$, in the associated diagram. 
The weighting factor is $N^{F}$. If the diagram is connected (i.e., if it is a map), this can  be expressed in terms of the genus of the Riemann surface that
the diagram maps to by using Euler's formula. Since the number of vertices
of a diagram is $n$ and the number of edges is $2n$ (= the number of couples
in a Wick coupling), we have $2-2g = n -2n + F$ from which we deduce that the
weighting factor for a 4-valent, n-vertex, g-map is $N^{2-2g+n}$. We can 
directly extend this to the case of {\it disconnected} diagrams since the
Euler characteristic is additive with respect to disjoint unions. This 
generalization results in the possibility that $g$ can become negative. 
Thus, to each diagram we can associate a unique integer $g$ and then we will
refer to that diagram as a $g$-diagram. 

The beautiful connection between the quantity $ \left< ( \mbox{Tr } M^{4} )^{n} \right>$ and the combinatorics of g-diagrams described in \cite{BIZ} (and discussed above) is summarized with the following formula:
\begin{equation} \label{genfcnA}
\left< \left( \mbox{Tr } M^{4}\right)^{n} \right> = 
\sum_g \#\{\mbox{4-valent, n-vertex, g-diagrams}\}N^{2-2g+n}.
\end{equation}
We remark that this is a finite sum, and the reader may easily verify
that the nonzero contributions to this sum come from $ 1-n \le g \le
[(n+1)/2]$, where $[\ell]$ is the closest integer to $\ell$ less than or
equal to $\ell$.  A straightforward extension of the above analysis to the case
of mixed  moments shows that

\begin{equation} \label{mixedgenfcnA}
\left< \prod_{j=1}^\nu \left( \mbox{Tr } M^{j}\right)^{n_j} \right> = 
\sum_g \#\{\mbox{g-diagrams with $n_j$ $j$-valent vertices; 
$j = 1,\dots,\nu$}\}N^{2-2g+\sum_{j=1}^\nu (\frac{j}{2}-1) n_j}.
\end{equation}
(Again we observe that this is a finite sum.)

\subsection{Asymptotics of the partition function and enumeration of g-maps.}

We turn now to the proof of Theorem \ref{GeomAsymp}.  The starting point
is the asymptotic expansion in Theorem \ref{I.002thm}, rewritten in the
form 
$$
\ \ \ \log{ \left(\hat{Z}_{N} \right)} - N^{2}
e_{0}({\bf t}) - e_{1}({\bf t}) = \sum_{g=2}^\infty e_g({\bf t}) 
\left(\frac{1}{N}\right)^{2g-2}.
$$
This is, as usual, shorthand:  the right hand side is an asymptotic
expansion in {\it inverse} powers of $N^{2}$.
Exponentiating both sides of this equation we deduce
that $Z_{N}e^{-N^{2} e_{0} - e_{1}}$ has an asymptotic expansion in
inverse powers of $N^{2}$.  Some algebraic manipulation shows that  

\begin{equation} \label{Zhat}
\hat{Z}_{N} = \exp\left(N^2 e_0 + e_1\right) \left\{1 + \sum_{h=1}^\infty 
 \rho_h \left(\frac{1}{N^2}\right)^h \right\}
\end{equation}
where
$$
\rho_h = \sum_{k=1}^h \frac{1}{k!} \sum_{\ell_1 + \ldots +\ell_k = h, 
\ell_j > 0} \prod_{j=1}^k e_{\ell_j +1}.
$$ 
Formula (\ref{Zhat}) may appear to be nothing more than a simple
calculational re-representation of $\hat{Z}_{N}$.  However (\ref{Zhat})
explains concretely in what sense $\hat{Z}_{N}$ itself possesses an
asymptotic representation.

It follows from Theorem \ref{I.002thm} that the 
asymptotic series on the right hand side of (\ref{Zhat}) can be differentiated 
term-by-term for ${\bf{t}} \in \mathbb{T}(T,\gamma)$, and the 
ensuing derivatives may be evaluated at $\bf{t}=0$.
Having established this we now rewrite (\ref{Zhat}) as 

\begin{equation} \label{ZHAT}
\hat{Z}_{N}({\bf t}) = \exp\left(N^2 e_0({\bf t}) + e_1({\bf t}) 
+ \frac{1}{N^2} e_2({\bf t})  + \dots + \frac{1}{N^{2h-2}} e_h({\bf t}) 
+ \dots \right).
\end{equation}
In light of the above discussion, it is clear what is meant by this
asymptotic expansion, which is valid for ${\bf t}$ satisfying the regularity 
conditions of Theorem \ref{I.002thm}.

Differentiating (\ref{ZHAT}) $n_j$ times in each $t_j$ and
taking the limit $\bf{t} \to 0$, we may now conclude, by comparison with
(\ref{mixedgenfcnA}), that

\begin{eqnarray} \label{Valid1}
 & & 
\left.
 \frac{\partial^n}{\partial t_1^{n_1}\dots\partial t_\nu^{n_\nu}}
 Z_{N}({\bf t}) \right|_{{\bf t} = {\bf 0}} = 
 \left< \prod_{j=1}^\nu \left( \mbox{Tr } M^{j}\right)^{n_j} 
\right> N^{-\sum_{j=1}^\nu (\frac{j}{2}-1) n_j} \\
\nonumber
& & =  \sum_g
\#  \{\mbox{g-diagrams  with $n_j$ $j$-valent vertices; 
$j = 1,\dots,\nu$}\} N^{2-2g}
\\
\nonumber
& &  = 
(-1)^n \frac{\partial^n}{\partial t_1^{n_1}\dots\partial t_\nu^{n_\nu}}
\left[\exp \left(\sum_{g=0}^\infty N^{2-2g}
e_{g}({\bf t})\right)\right]_{\bf{t}=0},
\end{eqnarray}
where $n = n_1 + \dots + n_\nu$.  

The string of equalities in
(\ref{Valid1}) states that the finite sum over $g$ in the second line
possesses an asymptotic expansion in inverse powers of $N^{2}$.  We
learn that all of the coefficients in this final asymptotic expansion must
vanish from some point on.  Much more information can be obtained in
this manner from (\ref{Valid1}).  For example, by examining (\ref{Valid1})
for the case of a single derivative, one may verify that for $\ell$ even,
\begin{eqnarray}
\label{1Vert.24}
& &
\frac{\partial}{\partial t_{\ell}} e_{0}({\bf 0}) = -\# \left\{ \mbox{
genus 0 diagrams with 1 $\ell$-valent vertex} \right\} \\
\label{1Vert.25}
& &
\frac{\partial}{\partial t_{\ell}} e_{1}({\bf 0}) = -\# \left\{
\mbox{ genus 1 diagrams with 1 $\ell$-valent vertex} \right\} \\
\label{1Vert.26}
& & \frac{\partial}{\partial t_{\ell}} e_{g}({\bf 0}) = 0 \ \ \ \mbox{ for } g
> \frac{1}{2}\left\{ \begin{array}{cc}
\ell/2 & \mbox{ if $\ell/2$ is even} \\
\ell/2 - 1 & \mbox{ if $\ell/2$ is odd} \\
\end{array} \right.  .
\end{eqnarray}
We observe that in (\ref{1Vert.24}) and (\ref{1Vert.25}), the
diagrams have a single vertex, and so we may say that (\ref{1Vert.24})
and (\ref{1Vert.25}) enumerate 1-vertex g-maps.

Proceeding further, one may examine (\ref{Valid1}) for the case of
higher order derivatives.  Tedious calculations may be carried out which
show that the $e_{g}$'s are generating functions for g-maps (they
enumerate those diagrams that are connected).  However, this approach is
quite complicated, and we adopt a different approach, which is 
to use (\ref{Valid1}) to discuss
the relation between the logarithm of the partition function 
and connected diagrams, or g-maps, that was alluded to in (\ref{logconn}).
For the sake of clarity we will present this just in the case of 
${\bf t} = (0,0,0,t_{4}=t,0,\ldots,0)$; it will be evident from the
discussion that the extension to more general
valences is straightforward.

We observe that, from our earlier discussion of the relation between
diagrams and Wick orderings, that:

\begin{eqnarray} 
\label{QNDEF}
\sum_g
&\# & \{\mbox{4-valent, n-vertex, g-diagrams}\} N^{2-2g} 
\, \, = \, \, \sum_{\omega \in W_{4n}} N^{\chi(\omega)} 
:= Q_n(N).
\end{eqnarray}
$Q_n(N)$ is a Laurent polynomial which we will refer to as the {\it diagram
polynomial} for 4-valent, n-vertex diagrams. We introduce an {\it exponential 
generating function} \cite{Harary} for diagram polynomials:
\begin{eqnarray}
\label{ExpGFII}
& & 
G(t) := \sum_{k=1}^\infty \frac{t^k}{k!} Q_k(N).
\end{eqnarray}
This is a formal generating function in the same sense as was discussed for
(\ref{genfcn}).  It is by no means a convergent series, but rather is
interpreted in the sense that one is allowed to evaluate finitely many
derivatives of both sides at $t=0$.  (Below we will indicate how to
remove the formality of these arguments by truncating the series.)   We
also introduce an exponential 
generating function for {\it connected} 4-valent, diagrams (i.e. maps):
$$
C(t) := \sum_{k=1}^\infty \frac{t^k}{k!} P_k(N),
$$
where 
\begin{eqnarray}
\label{PNDEF}
P_n(N) = \sum_{\omega \in W_{4n}^{conn}} N^{\chi(\omega)} 
\end{eqnarray}
and $W_{4n}^{conn}$ is defined to be the subset of $W_{4n}$ consisting of Wick 
couplings that correspond to connected diagrams.
We observe that 

\begin{eqnarray*}
C(t)C(t) &=& \sum_{n=2}^\infty \sum_{k + \ell = n} \frac{t^n}{k! \ell !} 
P_k(N)P_\ell(N) \\
&=& \sum_{n=2}^\infty \frac{t^n}{n!} \sum_{k=1}^{n-1} \left(\begin{array}{c}
n \\ k\end{array}\right) P_k(N)P_{n-k}(N).
\end{eqnarray*}
This series is a generating function for diagram polynomials of ordered pairs
$(\omega_1, \omega_2)$ of connected Wick couplings. There is a geometrical 
explanation for the appearance of the  combinatorial coefficient 
$\left(\begin{array}{c} n \\ k\end{array}\right)$  in this expression.  It is a
consequence of the fact that the labels $1, \dots n$ get distributed over
$\omega_1 \cup \omega_2$ and if $|\omega_1| = k$ there are 
$\left(\begin{array}{c} n \\ k\end{array}\right)$ different ways of doing this.

Next, one may prove that $C(t) C(t) /2$ is the exponential generating
function for 4-valent diagrams which consist of exactly two connected
components.  In other words, one may define an exponential generating
function for 4-valent diagrams consisting of exactly two connected
components, and prove that this generating function coincides,
term-by-term, with $C(t) C(t)/2$.  Subsequently, and in the same way, 
one may prove that $C^{d}(t)/d!$ is the exponential generating function 
for 4-valent, n-vertex diagrams with exactly $d$ connected components.
One may finally conclude that, in the sense of formal generating functions,
\begin{eqnarray} \label{Valid2}
G(t) &=&
\sum_{d=1}^\infty \frac{1}{d!} C^d(t) \ \ \ = 
\exp\left( C(t) \right) - 1.
\end{eqnarray}

To get around the formality of these generating functions, we 
repeat most of the argument from (\ref{ExpGFII}) to (\ref{Valid2}),
truncating the infinite series involved.  So we define
\begin{eqnarray}
\label{TruncEGF}
G_{m}(t) = \sum_{k=1}^{m} \frac{t^{k}}{k!} Q_{k}(N), & & 
 C_{m}(t) = \sum_{k=1}^{m} \frac{t^{k}}{k!} P_{k}(N), 
\end{eqnarray}
with $Q_{k}$ and $P_{k}$ defined in (\ref{QNDEF}) and (\ref{PNDEF}),
respectively.  Next we have that
\begin{eqnarray}
\frac{1}{2} C_{m}(t)^{2} = \sum_{k=2}^{m} \frac{t^{k}}{k!}
P_{k}^{(2)}(N) + {\mathcal{O}}\left( t^{m+1} \right),
\end{eqnarray}
where the sum on the right hand side is the truncated exponential
generating function for 4-valent diagrams consisting of exactly two
connected components (i.e. $P_{k}^{(2)}(N)$ is the diagram polynomial
for 4-valent, k-vertex diagrams with two connected components).
Similarly, the reader may verify that 
\begin{eqnarray}
\frac{1}{d!} C_{m}(t)^{d} = \sum_{k=d}^{m}  \frac{t^{k}}{k!}
P_{k}^{(d)}(N) + {\mathcal{O}} \left( t^{m+1} \right),
\end{eqnarray}
where on the right hand side, $P_{k}^{(d)}(N)$ is the diagram polynomial
for 4-valent, k-vertex diagrams with d connected components.
Putting this all together, we have shown that
\begin{eqnarray}
G_{m}(t) = e^{C_{m}(t)} + {\mathcal{O}}\left( t^{m+1} \right).
\end{eqnarray}
And since $m$ was arbitrary, this is the proper sense in which one may
interpret (\ref{Valid2}).

\smallskip
We have established the following set of relations:
\begin{eqnarray}
& & 1 + G_{m}(-t) = Z_{N}(t) + {\mathcal{O}}\left(t^{m+1} \right), \\
& & \exp{ C_{m}(-t)} = 1 + G_{m}(-t) + {\mathcal{O}}\left( t^{m+1} \right).
\end{eqnarray}
We therefore have
\begin{eqnarray}
C_{m}(-t) = \log{ \left( Z_{N} + {\mathcal{ O}}\left( t^{m+1} \right)
\right) }. 
\end{eqnarray}
We may now differentiate this $n$ times (for  $n \le m$) to obtain
\begin{eqnarray}
\left. \frac{\partial^{n}}{\partial t^{n}} C_{m}(-t) \right|_{t=0} =
P_{n}(N) = 
\left. \frac{\partial^{n}}{\partial t^{n}} \log{Z_{N}(t)} \right|_{t=0}.
\end{eqnarray}
So we have proven that $\log{Z_{N}}$ is the generating function for {\it
connected} diagrams.  Now using our asymptotic expansion for
$\log{Z_{N}}$, we have 
\begin{eqnarray}
\left.
(-1)^{n} \frac{\partial^{n}}{\partial t^{n}} \sum_{g=0}^{\infty} N^{2 -
2g} e_{g}(t) \right|_{t=0} & = & P_{n}(N) \\
\nonumber
& = &\sum_{\omega \in
W_{4n}^{\mbox{\tiny{conn}}} } N^{\chi(\omega)} = \sum_{g} N^{2 - 2g} \#
\{ \mbox{4-valent, n-vertex, g-maps} \}.
\end{eqnarray}
We have now shown that 
$e_g(t) = E_g(t)$ in the sense that 
$$
(-1)^n \frac{\partial^n}{\partial t^n} e_g(0)
=
\#  \{\mbox{4-valent, n-vertex, g-maps}\}. 
$$
By similar reasoning one can fully establish Theorem \ref{GeomAsymp},
and we will not present these details here.

We refer the reader to \cite{Harary} for a fuller discussion of the
general cumulant relations as they pertain to graphical enumeration.

\section{Asymptotics for orthogonal polynomials via Riemann--Hilbert
methods} \label{Sec:2}

The amazing fact, due to Gaudin and Mehta \cite{GauMeh}, is that
the mean density $\rho_{N}^{(1)}$ can be expressed in closed form in
terms of the orthogonal polynomials $\set{p_{j}(\cdot ; N,{\bf t}\
)}$.
  \begin{eqnarray}
  \label{II.001}
& & \rho_N^{(1)}(\lam ) =
\frac{1}{N}e^{-NV(\lam )} \sum^{N-1}_{k=0} p_{k}(\lam )^2\\
& & \hspace{0.5in} = \frac{1}{N}e^{-NV} \Bigl[ p'_{N}(\lambda)\,
p_{N-1}(\lambda) - p_{N}(\lambda)\, p'_{N-1}(\lambda)\Bigr]
\frac{\gam^{(N)}_{N-1}}{\gam_N^{(N)}} .
  \end{eqnarray}
The second equality above follows from the first because of the
Christoffel-Darboux formula.  (See, for example, formula (3.2.3) in
\cite{Szego}.)

We obtain global asymptotics for $\rho_{N}^{(1)}$ by carrying out to
higher order the results of \cite{DKMVZ1}-\cite{DKMVZ3}.  Deift,
Kriecherbauer, McLaughlin, Venakides, and Zhou established very
detailed Plancherel-Rotach type asymptotics for polynomials
orthogonal with respect to $e^{-NV(x)}dx$, under some very general
assumptions on the function $V$, including the functions $V$
considered here. In \cite{DKMVZ2}, the authors demonstrated that
their method can be used to obtain a complete asymptotic expansion
for the polynomials. We have carried this procedure out to higher
order.

Using the global asymptotic expansion of $\rho_{N}^{(1)}$, together with (\ref{I*5}),
we deduce an asymptotic expansion for $\log{  Z_N({\bf t}\, )}$.

In this section we review the Riemann--Hilbert approach for
computing and establishing a global, uniform asymptotic expansion
for the orthogonal polynomials.  In \cite{DKMVZ1}-\cite{DKMVZ3},
the authors began with a matrix valued function $Y(z)$, defined by
  \[
Y = \begin{pmatrix} \frac{1}{\gam^{(N)}_N}\, p_{N} &
\frac{1}{2\pi_i\gam_{N}^{(N)}} \int \frac{p_{N}(s)e^{-NV(s)}}{s-z} ds\\
-2\pi i\gam^{(N)}_{N-1} p_{N-1} & -\gam_{N-1}^{(N)} \int
\frac{p_{N-1}(s)\, e^{-NV(s)}}{s-z} ds \end{pmatrix},
z\in\Cbb\backslash \Real .
  \]
The matrix $Y$ solves the following Riemann--Hilbert problem:
  \begin{itemize}
\item $Y$ analytic in $\Cbb\backslash\Real$,
\item $Y = \bigl( I + \, \Ocal (\frac{1}{z})\bigr)\, z^{N\sigma_3}$,
\item $Y$ has H\"older continuous boundary values $Y_\pm$ for $z\in\Real$,
\item $Y_{+} = Y_{-} \Bigl( \begin{smallmatrix} 1 & e^{-NV}\\ 0 &
1\end{smallmatrix}\Bigr)$, for $z \in \mathbb{R}$ .
  \end{itemize}
The authors then make two explicit transformations: $Y\to M\to M_1$.  Here
we will explain these two transformations.

\subsection{Equilibrium measure and first transformation
$\mathbf{Y}\mathbf{\to}\,\mathbf{M}$}.  A fundamental object in
the theory of random matrices, as well as approximation theory, is
the equilibrium measure, defined as follows:
  \[
\sup_{\text{Borel measures  }\mu ,\mu\geq 0, \int d\mu =1} \left[
-\int V(\lam ) d\mu (\lam ) +\int\int \log | \lam -\mu |\, d\mu
(\lam )\, d\mu (\eta )\right] .
  \]
There is a vast literature on this variational problem,
originating in the work of Gauss (see \cite{SaffTotik}).  Under general assumptions on $V$, the supremum is achieved
at a unique measure $\mu_{V}$, called the equilibrium measure. For
the external field $V$, considered here, it is a well-known fact
(see \cite{DKM}, or \cite{Difrancesco95}) that the equilibrium
measure is supported on finitely many intervals, with density that
is analytic on the interior of each interval, behaving at worst
like a square root at each endpoint.

We remind the reader of the definition of the set $\mathbb{T}(T,\gamma)$:
\begin{eqnarray}
& &
\mathbb{T}(T,\gamma) = \{ {\bf t} \in \mathbb{R}^{\nu}: |{\bf t} |
\le T, \ t_{\nu} > \gamma \sum_{j=1}^{\nu-1} |t_{j}| \} 
\end{eqnarray}

The following theorem is perhaps well understood.  A proof can be
constructed by combining the results of \cite{DKM} and \cite{KuijMcL1}.

\begin{thm}\label{EQMSTHM}  There is $T_{0}>0$  and $\gamma_{0} > 0$ 
such that for all $0 < T
< T_{0}$ and $\gamma > \gamma_{0}$, the following holds true: If ${\bf t} \in
\mathbb{T}(T,\gamma)$, then
  \begin{align*}
d\mu_{V} &= \psi \, d\lam ,\\
\psi (\lam ) &= \frac{1}{2 \pi}
\chi_{(\al ,\beta )} (\lam ) \sqrt{(\lam -\al )(\beta
-\lam )}\, h(\lam ),
  \end{align*}
where $h(\lam )$ is a polynomial of degree $\nu -2$, which is
strictly positive on the interval $[\alpha,\beta]$ (recall that the
external field $V$ is a polynomials of degree $\nu$). The polynomial
$h$ is defined by
  \[
h(z) = \frac{1}{2\pi i}\, \oint\, \frac{V'(s)}{\sqrt{(s-\al )}
\sqrt{(s-\beta )}} \, \frac{ds}{s-z}
  \]
where the integral is taken on a circle containing $(\al ,\beta )$ and $z$
in the interior, oriented counter-clockwise.

The endpoints $\al$ and $\beta$ are determined by the equations
  \begin{align*}
\int^{\beta}_{\al}\, \frac{V'(s)}{\sqrt{(s-\al )(\beta -s)}}\, ds &= 0\\
 \int^{\beta}_{\al}\, \frac{sV'(s)}{ \sqrt{(s-\al )(\beta -s)} }\, ds &=
2\pi .
  \end{align*}
  \end{thm}

The endpoints $\al ({\bf t}\, )$ and $\beta ({\bf t}\,
)$ are analytic functions of ${\bf t}$ which possess smooth extensions
to the closure of $\mathbb{T}(T,\gamma)$, and satisfy $-\alpha({\bf 0}) = \beta({\bf 0}) = 2$.
In addition, the coefficients of the
polynomial $h(\lam )$ are also analytic functions of ${\bf t}$ with
smooth extensions to $\overline{\mathbb{T}(T,\gamma)}$, with
  \[
h(\lam ,{\bf t}={\bf 0}) = 1.
  \]

In this paper we will assume that ${\bf t} \in \mathbb{T}(T, \gamma)$, for $T$
small enough, and $\gamma$ large enough, so that Theorem \ref{EQMSTHM} holds true.  In Section \ref{sec:MDOS} we define $T_{\beta}$ and $\gamma_{\beta}$ (as well as $T_{\alpha}$ and $\gamma_{\alpha}$)  so that if $T < \mbox{min}\{ T_{\beta}, T_{\alpha}\}$ and $\gamma > \mbox{max} \{\gamma_{\alpha}, \gamma_{\beta}\}$ then ${\bf t} \in \mathbb{T}(T,\gamma)$ implies more:  the neighborhoods of $z=\beta$ and $z=\alpha$ defined in this section can be taken so large that their intersection contains a subinterval of the real axis. We will not
present a proof of Theorem \ref{EQMSTHM} here, but the basic idea is as follows:
for ${\bf t} = {\bf 0}$, the equilibrium
measure is explicitly known, $\psi(\lambda) = \chi_{(-2,2)}(\lambda)
\sqrt{4 - \lambda^{2}}/(2 \pi)$.  Taking $T$ small
implies that the endpoints of the support cannot stray too far from $\pm
2$; taking $t_{\nu}$ larger than $\gamma \sum_{j=1}^{\nu - 1}|t_{j}|$
implies that $h$ should be strictly positive on $\mathbb{R}$.

\bigskip
It will prove useful to adapt the following alternative
representation for the function $\psi$:
\begin{eqnarray} \label{psi}
\psi(\lambda) = \frac{1}{2 \pi i} R_{+}(\lambda) h(\lambda), \ \lambda
\in (\alpha, \beta),
\end{eqnarray}
where the function $R(\lambda)$ is defined via $R(\lambda)^{2} = (
\lambda - \alpha) ( \lambda - \beta)$, with $R(\lambda)$ analytic in
$\cb \setminus [\alpha, \beta]$, and normalized so that $R( \lambda)
\sim \lambda$ as $\lambda \to \infty$.  The subscript $\pm$ in $R_{\pm}(\lambda)$
denotes the boundary value obtained from the upper (lower) half plane.

\vskip 0.25in

Following \cite{DKMVZ1}, we define a function $g(z)$ as follows,
\begin{eqnarray}
\label{GFdef}
g(z) = \int^{\beta}_{\al} \log (z-s)\, \psi (s)ds, \quad
z\in\Cbb\backslash (-\infty ,\beta ],
\end{eqnarray}
where for each $s$, $\log (z-s)$ is chosen to be branched along $(-\infty
,s)$, $\log (z-s)>0$ if $z>s$.

We now list several important properties of the function $g(z)$:
  \renewcommand{\theenumi}{\arabic{enumi}.}
  \begin{enumerate}
\item[$\bullet$ 1] $g$ is analytic for $z\in\Cbb\backslash (-\infty ,\beta ]$, with
continuous boundary values $g_{\pm}(z)$ on $(-\infty ,\beta ]$ .
\item[$\bullet$ 2] There is a constant $\ell$ such that for $z\in [\al ,\beta ]$,
  \[  g_+ + g_- - V(z) = \ell, \]
and for $z\in\Real\backslash [\al ,\beta ]$,
  \[  g_+ +g_- - V < \ell . \]
  \end{enumerate}
Indeed, for $z>\beta$,
\begin{eqnarray}
\label{II.002}
g_+ + g_- -V-\ell = 2 g - V - \ell = -\int^z_{\beta} Rhds,
\end{eqnarray}
and for $z<\al$,
  \[  g_+ + g_- -V-\ell = - \int^z_{\al} Rhds.  \]
  \begin{enumerate}
\item[$\bullet$ 3] For $z\in [\al ,\beta ]$,
  \[  g_+ - g_- = 2\pi i\int_z^{\beta} \psi (s)\, ds , \]
and this function possesses an analytic continuation to a
neighborhood of $(\al ,\beta )$.
\item[$\bullet$ 4] For $z>\beta$,
\[g_+-g_- =0, \]
and for $z<\al$, \[g_+ -g_- =2\pi i.\]
  \end{enumerate}

 Using the function $g(z)$, and the constant $\ell$, we define
(cf. Theorem 5.57 in \cite{DKMVZ2}) $M(z)$ via
  \[
Y(z) = e^{N \frac{\ell}{2} \sigma_3}\, M(z)\,
e^{N(g-\frac{\ell}{2})\sigma_3}, \quad z\in\Cbb\backslash\Real .
  \]

\subsection{Second Transformation $\mathbf{M}\mathbf{\to}\mathbf{M_1}$}

Next, we define (cf. Section 6 of \cite{DKMVZ2}) a lens-shaped
region around $(\al ,\beta )$ (see Figure \ref{KF2})

\vskip0.2cm
\font\thinlinefont=cmr5
\centerline{\beginpicture
\setcoordinatesystem units <.5cm,.5cm>
\linethickness=1pt
\setshadesymbol ({\thinlinefont .})
\setlinear
%
%
\linethickness= 0.500pt
\setplotsymbol ({\thinlinefont .})
\circulararc 87.206 degrees from 19.050 16.510 center at 12.700  9.842
%
%
\linethickness= 0.500pt
\setplotsymbol ({\thinlinefont .})
\circulararc 87.206 degrees from  6.350 16.510 center at 12.700 23.178
%
%
\linethickness= 0.500pt
\setplotsymbol ({\thinlinefont .})
\putrule from  2.540 16.510 to  6.350 16.510
%
%
\linethickness= 0.500pt
\setplotsymbol ({\thinlinefont .})
\putrule from 19.050 16.510 to 22.860 16.510
%
%
\linethickness= 0.500pt
\setplotsymbol ({\thinlinefont .})
\plot 12.700 19.050 12.383 19.209 /
%
%
\linethickness= 0.500pt
\setplotsymbol ({\thinlinefont .})
\plot 12.700 19.050 12.383 18.891 /
%
%
\linethickness= 0.500pt
\setplotsymbol ({\thinlinefont .})
\putrule from  6.350 16.510 to 19.050 16.510
%
%
\linethickness= 0.500pt
\setplotsymbol ({\thinlinefont .})
\plot 12.700 16.510 12.383 16.669 /
%
%
\linethickness= 0.500pt
\setplotsymbol ({\thinlinefont .})
\plot 12.700 16.510 12.383 16.351 /
%
%
\linethickness= 0.500pt
\setplotsymbol ({\thinlinefont .})
\plot 12.383 14.129 12.700 13.970 /
%
%
\linethickness= 0.500pt
\setplotsymbol ({\thinlinefont .})
\plot 12.700 13.970 12.383 13.811 /
%
%
\linethickness= 0.500pt
\setplotsymbol ({\thinlinefont .})
\plot  4.445 16.510  4.128 16.669 /
%
%
\linethickness= 0.500pt
\setplotsymbol ({\thinlinefont .})
\plot  4.445 16.510  4.128 16.351 /
%
%
\linethickness= 0.500pt
\setplotsymbol ({\thinlinefont .})
\plot 21.590 16.510 21.273 16.669 /
%
%
\linethickness= 0.500pt
\setplotsymbol ({\thinlinefont .})
\plot 21.590 16.510 21.273 16.351 /
%
%
\put{$\alpha$} [lB] at  5.800 15.875
%
%
\put{$\beta$} [lB] at 19.050 15.875
%
%
\put{$\Sigma _1$} [lB] at 15.240 18.891
%
%
\put{$\Sigma _2$} [lB] at 15.240 16.828
%
%
\put{$\Sigma _3$} [lB] at 15.240 14.828
%
%
\put{$\Sigma _5$} [lB] at 22.225 16.828
%
%
\put{$\Sigma _4$} [lB] at  3.175 16.828
%
%
\linethickness=0pt
\putrectangle corners at  2.515 19.837 and 22.885 13.786
\endpicture}
\vskip0.cm
\centerline{Figure \refstepcounter{equation}\theequation\label{KF2}.
The contour $\Sigma _S$.}
\vskip0.2cm

and define $M_1$ as follows:
  \begin{itemize}
\item For $z$ outside the upper and lower lenses,
\begin{eqnarray}
\label{M1DEF1}
 M_1 = M .
\end{eqnarray}
\item For $z$ within the upper lens,
\begin{eqnarray}
\label{M1DEF2}
M_1 = M \begin{pmatrix} 1 & 0\\ -e^{-N(g_+-g_-)} & 1\end{pmatrix}.
\end{eqnarray}
\item For $z$ in the lower lens,
\begin{eqnarray}
\label{M1DEF3}
M_1 = M \begin{pmatrix} 1 & 0\\ e^{N(g_+-g_-)} & 1\end{pmatrix}.
\end{eqnarray}
(Recall that $g_+ - g_-$ possesses an analytic continuation to such a
lens-shaped region around $(\al ,\beta )$.)
  \end{itemize}

With these two transformations, we arrive at a matrix-valued
function $M_1$, satisfying the following Riemann--Hilbert problem.

The problem is to determine a $2\times 2$ matrix valued function
$M_1$ satisfying
  \begin{enumerate}
\item $M_1$ analytic in $\Cbb\backslash \Sigma_{M_1}$.
\item $M_1 = I + \, \Ocal (\frac{1}{z})$, $z\to\infty$.
\item $M_1$ possesses continuous boundary values $(M_1)_{\pm}$ for
$z\in\Sigma_{M_1}$.
\item (Jump relation)\quad $(M_1)_{+} = (M_1)_{-} V_M$,
\end{enumerate}
where
\begin{eqnarray}
& & V_{M} = \pmtwo
{1} {e^{N ( g_{+} + g_{-} - V - \ell)}} {0} {1}, \ \mbox{ for } z \in
\Sigma_{4} \cup \Sigma_{5}, \\
& & V_{M} = \pmtwo
{1} {0}{e^{-N( g_{+} - g_{-})}}
{1}, \ \mbox{ for } z \in \Sigma_{1}, \\
& & V_{M} = \pmtwo
{1}                              {0}
{e^{N( g_{+} - g_{-})}}          {1}, \ \mbox{ for } z \in \Sigma_{3}, \\
& & V_{M} = \pmtwo
{0} {1}
{-1} {0}, \ \mbox{ for } z \in \Sigma_{2},
\end{eqnarray}

\vskip 0.3in

\subsection{The construction of a global approximation to $M_{1}$}

The next step in the Riemann Hilbert approach is to construct an
explicit approximation to $M_1$, which we will denote $M^{(A)}_1$.
The definition of $M_1^{(A)}$ also is adapted directly from
\cite{DKMVZ1}-\cite{DKMVZ3}. Let $B^{\al}_{\del}$ denote a disc of
radius $\del$ centered at $\al$, and let $B^{\beta}_{\del}$ denote
a disc of radius $\del$ centered at $\beta$.  We will define
$M_1^{(A)}$ outside $B^{\al}_{\del}\cup B^{\beta}_{\del}$ as
follows:
\begin{eqnarray}
\label{M1Ap1}
M_1^{(A)}(z) &=& \begin{pmatrix} \frac{\gam +\gam^{-1}}{2}
&\frac{\gam -\gam^{-1}}{2i}\\ \frac{\gam -\gam^{-1}}{-2i}
&\frac{\gam +\gam^{-1}}{2}
\end{pmatrix} , \\
\label{GAMDEF}
\gam (z) &=& \frac{(z-\beta )^{\frac{1}{4}}}{(z-\al )^{\frac{1}{4}}} .
\end{eqnarray}

To define $M_{1}^{(A)}(z)$ for $z$ within $B^{\beta}_{\del}$, we
require the following auxiliary function
\begin{eqnarray}
\label{Phich1a} & & \Phi_{\beta}(z)  \equiv   \left(\frac{3 N
}{4}\right)^{2/3} \left( -
2g(z) + V(z) + \ell  \right)^{2/3}, \ \ \ z \in B_{\del}^{\beta}.  
\end{eqnarray}
 As $h(\beta) \neq 0$,
$- \left( 2g(z) - V(z) - \ell \right)$ behaves like
$(z - \beta)^{3/2}$ for $z$ near $\beta$ (cf. relations (\ref{II.002})),
and we may choose the branch so that $\Phi_{\beta}(z) $
is analytic in a neighborhood of $\beta$, and
\begin{eqnarray}
\label{Phich1b} \\
\nonumber \Phi_{\beta}(z) = c (z - \beta) + O\left(\left( z -
\beta \right)^2 \right), \
c > 0, \ z \to \beta. \ 
\end{eqnarray}

Let the radius $\del$ of the disc $B_{\del}^{\beta}$ be chosen
sufficiently small so that $\Phi_{\beta}$ maps $B_{\del}^{\beta}$
injectively onto a neighborhood $D'$ of $0$. The decomposition of
$D_{\epsilon,\beta}$ into four regions is as in \cite{DKMVZ3}:
$D_{\epsilon,\beta} = \mbox{I} \cup \mbox{II} \cup \mbox{III} \cup
\mbox{IV} $, as shown in Figure \ref{FIG4} below.

\font\thinlinefont=cmr5 \centerline{\beginpicture
\setcoordinatesystem units <.75cm,.75cm> \linethickness=1pt
\setshadesymbol ({\thinlinefont .}) \setlinear
%
%
\linethickness= 0.500pt \setplotsymbol ({\thinlinefont .})
\ellipticalarc axes ratio  2.714:2.714  360 degrees
    from  7.794 16.510 center at  5.080 16.510
%
%
\linethickness= 0.500pt \setplotsymbol ({\thinlinefont .})
\ellipticalarc axes ratio  1.587:2.381  360 degrees
    from 14.287 16.510 center at 12.700 16.510
%
%
\linethickness= 0.500pt \setplotsymbol ({\thinlinefont .}) \plot
12.700 16.510 10.636 18.891 /
%
%
\linethickness= 0.500pt \setplotsymbol ({\thinlinefont .})
\putrule from 10.160 16.510 to 15.399 16.510
%
%
\linethickness= 0.500pt \setplotsymbol ({\thinlinefont .}) \plot
12.700 16.510 10.636 14.129 /
%
%
\linethickness= 0.500pt \setplotsymbol ({\thinlinefont .})
\putrule from  1.905 16.510 to  8.414 16.510 \linethickness=
0.500pt \setplotsymbol ({\thinlinefont .})
%
%
\plot  8.731 17.621      8.808 17.664
     8.880 17.703
     8.947 17.739
     9.010 17.771
     9.124 17.826
     9.223 17.869
     9.310 17.901
     9.388 17.923
     9.459 17.935
     9.525 17.939
     9.591 17.935
     9.662 17.923
     9.740 17.901
     9.827 17.869
     9.926 17.826
    10.040 17.771
    10.103 17.739
    10.170 17.703
    10.242 17.664
    10.319 17.621
    /
%
%
\plot 10.066 17.690 10.319 17.621 10.128 17.801 /
\linethickness= 0.500pt \setplotsymbol ({\thinlinefont .})
%
%
\plot  5.080 16.510      5.053 16.432
     5.028 16.357
     5.002 16.284
     4.978 16.214
     4.954 16.147
     4.931 16.082
     4.886 15.961
     4.843 15.848
     4.802 15.744
     4.763 15.649
     4.725 15.562
     4.689 15.481
     4.653 15.408
     4.618 15.341
     4.583 15.280
     4.515 15.172
     4.445 15.081
     4.360 14.989
     4.256 14.895
     4.130 14.795
     4.057 14.743
     3.977 14.689
     3.889 14.631
     3.794 14.571
     3.689 14.508
     3.575 14.441
     3.451 14.370
     3.386 14.332
     3.317 14.294
     3.246 14.255
     3.172 14.214
     3.096 14.172
     3.016 14.129
    /
\linethickness= 0.500pt \setplotsymbol ({\thinlinefont .})
%
%
\plot  5.080 16.510      5.053 16.588
     5.028 16.663
     5.002 16.736
     4.978 16.806
     4.954 16.873
     4.931 16.938
     4.886 17.059
     4.843 17.172
     4.802 17.276
     4.763 17.371
     4.725 17.458
     4.689 17.539
     4.653 17.612
     4.618 17.679
     4.583 17.740
     4.515 17.848
     4.445 17.939
     4.360 18.031
     4.256 18.125
     4.130 18.225
     4.057 18.277
     3.977 18.331
     3.889 18.389
     3.794 18.449
     3.689 18.512
     3.575 18.579
     3.451 18.650
     3.386 18.688
     3.317 18.726
     3.246 18.765
     3.172 18.806
     3.096 18.848
     3.016 18.891
    /
%
%
\put{$\zeta = \Phi_{\beta}(z)$} [lB] at  8.00 18.215
%
%
\put{$0$} [lB] at 12.700 16.034
%
%
\put{$\mbox{I}'$} [lB] at 13.018 17.304
%
%
\put{$\mbox{IV}'$} [lB] at 12.859 15.081
%
%
\put{$\beta$} [lB] at  5.239 16.034
%
%
\put{I} [lB] at  5.715 17.621
%
%
\put{II} [lB] at  3.334 17.145
%
%
\put{III} [lB] at  3.334 15.399
%
%
\put{IV} [lB] at  5.715 14.922
%
%
\put{$D_{\epsilon,\beta}$} [lB] at  3.651 13.176
%
%
\put{$D'$} [lB] at 14.129 13.652
%
%
\put{$\mbox{III}'$} [lB] at 11.430 15.875
%
%
\put{$\mbox{II}'$} [lB] at 11.430 16.828 \linethickness=0pt
\putrectangle corners at  1.880 19.238 and 15.424 13.119
\endpicture}
\centerline{Figure
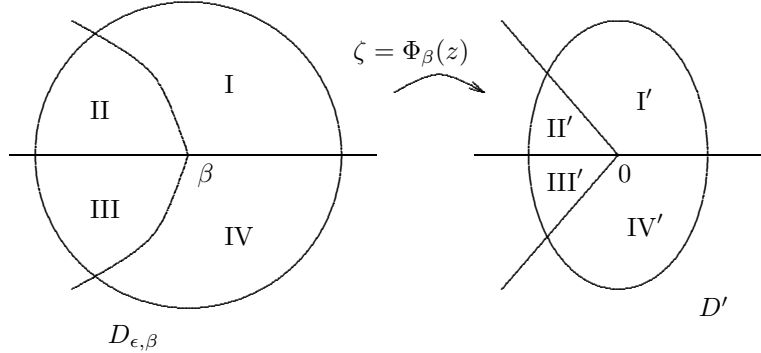
\refstepcounter{equation}\theequation\label{FIG4}. Decompositions
of $D_{\epsilon,\beta}$ and $D'$.}
\bigskip

The four rays on the right of Figure \ref{FIG4} divide the
$\zeta$--plane into four sectors, $0 < \arg(\zeta) < \frac{2
\pi}{3}$, $\frac{2 \pi}{3} < \arg(\zeta) < \pi$, $-\pi <
\arg(\zeta) < \frac{-2 \pi}{3}$ and $\frac{-2 \pi}{3} <
\arg(\zeta) < 0$.  The intersection of these sectors with $D'$
defines the regions $\mbox{I}'$, $\mbox{II}'$, $\mbox{III}'$ and
$\mbox{IV}'$.  (The precise angles separating the sectors are not
important:  we could, for example, replace $\frac{2 \pi}{3}$ by
any angle strictly between $0$ and $\pi$, etc.) Finally, the
regions I, II, III and IV are defined to be the pre--images under
$\Phi(z)$ of $\mbox{I}'$, $\mbox{II}'$, $\mbox{III}'$ and
$\mbox{IV}'$.  We define the contour $\gamma_{\sigma}$ in the $\zeta$-plane to be the four rays shown in Figure \ref{FIG4}, with orientation inherited from the original contour $\Sigma_{S}$ and the mapping $\Phi_{\beta}$.

We will need the following auxiliary functions in order to define the
global approximation.

Denote
\begin{eqnarray}
\label{K5.7} \omega := e^{\frac{2 \pi i}{3}},
\end{eqnarray}
and define
\begin{eqnarray}
\label{K5.8} \Psi^{\sigma}: {\mathbb C} \setminus \gamma _{\sigma}
\rightarrow {\mathbb C}^{2 \times 2},
\end{eqnarray}
\begin{eqnarray}
\label{K5.9} & &
\Psi^{\sigma}(\zeta) = \left\{
\begin{array}{ll}
\left(\begin{array}{cc} Ai(\zeta)&Ai(\omega^2
\zeta)\\Ai'(\zeta)&\omega^2 Ai'(\omega^2 \zeta)
\end{array}\right)
e^{- \frac{ \pi i}{6} \sigma_3} & , \mbox{ for } \zeta \in I,
\\
\left(\begin{array}{cc}
Ai(\zeta)&  Ai(\omega^2 \zeta)\\
Ai'(\zeta)& \omega^2 Ai'(\omega^2 \zeta)
\end{array} \right)
e^{-\frac{\pi i}{6} \sigma_3} \left(\begin{array}{cc} 1&0\\-1&1
\end{array} \right)
& , \mbox{ for } \zeta \in II,
\\
\left(\begin{array}{cc}
Ai(\zeta)&-\omega^2 Ai(\omega \zeta)\\
Ai'(\zeta)&-Ai'(\omega \zeta)
\end{array} \right)
e^{-\frac{\pi i}{6} \sigma_3} \left(\begin{array}{cc} 1&0\\1&1
\end{array} \right)
& , \mbox{ for } \zeta \in III,
\\
\left(\begin{array}{cc}
Ai(\zeta)& -\omega^2 Ai(\omega \zeta)\\
Ai'(\zeta)& -Ai'(\omega \zeta)
\end{array} \right)
e^{-\frac{\pi i}{6} \sigma_3} & , \mbox{ for } \zeta \in IV.
\end{array}
\right.
\end{eqnarray}
Here $\sigma_{3}$ is the Pauli matrix
$\pmtwo
{ 1} {0}
{0} {-1}$.

The next auxiliary quantity which we define is the analytic matrix
valued function $E_{\beta}(z)$, defined as follows.
\begin{eqnarray}
\label{2.018}
 & &  E_{\beta}(z) =  \sqrt{\pi} e^{i \pi /6} \pmtwo
    {\gamma^{-1}}
    { -\gamma}
    { -i \gamma^{-1}}
    {-i \gamma}
     \Phi_{\beta}^{\sigma_{3}/4}.
\end{eqnarray}
(At first blush, this matrix valued function is analytic on a
domain which consists of a disc centered at $z = \beta$, with the
real axis deleted.  However, a straightforward argument (as in
\cite[Section 7]{DKMVZ2}) shows that $E$ extends to an analytic
function on the disc $B_{\delta}^{\beta}$.)

We are now ready to define $M_1^{(A)}$ within $B^{\beta}_{\del}$:
\begin{eqnarray}
\label{II.014} M_{1}^{(A)}(z) = E_{\beta}(z) \
\Psi^{\sigma}(\Phi_{\beta}(z) ) \ e^{2 \sigma_{3}
\Phi_{\beta}^{3/2}/3 }, \ \ \mbox{ for } z \in B_{\delta}^{\beta}.
\end{eqnarray}

\vskip 0.3in We now proceed to the definition of the approximation
$M_{1}^{(A)}(z)$ within $B_{\del}^{\al}$.  As the construction is
entirely analogous to the construction of $M_{1}^{(A)}$ in
$B_{\del}^{\beta}$, we will only present the formulae.  More
details on this construction can be found in \cite[Section 7]
{DKMVZ2} , or \cite[Subsection 4.4]{DKMVZ3}. We define the
transformation $\Phi_{\alpha}(z)$, analogous to $\Phi_{\beta}$
defined in (\ref{Phich1a}) above,
\begin{eqnarray}
\label{Phial1a} & & \Phi_{\alpha}(z) \equiv \left(\frac{3 N
}{4}\right)^{2/3} \left(
    2g(\alpha) - V(\alpha) - \ell
- \left(
2g(z) - V(z) - \ell \right)  \right)^{2/3}, \ \ \ z \in B_{\delta}^{\al}.  
\end{eqnarray}
 As $h(\al) \neq 0$,
$2g(\alpha) - V(\alpha) - \ell - \left( 2g(z) - V(z) - \ell
\right)$ behaves like $(z - \al)^{3/2}$ for $z$ near $\al$ (cf.
relations (\ref{II.002})), and we may choose the branch so that
$\Phi_{\al}(z) $ is analytic in a neighborhood of $\al$, and
\begin{eqnarray}
\label{Phial1b} \\
\nonumber \Phi_{\al}(z) = - c (z - \al) + O\left(\left( z - \al
\right)^2 \right), \
c > 0, \ z \to \al. \ 
\end{eqnarray}
Next we define $E_{\alpha}(z)$ as follows:
\begin{eqnarray}
& &  E_{\al}(z) =  \sqrt{\pi} e^{i \pi /6} \pmtwo
    {\gamma}
    { -\gamma^{-1}}
    { i \gamma}
    {i \gamma^{-1}}
     \Phi_{\alpha}^{\sigma_{3}/4}.
\end{eqnarray}
We now define the approximation $M_{1}^{(A)}$ within
$B_{\del}^{\al}$:
\begin{eqnarray}
\label{II.018} M_{1}^{(A)}(z) = E_{\al}(z) \ \Psi^{\sigma} \
(\Phi_{\al}(z)) \ e^{2 \sigma_{3} \Phi_{\al}^{3/2} /3} \sigma_{3},
\ \ \mbox{ for } z \in B_{\del}^{\al}.
\end{eqnarray}

\vskip 0.3in

\subsection{The Riemann--Hilbert problem for the error.}

The next step in the Riemann--Hilbert approach is to define the ratio of
$M_1$ to $M_1^{(A)}$:
\begin{eqnarray}
\label{SDEF}
 S = M_1 (z) \Bigl( M_1^{(A)}(z)\Bigr)^{-1}, \quad z\in\Cbb\backslash
\Sigma_{M_1} .
\end{eqnarray}
It then follows that $R(z)$ satisfies the following Riemann--Hilbert problem:

\begin{prob}
\label{RHR}
 The problem is to determine a $2\times 2$ matrix
$S(z)$ satisfying
  \begin{enumerate}
\item[(a)] $S(z)$ analytic in $\Cbb\backslash \Sigma_{S}$,
\item[(b)] $S(z)$ possesses continuous boundary values $S_{\pm}(z)$ for
$z\in\Sigma_S$,
\item[(c)] $S = I+\Ocal (1/z), \; z\to\infty$,
\item[(d)] $S_+ = S_{-} V_{S}, \; z\in\Sigma_S$,
  \end{enumerate}
where $\Sigma_S$ is defined in Figure \ref{KF10}.
\end{prob}
\vskip0.2cm
\font\thinlinefont=cmr5 \centerline{\beginpicture
\setcoordinatesystem units <.5cm,.5cm> \linethickness=1pt
\setshadesymbol ({\thinlinefont .}) \setlinear
%
%
\linethickness= 0.500pt \setplotsymbol ({\thinlinefont .})
\ellipticalarc axes ratio  2.540:2.540  360 degrees
    from 10.160 16.510 center at  7.620 16.510
%
%
\linethickness= 0.500pt \setplotsymbol ({\thinlinefont .})
\ellipticalarc axes ratio  2.540:2.540  360 degrees
    from 20.320 16.510 center at 17.780 16.510
%
%
\linethickness= 0.500pt \setplotsymbol ({\thinlinefont .})
\putrule from  9.366 18.415 to 16.034 18.415
%
%
\linethickness= 0.500pt \setplotsymbol ({\thinlinefont .})
\putrule from  9.366 14.605 to 16.034 14.605
%
%
\linethickness= 0.500pt \setplotsymbol ({\thinlinefont .}) \plot
12.859 18.415 12.541 18.574 /
%
%
\linethickness= 0.500pt \setplotsymbol ({\thinlinefont .}) \plot
12.859 18.415 12.541 18.256 /
%
%
\linethickness= 0.500pt \setplotsymbol ({\thinlinefont .}) \plot
12.859 14.605 12.541 14.764 /
%
%
\linethickness= 0.500pt \setplotsymbol ({\thinlinefont .}) \plot
12.859 14.605 12.541 14.446 /
%
%
\linethickness= 0.500pt \setplotsymbol ({\thinlinefont .}) \plot
4.128 16.510  4.128 16.510 /
%
%
\linethickness= 0.500pt \setplotsymbol ({\thinlinefont .}) \plot
4.128 16.510  3.810 16.669 /
%
%
\linethickness= 0.500pt \setplotsymbol ({\thinlinefont .}) \plot
4.128 16.510  3.810 16.351 /
%
%
\linethickness= 0.500pt \setplotsymbol ({\thinlinefont .})
\putrule from 21.590 16.510 to 21.273 16.510
%
%
\linethickness= 0.500pt \setplotsymbol ({\thinlinefont .}) \plot
21.590 16.510 21.273 16.669 /
%
%
\linethickness= 0.500pt \setplotsymbol ({\thinlinefont .}) \plot
21.590 16.510 21.273 16.351 /
%
%
\linethickness= 0.500pt \setplotsymbol ({\thinlinefont .}) \plot
17.780 19.050 17.462 19.209 /
%
%
\linethickness= 0.500pt \setplotsymbol ({\thinlinefont .}) \plot
17.780 19.050 17.462 18.891 /
%
%
\linethickness= 0.500pt \setplotsymbol ({\thinlinefont .}) \plot
7.620 19.050  7.303 19.209 /
%
%
\linethickness= 0.500pt \setplotsymbol ({\thinlinefont .}) \plot
7.620 19.050  7.303 18.891 /
%
%
\linethickness= 0.500pt \setplotsymbol ({\thinlinefont .})
\putrule from  1.905 16.510 to  5.080 16.510
%
%
\linethickness= 0.500pt \setplotsymbol ({\thinlinefont .})
\putrule from 20.320 16.510 to 23.495 16.510
%
%
\linethickness= 0.500pt \setplotsymbol ({\thinlinefont .}) \plot
7.541 16.589  7.699 16.431 /
%
%
\linethickness= 0.500pt \setplotsymbol ({\thinlinefont .}) \plot
7.541 16.431  7.699 16.589 /
%
%
\linethickness= 0.500pt \setplotsymbol ({\thinlinefont .}) \plot
17.701 16.589 17.859 16.431 /
%
%
\linethickness= 0.500pt \setplotsymbol ({\thinlinefont .}) \plot
17.701 16.431 17.859 16.589 /
%
%
\put{$\Sigma_{S,1}$} [lB] at  2.540 16.828
%
%
\put{$\Sigma_{S,5}$} [lB] at  5.80 19.250
%
%
\put{$\Sigma_{S,2}$} [lB] at 13.970 18.733
%
%
\put{$\Sigma_{S,3}$} [lB] at 13.970 14.922
%
%
\put{$\Sigma_{S,4}$} [lB] at 22.225 16.828
%
%
\put{$\Sigma_{S,6}$} [lB] at 19.050 19.050
%
%
\put{$_{\alpha}$} [lB] at  7.261 15.975
%
%
\put{$_{\beta}$} [lB] at 17.680 15.975 \linethickness=0pt
\putrectangle corners at  1.880 19.279 and 23.520 13.955
\endpicture}
\centerline{Figure
\refstepcounter{equation}\theequation\label{KF10}. The contour
$\hat{\Sigma}_S$.} \vskip.2cm

The jump matrix $V_{S}$ is defined as follows:
\begin{eqnarray}
\label{II.020}
 & & V_{S,1} = M_{1}^{(A)} \pmtwo {1} {e^{ N ( g_{+}
+ g_{-} - V - \ell)}} {0}{1}
\left( M_{1}^{(A)} \right)^{-1}  , \ \ z \in \Sigma_{S,1},\\
\label{II.021}
& & V_{S,2} = M_{1}^{(A)} \pmtwo {1} {0}
{e^{-N(g_{+}-g_{-})}}{1} \left( M_{1}^{(A)} \right)^{-1} , \ \ z \in \Sigma_{S,2},\\
\label{II.022}
& & V_{S,3} = M_{1}^{(A)} \pmtwo {1} {0}
{e^{N(g_{+}-g_{-})}}{1} \left( M_{1}^{(A)} \right)^{-1} , \ \ z \in \Sigma_{S,3},\\
\label{II.023}
& & V_{S,4} = M_{1}^{(A)} \pmtwo {1} {e^{ N ( g_{+}
+ g_{-} - V - \ell)}} {0}{1}
\left( M_{1}^{(A)} \right)^{-1} , \ \ z \in \Sigma_{S,4},\\
\label{II.024}
& & V_{S,5} = \left( M_{1}^{(A)} \right)_{-} \
\left( M_{1}^{(A)} \right)_{+}^{-1}, \ \ z \in \Sigma_{S,5},\\
\label{II.025}
& & V_{S,6} = \left(  M_{1}^{(A)} \right)_{-} \
\left( M_{1}^{(A)} \right)_{+}^{-1}, \ z \in \Sigma_{S,6}.
\end{eqnarray}

The fact of the matter is that
  \begin{enumerate}
\item[1.] On the upper lens, lower lens, and the real axis component of
$\Sigma_S$
  \begin{eqnarray} \label{II.026a}
\| V_S-I\|_{L^{\infty}},\; \| V_S-I\|_{L^2} \leq Ce^{-dN},\; C,d>0.
  \end{eqnarray}
This can be deduced from (\ref{II.020})-(\ref{II.024}), because
(1) the global approximation $M_{1}^{(A)}$, as well as its
inverse, is uniformly bounded on these portions of the contour,
(2) the off-diagonal entries of the middle matrices on the right
hand side of (\ref{II.020})-(\ref{II.024}) are uniformly
exponentially decaying in $N$ on these portions of the contour,
and (3) for $z \to \infty$, $z\in \Sigma_{S,1} \cup \Sigma_{S,4}$,
one has $e^{N(g_{+}+g_{-} - V - \ell)} \le C e^{- d N ( |z| + 1)}$
(for the complete argument, see the proof of \cite[Proposition
7.64]{DKMVZ2}).
\item[2.] On the circle $\Sigma_{S,6}$, we have the following complete
asymptotic expansion:
  \begin{align}\label{II.027}
V_{S,6} &= I + \frac{1}{N}V^{\beta}_1 + \frac{1}{N^2} V^{\beta}_2 +
\cdots \\
\label{II.030a}
V_1^{\beta} &= \frac{ 5 (z - \alpha)^{1/2}}{72  (z-\beta)^{1/2}
\int_{\beta}^{z} R(s) h(s) ds } \ \pmtwo {-1} {i} {i} {1}
\\ \nonumber
& \ \ \ +\frac{7 ( z - \beta)^{1/2}}{72 ( z - \alpha)^{1/2}
\int_{\beta}^{z}R(s) h(s) ds } \pmtwo {1}{i}
{i}{-1}, \\
V_2^{\beta} &= \frac{35}{2592} \left( \int_{\beta}^{z} R(s) h(s)
ds \right)^{-2} \pmtwo {-1}{12 i}{-12i}{-1}.
\end{align}
More generally, we have for $k$ even, $k \ge 2$:
\begin{align}
\label{II.032a}
V_{k}^{\beta} &= \left( \frac{ 1}{2} \int_{\beta}^{z} R(s) h(s) ds \right)^{-k}
\pmtwo{\frac{t_{k} + s_{k}}{2}}
{\frac{i(s_{k}-t_{k})}{2}}
{\frac{-i(s_{k}-t_{k})}{2}}
{\frac{t_{k}+s_{k}}{2}},
  \end{align}
and for $k$ odd, $k \ge 1$:
\begin{align}
V_{k}^{\beta} &= \frac{ 2^{k-1} t_{k} (z - \alpha)^{1/2}}{ (z-\beta)^{1/2}
\left(  \int_{\beta}^{z} R(s) h(s) ds \right)^{k} } \
\pmtwo {-1} {i} {i} {1}
\\ \nonumber
& \ \ \ +\frac{ 2^{k-1} s_{k} ( z - \beta)^{1/2}}{( z - \alpha)^{1/2}
\left(   \int_{\beta}^{z}R(s) h(s) ds\right)^{k} } \pmtwo {1}{i}
{i}{-1},
\end{align}
where
\begin{eqnarray}
\label{K5.26}
s_0 &=& t_0  =1, \\
s_k &=& \frac{\Gamma(3k+\frac{1}{2})}{54^k k! \Gamma(k + \frac{1}{2})}, \;
t_k = -\frac{6k+1}{6k-1}s_k \;
\mbox{ for } k \geq 1,
\end{eqnarray}

\item[3.] On the circle $\partial B^{\al}_{\del}$, we have the following
complete asymptotic expansion:
  \begin{align}\label{II.030}
V_{S,5} &= I + \frac{1}{N} V^{\al}_1 + \frac{1}{N^2} V^{\al}_2 + \cdots
\\
\label{II.037}
V_1^{\al} &= \frac{5 ( z - \beta)^{1/2}}{72  ( z - \alpha)^{1/2}
\int_{\alpha}^{z} R(s) h(s) ds } \pmtwo
{-1} {-i}
{-i} {1} + \\
\nonumber
& \ \ \ + \frac{7 ( z - \alpha)^{1/2}}{72  ( z - \beta)^{1/2}
\int_{\alpha}^{z} R(s) h(s) ds } \pmtwo {1} {-i} {-i} {-1},\\
\label{II.032} V_2^{\al} &= \frac{35}{2592} \left(
\int_{\alpha}^{z} R(s) h(s) ds \right)^{-2} \pmtwo{-1}{-12 i}{12 i}{-1}.
  \end{align}
  \end{enumerate}

More generally, we have for $k$ even, $k \ge 2$,
\begin{eqnarray}
\label{II.039}
V_{k}^{\alpha} = \frac{2^{k-1}}
{\left(  \int_{\alpha}^{z} R(s) h(s) ds \right)^{k}}
\pmtwo{s_{k}+t_{k}}{i(t_{k}-s_{k})}
{i(s_{k} - t_{k})}{s_{k}+t_{k}},
\end{eqnarray}
and for $k$ odd, $k \ge 1$, we have
\begin{eqnarray}
V_{k}^{\alpha} = \frac{2^{k-1}s_{k} (z - \beta)^{1/2}}
{(z-\alpha)^{1/2}\left(  \int_{\alpha}^{z} R(s) h(s) ds \right)^{k}}
\pmtwo{-1}{-i}
{-i}{1}  \\
\nonumber
+ \frac{2^{k-1}t_{k} (z - \alpha)^{1/2}}
{(z-\beta)^{1/2}\left(  \int_{\alpha}^{z} R(s) h(s) ds \right)^{k}}
\pmtwo{-1}{i}
{i}{1},
\end{eqnarray}

The reader may verify these asymptotic expansions using the
definitions (\ref{II.024})-(\ref{II.025}) of $V_{S}$, the
definitions (\ref{II.014}) and (\ref{II.018}) of $M_{1}^{(A)}$,
together with the well-known asymptotic expansions for the Airy
function and its derivative.  In \cite{DKMVZ2}, the authors
present a complete proof of this fact (see, Lemma 7.34 and its
proof), and we leave the reader to carry out the analogous
arguments in the present case.

\subsection{The solution $S$ of the Riemann--Hilbert problem \ref{RHR}
and its asymptotic expansion}

In this subsection we will present a formula for the solution $S$
of the Riemann--Hilbert problem \ref{RHR}, as well as an
asymptotic expansion for $S$ in powers of $N^{-1}$.  To do so, we
will require the integral operators $C_{\pm}$, defined as follows:
\begin{eqnarray}
& & C_{\pm} : L^{2} ( \Sigma_{S}) \to L^{2} (\Sigma_{S}), \\
& & C_{\pm}(f)(z) = \lim_{z' \to z, \ z \in \ \pm \mbox{ side of }
\Sigma_{S}} \frac{1}{2 \pi i} \int_{\Sigma_{S}} \frac{f(s)}{s-z'}
ds, \ \ z \in \Sigma_{S}.
\end{eqnarray}
The relevant integral operator $C_{V_{S}} : L^{2}(\Sigma_{S}) \to
L^{2}(\Sigma_{S})$ is defined via
\begin{eqnarray}
C_{V_{S}}(f) = C_{-} \left[ f \left( V_{S} - I \right) \right].
\end{eqnarray}

The asymptotic expansions described in
(\ref{II.027})-(\ref{II.032}), together with the estimates
(\ref{II.026a}), and the well known boundedness of the integral
operators $C_{\pm}$ imply that the integral operator $C_{V_{S}}$
is bounded, with operator norm satisfying
\begin{eqnarray}
\| C_{V_{S}} \|_{L^{2}(\Sigma_{S}) \to L^{2}(\Sigma_{S})} \le
\frac{C}{N}.
\end{eqnarray}
This implies that ${\mathbb I} - C_{V_{S}}$ can be inverted by
Neumann series for $N$ sufficiently large.  Therefore, we may
define
\begin{eqnarray}
\label{MUSDef}
\mu_{S} = \left( {\mathbb I} - C_{V_{S}} \right)^{-1}
C_{V_{S}}(I).
\end{eqnarray}
Then, as in \cite[Theorem 7.70]{DKMVZ2}, the explicit formula for
the solution $S$ to the Riemann--Hilbert problem (\ref{RHR}) is
\begin{eqnarray}
S(z) = I + \frac{1}{2 \pi i} \int_{\Sigma_{S}} \frac{\left(I +
\mu_{S} \right) \left(V_{S} - I \right)}{s - z} ds, \ \ z \in \cb
\setminus \Sigma_{S}.
\end{eqnarray}
Furthermore, as shown in Corollary 7.77 and Theorem 7.81 in
\cite{DKMVZ2}, $S$ is uniformly bounded in $\cb \setminus
\Sigma_{S}$, and possesses a complete asymptotic expansion in
powers of $N^{-1}$ (in Section 7 of \cite{DKMVZ2}, the error matrix is
called $R(z)$, whereas here the error matrix is called $S$).

The terms in the asymptotic expansion may be obtained in a number
of ways.  For example, one may just compute the Neumann expansion
of $\mu_{S}$, and keep only those terms which arise due to
integrations around the circles $\partial B_{\delta}^{\al}$ and
$\partial B_{\delta}^{\beta}$. Alternatively, one may posit an
asymptotic expansion for $S$ in powers of $N^{-1}$, plug this
expansion into the Riemann--Hilbert problem \ref{RHR}, and obtain
a sequence of Riemann--Hilbert problems for the coefficients in
the expansion, which may be solved iteratively.  The proof of
Theorem 7.81 in \cite{DKMVZ2} involves (1) defining an integral
operator $C_{V_{S}}^{(l)}$ associated only to the circles
$\partial B_{\delta}^{\al}$ and $\partial B_{\delta}^{\beta}$,
using only the first $l$ terms in the expansion of $V_{S}$ on
these circles, (2) building an explicit approximation $S_{l}(z)$
by using the first $l$ terms in a Neumann series for $ \left(
{\mathbb I} - C_{V_{S}}^{(l)} \right)^{-1}$, and (3) comparing
this approximation to the true solution $S(z)$.  The ensuing
Riemann--Hilbert problem for the error is shown to have a solution
which is uniformly bounded by the first neglected term in the
asymptotic expansion, etc.

The result of these calculations is the following asymptotic
expansion for the solution $S$:
\begin{eqnarray}
\label{SASSYM}
& & S(z) \sim I + \sum_{j=1}^{\infty} N^{-j} S_{j}(z),
\end{eqnarray}
where $S_{j}(z)$ is piecewise analytic, and can be defined iteratively
via the following sequence of additive Riemann--Hilbert problems.
\begin{prob}
\label{RHRec}
 The first problem is to determine a $2\times 2$ matrix
$S_{1}(z)$ satisfying
  \begin{enumerate}
\item[(a)] $S_{1}(z)$ analytic in $\Cbb\backslash (\partial
B_{\delta}^{\al} \cup \partial B_{\delta}^{\beta})$,
\item[(b)] $S_{1}(z)$ possesses continuous boundary values $\left( S_{1} \right)_{
\pm}(z)$ for $z\in \partial B_{\delta}^{\al} \cup \partial B_{\delta}^{\beta}$,
\item[(c)] $S_{1} = \Ocal (1/z), \; z\to\infty$,
\item[(d)] $\left(S_{1}\right)_{+} - \left( S_{1} \right)_{ -} = V_{1}, \; z\in \partial
B_{\delta}^{\al} \cup \partial B_{\delta}^{\beta}$,
  \end{enumerate}
\end{prob}
\vskip 0.2in
\begin{prob}
\label{RHReck}
Having determined $S_{1}$, we now determine $S_{k}$, $k \ge 2$ satisfying
\begin{enumerate}
\item[(a)] $S_{k}(z)$ analytic in $\Cbb\backslash (\partial
B_{\delta}^{\al} \cup \partial B_{\delta}^{\beta})$,
\item[(b)] $S_{k}(z)$ possesses continuous boundary values $\left( S_{k} \right)_{
\pm}(z)$ for $z\in \partial B_{\delta}^{\al} \cup \partial B_{\delta}^{\beta}$,
\item[(c)] $S_{k} = \Ocal (1/z), \; z\to\infty$,
\item[(d)] $\left(S_{k}\right)_{+} - \left( S_{k} \right)_{ -} =
V_{k} + \sum_{j=1}^{k-1} \left( S_{j} \right)_{-} \ V_{k-j}\ \ \ , \; z\in \partial
B_{\delta}^{\al} \cup \partial B_{\delta}^{\beta}$.
  \end{enumerate}
\end{prob}

Each of these Riemann--Hilbert problems possesses a unique solution,
explicitly representable by contour integration.  It is interesting and
useful to
note that in fact each of these contour integral representations can be
evaluated explicitly.  Thus, for example, we include an explicit
representation for $S_{1}(z)$ in Appendix A.

Some very important properties of the sequence $\{ S_{k} \}_{k \ge 0}$
are established in the following Lemma.  We will need the Pauli matrices
\begin{eqnarray}
& & \hspace{0.2in}
\sigma_{3} = \pmtwo{1}{0}{0}{-1}, \hspace{0.5in}
\sigma_{2} = \pmtwo{0}{-i}{i}{0}, \hspace{0.5in}
\sigma_{1} = \pmtwo{0}{1}{1}{0}.
\end{eqnarray}

\begin{lem}
\label{Lem:Sym}
For each $k \in {\mathbb{Z}_{ \ge 0}} $, there are complex valued
functions $s_{k}(z)$ so that the following holds.
\begin{enumerate}
\item[1.]
For $k$ even, the function $S_{k}(z)$, defined to be the unique solution
to the Riemann-Hilbert problem \ref{RHReck} satisfies the symmetry
conditions
\begin{eqnarray}
\label{II.051}
S_{k}(z) = s_{k}^{(1)}(z) I + s_{k}^{(2)}(z) \sigma_{2}.
\end{eqnarray}
\item[2.]
For $k$ odd, the function $S_{k}$ satisfies the symmetry conditions
\begin{eqnarray}
\label{II.052}
S_{k}(z) = s_{k}^{(1)}(z) \sigma_{3} + s_{k}^{(2)}(z) \sigma_{1}.
\end{eqnarray}
\end{enumerate}
The functions $s_{k}^{(j)}(z)$, $j=1,2$, are piecewise analytic
functions of $z$, explicitly computable in terms of the parameters ${\bf
t}$.

\end{lem}

In words:  for $k$ odd, $S_{k}(z)$, is a symmetric, trace 0 matrix.  For
$k$ even, $S_{k}$ is the sum of a multiple of the identity matrix, and a
skew-symmetric matrix.

\begin{Proof}

For $k=1$, the Lemma is seen to be true by studying the solution to the
Riemann-Hilbert problem \ref{RHRec}:
\begin{eqnarray}
S_{1}(z) = \frac{1}{2 \pi i} \int \frac{V_{1}(s)}{s-z} ds,
\end{eqnarray}
where the integral is taken along the boundaries of the discs
$B_{\delta}^{\alpha}$ and $B_{\delta}^{\beta}$, oriented clockwise.
Since $V_{1}$, as defined in (\ref{II.030a}) and (\ref{II.037}) is
clearly symmetric trace zero, $S_{1}$ inherits the same property.

We now argue inductively.  Suppose that $k$ is even, and the Lemma
holds true up to $k-1$.  The reader may verify that the right hand
side of the jump condition (d) in the Riemann--Hilbert problem
\ref{RHReck} is of the form
\begin{eqnarray}
\label{niceform}
{\nu}_{k}^{(1)}(z) I + {\nu}_{k}^{(2)}(z) \sigma_{2},
\end{eqnarray}
where ${\nu}_{k}^{(j)}(z)$, $j=1,2$, are scalar complex valued
functions of $z$.  Indeed, for $k$ even, $V_{k}(z)$, as defined in
(\ref{II.032a}) and (\ref{II.039}) are of this form.  Moreover, each
term appearing in the sum on the right hand side of the jump condition
(d) in the Riemann--Hilbert problem \ref{RHReck} is also of the same
form (this can be deduced from the form of $S_{j}$ and $V_{k-j}$ for
each value of $j \in \{1, \ldots, k-1\}$, given that $k$ is even).  
There is an explicit formula for the solution to this additive
Riemann--Hilbert problem, which is given by the Cauchy transform of
the right hand side of jump condition (d).  Since every term is of
the form (\ref{niceform}), the Cauchy transform is also of this form,
and so (\ref{II.051}) holds true for $k$ even.

The argument for $k$ odd follows the same line of reasoning, and we
leave the details to the reader.

\end{Proof}

At this point we have established the following formula for the solution
$Y$ to the original Riemann--Hilbert problem stated just before
Subsection 2.1:
\begin{eqnarray}
\label{YREP}
Y(z) = e^{ N \ell \sigma_{3}/2} S(z) M_{1}^{(A)}(z) T(z) e^{N (g(z) -
\ell/2) \sigma_{3}},
\end{eqnarray}
where $\ell$ is the Lagrange multiplier associated to the variational
problem, $S(z)$ is the solution to the Riemann--Hilbert problem
\ref{RHR} for the error, $M_{1}^{(A)}(z)$ is the explicit approximation
defined in (\ref{M1Ap1}), (\ref{II.014}), and (\ref{II.018}), $T(z)$ is
either the identity matrix $I$ for $z$ outside the lens shaped regions,
or one of two explicit triangular factors appearing in (\ref{M1DEF2})
and (\ref{M1DEF3}), and $g(z)$ is the function defined in
(\ref{GFdef}).  It is straightforward to verify that with the exception
of $S(z)$, all terms appearing in (\ref{YREP}) depend analytically on
the times ${\bf t}$ for ${\bf t}$ in any open subset of $\mathbb{T}(T,\gamma)$,
and in fact extend as infinitely differentiable functions to all of
$\mathbb{T}(T,\gamma)$.  It is also true that the function $S(z)$
enjoys the same analyticity properties.  This is made precise in the
following Theorem.
\begin{thm}
The matrix valued function $S$ is infinitely
differentiable in the times ${\bf t}$ for ${\bf t} \in \mathbb{T}(T,\gamma)$,
and derivatives of $S$ again possess an asymptotic expansion in $N$,
obtained by differentiating the original asymptotic expansion
for $S$.  The coefficients in this asymptotic expansion
depend analytically on ${\bf t}$ for ${\bf t} \in
\mathbb{T}(T,\gamma)$  (with smooth extension to the closure), 
for $T$ sufficiently small.
\end{thm}

\begin{Proof}
The proof hinges on the following fundamental facts: the jump matrix $V_{S}$ is
analytic in a vicinity of every point $z$ on the contour $\Sigma_{S}$,
it possesses an asymptotic expansion in powers of $N^{-1}$
in which each term is a piecewise analytic function
of $z$ and depends analytically on ${\bf t}$ for ${\bf
t} \in \mathbb{T}(T,\gamma)$; this asymptotic expansion may be differentiated term by term with
respect to the times ${\bf t} \in \mathbb{T}(T,\gamma)$, yielding asymptotic
expansions for arbitrary derivatives of $V_{S}$.  From this information,
it is straightforward to prove that $\mu_{S}$ defined in (\ref{MUSDef})
is an analytic function of ${\bf t}$ for ${\bf t} \in
\mathbb{T}(T,\gamma)$, has an analytic continuation to a neighborhood of the
contour $\Sigma_{S}$ in the $z$-variable, 
and possesses an asymptotic expansion (which can be
differentiated term by term) 
obtained directly from a Neumann series representation.
An outline of the argument is as follows.

First, observe that the terms in the Neumann series representation of
$\mu_{S}$ are individually analytic functions of $z$ in a vicinity of
every point of the contour $\Sigma_{S}$, and they are
analytic function of
${\bf t}$ for ${\bf t} \in \mathbb{T}(T,\gamma)$.  Second, the series is
uniformly convergent.  This follows because (1) each Neumann iterate can be
expressed as an integral on a {\it deformed} contour which is bounded
away from the contour $\Sigma_{S}$ and (2) the $j$-th iterate may then be
bounded by $C (D/N)^{j}$, where the constants $C$ and $D$ are
independent of $j$.  For derivatives of $\mu_{S}$, similar calculations,
using the fact that derivatives of $V_{S}$ possess asymptotic
expansions, show that an $L$th order derivative of the $j$th iterate may be bounded by $\tilde{C}
j^{L} (\tilde{D}/N)^{j}$, where $\tilde{C}$ and $\tilde{D}$ are
independent of $j$, but may depend on $L$.  This shows that the Neumann
series for $\mu_{S}$ may be differentiated term by term.

The proof that derivatives of $S$ possess
expansions which may be obtained by differentiating the original
expansion for $S$ term by term now follows from the uniform convergence of
the Neumann series for $\mu_{S}$ and its derivatives.
\end{Proof}

\section{The mean density of eigenvalues:  exact formula and
asymptotics}
\label{sec:MDOS}

The previous section explains (and extends to higher orders) the work in \cite{DKMVZ1}-\cite{DKMVZ3},
in which techniques for the asymptotic analysis of
Riemann-Hilbert problems, originally developed for the analysis
of singular limits of integrable partial differential equations
by Deift and Zhou (and further developed by many researchers),
are used to
establish an asymptotic description for the polynomials
$p_{N}(z)$ and
$p_{N-1}(z)$ orthogonal with respect to the $N$-dependent measure
(\ref{OPMeas}).  There are many consequences of the complete asymptotic
description for the orthogonal polynomials.  For a number of such
applications, we refer the reader to \cite{DKMVZ1}-\cite{DKMVZ3}, where
detailed asymptotic expansions for (1) the leading coefficients of of the
polynomials, (2) the recurrence coefficients, (3) the zeros of the
polynomials were obtained, along with a proof of universality of spacings
of eigenvalues of random matrices for a wide class of unitarily invariant
probability measures on Hermitian matrices.  In this section our goal is
to establish formula (\ref{3.001}), and prove that this formula for the
mean density of eigenvalues holds true for $z$ in a fixed sized
neighborhood of
$[z^{*}, \beta]$ for $N$ sufficiently large.  The parameter $z^{*}$ will
be chosen below.  The formula (\ref{3.001}) holds true in a large
neighborhood containing the point $\beta$, and there is an analogous
formula, (\ref{3.003}) which holds true in a fixed size neighborhood of
the interval $[\alpha, z^{*}]$, and the two formulae hold simultaneously
on an overlap region.  In order to achieve this overlap region, we will need to introduce $T_{\beta}$, $\gamma_{\beta}$, $T_{\alpha}$, and $\gamma_{\alpha}$, which are values of $T$ and $\gamma$ from Theorem \ref{EQMSTHM} so that if $T<\mbox{min} \{T_{\beta} , T_{\alpha} \}$ and $\gamma>\mbox{max} \{\gamma_{\beta}, \gamma_{\alpha}\}$, then the neighborhoods $D_{\epsilon, \beta}$ and $D_{\epsilon, \alpha}$ may be taken large enough to have nontrivial overlap.

Using formula (\ref{3.001})
we will obtain a complete asymptotic expansion for the mean density of
eigenvalues, and then in the following section we will evaluate an
asymptotic expansion of integrals of the form
\begin{eqnarray}
\label{3.001INT}
\int_{{\mathbb{R}}} f(\lambda) \rho_{N}^{(1)}(\lambda) d \lambda,
\end{eqnarray}
for general $C^{\infty}$ functions that grow no worse than polynomial at
$\infty$.

We begin with two important observations.

{\bf Observation 1}:  It is
possible to use the results of the previous section to compute an
asymptotic expansion for the mean density $\rho_{N}^{(1)}$ which is valid
for all
$\lambda \in (\alpha, \beta)$.  That expansion is as follows,
\begin{eqnarray}
\label{3.002BUL}
& &
\rho_{N}^{(1)}(\lambda) = \psi(\lambda) + \frac{1}{4 \pi N} \left(
\frac{1}{\lambda - \beta} - \frac{1}{\lambda - \alpha}
\right) \cos{\left\{ N \int_{\lambda}^{\beta} \psi(s) ds  \right\} } \\
\nonumber
& &  \hspace{0.75in}+
\frac{1}{N^{2}} \left[ H(\lambda) + G(\lambda) \sin{\left\{N
\int_{\lambda}^{\beta} \psi(s) ds \right\} }\right] + \cdots
\end{eqnarray}
in which $H(\lambda)$ and $G(\lambda)$ are locally analytic functions
which are explicitly computable in terms of the original external field
$V(\lambda)$.  However, for the computation of integrals of the form
(\ref{3.001INT}), this expansion provides a piece of the puzzle, but
clearly cannot be used in a vicinity of the endpoints $\alpha$ and
$\beta$.  Indeed, the asymptotic expansion fails in a vicinity of
these endpoints, because the second term in (\ref{3.002BUL}) possesses
poles, whereas the mean density $\rho_{N}^{(1)}$ does not.  From this
observation we see that to compute asymptotic expansions of integrals of
the form (\ref{3.001INT}), we will need to to have an asymptotic
expansion for $\rho_{N}^{(1)}$ which is valid for $\lambda$ near these
endpoints.  It is a piece of good fortune that the representations which
we obtain are actually valid on regions which cover the entire interval
$[\alpha, \beta]$, and we will not need to evaluate integrals ``in the
bulk''.

{\bf Observation 2}:  It is straightforward to use the results of the
preceding section to prove that $\rho_{N}^{(1)}(\lambda)$ is
exponentially small (in $N$), and exponentially decaying (in $\lambda$)
for $\lambda \in {\mathbb{R}}
\setminus (\alpha-\delta, \beta+\delta)$, for any fixed $\delta > 0$.  We
will not present the details of those calculations, but leave them to the
avid reader.  The implication of this is that  the contribution to an
integral of the form (\ref{3.001INT}) from the set ${\mathbb{R}}
\setminus (\alpha-\delta, \beta+\delta)$ is exponentially small in $N$,
and hence negligible to all orders in $N$.  Therefore we must only
compute an asymptotic expansion for integrals of the form
\begin{eqnarray}
\label{3.002INT}
\int_{\alpha - \delta}^{\beta + \delta} f(\lambda)
\rho_{N}^{(1)}(\lambda) d \lambda,
\end{eqnarray}
in which the allowable functions $f$ are $C^{\infty}$ smooth, and
compactly supported within $(\alpha - \delta, \beta + \delta)$.

We now turn to the formula (\ref{II.001}) for
$\rho_{N}^{(1)}$, and re-express this fundamental quantity using the
Riemann--Hilbert analysis of the previous section.  We present two
functions, $\rho_{N}^{(1,\beta)}$ and $\rho_{N}^{(1,\alpha)}$
representing $\rho_{N}^{(1)}$; the first one is a valid representation
for $\rho_{N}^{(1)}$ in an interval
containing $z=\beta$, and the other, in an interval containing
$z = \alpha$.  The first such
formula is the following explicit, exact expression:
\begin{eqnarray}
\label{3.001}
& &
N \rho_{N}^{(1,\beta)}(z) = \left( \frac{\Phi_{\beta}'(z)}{4 \Phi_{\beta}(z)} -
\frac{\gamma'(z)}{\gamma(z)} \right) \left[ 2 \
Ai\left(\Phi_{\beta}(z) \right) \
Ai^\prime\left( \Phi_{\beta}(z)\right)  \right] \\
\nonumber
& & \hspace{0.75in}
+ \Phi_{\beta}'(z)
\left[ \Big(Ai^\prime\left(\Phi_{\beta}(z) \right)\Big)^{2} -
\Phi_{\beta}(z) \
\Big(Ai\left( \Phi_{\beta}(z) \right)\Big)^{2} \right] \\
\nonumber
& & \hspace{0.75in}
+ \frac{i}{2} \left[
\Big( S' \ B \ \Psi\left(\Phi_{\beta}(z) \right) \Big)_{11} \
\Big( S \ B \ \Psi\left(\Phi_{\beta}(z) \right) \Big)_{21} \right. \\
& & \nonumber \hspace{1.25in} \left.
 - \
\Big( S \ B \ \Psi\left(\Phi_{\beta}(z) \right) \Big)_{11} \
\Big( S' \ B \ \Psi\left(\Phi_{\beta}(z) \right) \Big)_{21}  \right],
\end{eqnarray}
where
\begin{eqnarray}
\label{III.005} & & \Psi(\zeta) = \left\{
\begin{array}{ll}
\left(\begin{array}{cc} Ai(\zeta)&Ai(\omega^2
\zeta)\\Ai'(\zeta)&\omega^2 Ai'(\omega^2 \zeta)
\end{array}\right)
& , \mbox{ for } \zeta \in \mathbb C_+,
\\
\left(\begin{array}{cc}
Ai(\zeta)& -\omega^2 Ai(\omega \zeta)\\
Ai'(\zeta)& -Ai'(\omega \zeta)
\end{array} \right)
& , \mbox{ for } \zeta \in \mathbb C_-.
\end{array}
\right.
\end{eqnarray}
The transformation $\Phi_{\beta}(z)$ was defined in (\ref{Phich1a}),
$\gamma(z)$ was defined in (\ref{GAMDEF}),
and the
matrix $B$ is defined by
\begin{eqnarray}
\label{3.002}
 B(z) = \pmtwo {\gamma^{-1}(z)}{-\gamma(z)}
{-i\gamma^{-1}(z)}{-i \gamma(z)} \Phi_{\beta}(z)^{\sigma_{3}/4},
\end{eqnarray}
and $S(z)$, defined in (\ref{SDEF}),
is the solution to the Riemann--Hilbert problem \ref{RHR} for the
error, which possesses a complete asymptotic expansion (\ref{SASSYM}),
with coefficients $S_{k}$ solving the Riemann--Hilbert problems
\ref{RHRec} and \ref{RHReck}, and satisfying the symmetry conditions
described in Lemma \ref{Lem:Sym}.
\bigskip

Shortly we will present an analogous representation of $\rho_N^{(1)}$
near the left endpoint $\alpha$. The following lemma will be instrumental
in describing the relation between these two representations as well as
their respective domains of validity.

\begin{lem}\label{match}
There is a unique point $z^*$ in the interval $(\alpha, \beta)$ at which
$(\Phi_{\beta}(z^*)) = (\Phi_{\alpha}(z^*))$
\end{lem}
\begin{Proof}
It follows from formulas (\ref{Phich1a}), (\ref{Phial1a}) and (\ref{II.002})
that for $z < \beta$,
\begin{equation}\label{matchbe}
(-\Phi_{\beta}(z))^{3/2} = \left(\frac{3 N}{4}\right)\int_z^{\beta}
\frac{R_+}{i}hds;
\end{equation}
and for $z > \alpha$,
\begin{equation}\label{matchal}
(-\Phi_{\alpha}(z))^{3/2} = \left(\frac{3 N}{4}\right)\int_{\alpha}^z
\frac{R_+}{i}hds.
\end{equation}
It then follows from comparison to (\ref{psi}) that the integrals in
(\ref{matchal}) and (\ref{matchbe}) are both positive.

By construction (see (\ref{Phich1b}) and (\ref{Phial1b})) we know that
$\Phi_{\beta}(z)$ and $\Phi_{\alpha}(z)$ are negative real for
$\alpha < z < \beta$. The $z^*$ we seek must satisfy

$$
\int_{\alpha}^{z^*} Rhds  \pm \int_{z^*}^{\beta} Rhds = 0.
$$
We already know that
$$
\int_{\alpha}^z Rhds  + \int_z^{\beta} Rhds =\int_\alpha^\beta Rhds = 2\pi i.
$$
It follows that $z^{*}$ is uniquely determined, equivalently, by either of
the two equations:
$$
\int_{\alpha}^{z*} Rhds = \pi i = \int_{z*}^{\beta} Rhds
$$
\end{Proof}

\begin{lem}
\label{Lem:4p2}
There is a $T_{\beta}>0$ sufficiently small and $\gamma_{\beta}$ sufficiently large (both independent of $N$)
so that for $ {\bf t} \in \mathbb{T}( T_{\beta}, \gamma_{\beta})$, there exists a complex
neighborhood ${\bf B}_{\beta} $ of the interval $[z^*,2]$,
independent of $N$, so that formula (\ref{3.001}) holds true for
all $z \in {\bf B}_{\beta}$.
\end{lem}

\begin{Proof}
The proof is a straightforward, though lengthy algebraic
manipulation of the explicit relationships between the original
quantity $Y(z)$, and the error matrix $S(z)$.  The only analytical
issue is this: is the transformation $\Phi_{\beta}(z)$ an
invertible analytic transformation from some fixed size
neighborhood of the interval $[z^*,2]$ into its range, for ${\bf t}$
sufficiently small? But this is clear, because for ${\bf t}=0$,
the function $\Phi_{\beta}$ is quite simple,
\begin{eqnarray}
\nonumber
 \left. \Phi_{\beta}(z) \right|_{{\bf t} = {\bf 0}} =
\left( \frac{3 N}{4} \int_{2}^{z} \sqrt{ s - 2}\sqrt{s+2} \ ds
\right)^{2/3}.
\end{eqnarray}
The reader may easily verify that there exists a branch of this
function which is analytic on a fixed open neighborhood of the
interval $(-2, 2]$.  Now since $\Phi_{\beta}(z)$ depends
analytically on ${\bf t}$, it is clear that $\Phi_{\beta}(z)$ is
certainly an invertible analytic transformation from some fixed
size neighborhood of $[0,2]$ into its range.

The reader may verify that formula (\ref{3.001}) is true, by
starting with the formula (\ref{II.001}) for $\rho_{1}^{(N)}$, and
re-expressing this formula in terms of the solution $Y$ of the
original Riemann--Hilbert problem
\begin{eqnarray} \label{3.003a}
\rho_{1}^{(N)}(z) = \frac{e ^{-N V(z)}}{-2 \pi i N} \left[
Y_{11}'(z) Y_{21}(z) - Y_{11}(z) Y_{21}'(z) \right].
\end{eqnarray}
Next, observe that for $z \in \cb_{\pm} \cap \{ \mbox{lens shaped
region in Figure \ref{KF2}} \}$ , we have the following
representation for the solution $Y$ of the original
Riemann--Hilbert problem:
\begin{eqnarray}
\label{3.004a}
\nonumber
 Y(z) &=& \\
\nonumber & & \hspace{-0.5in}
e^{ \frac{ N \ell}{2} \sigma_{3} } S(z) \
E_{\beta}(z) \ \Psi^{\sigma}(\Phi_{\beta}(z))
e^{\frac{2}{3}\sigma_3\Phi_{\beta}(z)^{3/2}}
\pmtwo {1}{0} {\pm e^{\mp N(g_+ - g_-)}}{1}
\ e^{ N (g-\frac{ \ell}{2}) \sigma_{3}} \nonumber \\
\label{III.010}
    & & = e^{ \frac{ n \ell}{2} \sigma_{3} } S(z) \
E_{\beta}(z) \ \Psi(\Phi_{\beta}(z)) \ e^{ \left( \frac{N}{2} V(z)
- \pi i /6 \right)\sigma_{3}}.
\end{eqnarray}
Here $E_{\beta}(z)$ was defined in (\ref{2.018}), and is related
to the matrix valued function $B(z)$ defined in (\ref{3.002}) via
\begin{eqnarray}
\label{3.005a} E_{\beta}(z) = \sqrt{\pi} e^{i \pi / 6} B(z).
\end{eqnarray}
The simplification in (\ref{III.010}) follows from the identity
\begin{equation}
\pmtwo{1}{0}{\mp 1}{1}
e^{\frac{2}{3}\sigma_3\Phi_{\beta}(z)^{3/2}} \pmtwo {1}{0} {\pm
e^{\mp N(g_+ - g_-)}}{1} \ e^{ N (g-\frac{ \ell}{2}) \sigma_{3}} =
e^{ \frac{N}{2} V(z)\sigma_{3}},
\end{equation}
which can be deduced from the properties of $g(z)$ and $\Phi_\beta(z)$.
Next, from the definition (\ref{3.002}), one easily obtains the
identity
\begin{eqnarray}
\label{3.005}
 B'(z) = \left( \frac{\Phi_{\beta}'(z)}{4 \Phi_{\beta}(z)} -
\frac{ \gamma'(z)}{\gamma(z)} \right) \ B(z) \sigma_{3}.
\end{eqnarray}

The following identity for $\Psi$ is also equally easy to prove,
using the definition (\ref{III.005}) and the differential equation
satisfied by the Airy function:
\begin{eqnarray}
\label{3.006a}
 \Psi'(\zeta) =
\pmtwo{0}{1}{\zeta}{0} \Psi(\zeta).
\end{eqnarray}
Formula (\ref{3.001}) may now be verified by differentiating
(\ref{3.004a}), substituting into (\ref{3.003a}) and using the
differential relations (\ref{3.005}) and (\ref{3.006a}).

\end{Proof}

Formula (\ref{3.001}) may seem unwieldy due to its length.
However, this is not quite the case.  Indeed, the first two lines
of this formula do not contain $S(z)$ at all, and are explicit
combinations of Airy functions (composed with $\Phi_{\beta}(z)$)
together with other functions defined explicitly in Section
\ref{Sec:2}.  Moreover,  we will show that the third and fourth
line of (\ref{3.001}) possess asymptotic expansions in powers of
$N^{-1}$ in which each term is of the same explicit form as the
first two lines.  This is the content of Lemma \ref{Lem:3.3}
below.

There is, of course, an analogous formula which is valid on a
fixed size complex neighborhood of the interval $[-2,z^*]$.  We
summarize the analogous result in the following Lemma.

\begin{lem}
\label{Lem:4p3}
There is $T_{\alpha}>0$ sufficiently small and $\gamma_{\alpha}$ sufficiently large (both independent of $N$) so
that for $ {\bf t} \in \mathbb{T}( T_{\alpha},\gamma_{\alpha})$, there exists a complex
neighborhood of ${\bf B}_{\alpha}$ of the interval $[-2,z^*]$,
independent of $N$, so that the formula (\ref{3.003}) holds true
for all $z \in {\bf B}_{\alpha}$.
\begin{eqnarray}
\label{3.003}
& &
N \rho_{N}^{(1,\alpha)}(z) =
- \left( \frac{\Phi_{\alpha}'(z)}{4 \Phi_{\alpha}(z)} +
\frac{\gamma'(z)}{\gamma(z)} \right) \left[ 2 \
Ai\left(\Phi_{\alpha}(z) \right) \
Ai^\prime\left( \Phi_{\alpha}(z)\right)  \right] \\
\nonumber
& & \hspace{0.75in}
- \Phi_{\alpha}'(z)
\left[ \Big(Ai^\prime\left(\Phi_{\alpha}(z) \right)\Big)^{2} -
\Phi_{\alpha}(z) \
\Big(Ai\left( \Phi_{\alpha}(z) \right)\Big)^{2} \right] \\
\nonumber
& & \hspace{0.75in}
+ \frac{i}{2} \left[
\Big( S' \ B \ \Psi\left(\Phi_{\alpha}(z) \right) \Big)_{11} \
\Big( S \ B \ \Psi\left(\Phi_{\alpha}(z) \right) \Big)_{21} \right. \\
& & \nonumber \hspace{1.25in} \left.
 - \
\Big( S \ B \ \Psi\left(\Phi_{\alpha}(z) \right) \Big)_{11} \
\Big( S' \ B \ \Psi\left(\Phi_{\alpha}(z) \right) \Big)_{21}  \right],
\end{eqnarray}
\end{lem}
where in this case
\begin{eqnarray}
 B(z) = \pmtwo {\gamma(z)}{-\gamma^{-1}(z)}
{i\gamma(z)}{i \gamma^{-1}(z)} \Phi_{\alpha}(z)^{\sigma_{3}/4}.
\end{eqnarray}

Next, we will compute an asymptotic expansion for
$\rho_{N}^{(1)}$, by using formulae (\ref{3.001}), (\ref{3.003}),
and the asymptotic expansion for $S(z)$ described in Section
\ref{Sec:2}.

\begin{lem}
\label{Lem:3.3} If ${\bf t} \in \mathbb{T}(T_{\beta},\gamma_{\beta} )$, then there is a
fixed size neighborhood of the interval $[z^{*},
\beta]$, the following asymptotic expansion holds true:
\begin{eqnarray}\label{3.004}
& & \frac{i}{2N} \left[ \Big( S' \ B \ \Psi\left(\Phi_{\beta}(z)
\right) \Big)_{11} \
\Big( S \ B \ \Psi \left(\Phi_{\beta}(z) \right) \Big)_{21} \right. \\
& & \nonumber \hspace{1.25in} \left.
 - \
\Big( S \ B \ \Psi\left(\Phi_{\beta}(z) \right) \Big)_{11} \ \Big(
S' \ B \ \Psi\left(\Phi_{\beta}(z)
\right) \Big)_{21}  \right] \\
\nonumber
& & \hspace{1.30in} = \sum_{j {\mbox{ {\rm{even,}} }} j
\ge 2} N^{-j} \tilde{a}_{j}(z) \Psi_{11}^{2}(\Phi_{\beta})
\frac{\sqrt{\Phi_{\beta}}}{\gamma(z)^{2}} \\
\nonumber
& & \hspace{1.5in}+ \sum_{j {\mbox{ {\rm{even, }}}} j \ge
2} N^{-j} \tilde{b}_{j}(z) \frac{ \Psi_{21}^{2}(\Phi_{\beta}(z))
\gamma(z)^{2}}{\sqrt{\Phi_{\beta}(z)} } \\
\nonumber
& & \hspace{1.5in} + \sum_{j \mbox{{\rm{ odd, }}}j
\ge3}^{\infty} N^{-j} \tilde{c}_{j}(z) \Psi_{11}(\Phi_{\beta}(z))
\Psi_{21}(\Phi_{\beta}(z)),
\end{eqnarray}
where for each $j$, the coefficient functions $\tilde{a}_{j}(z)$,
$\tilde{b}_{j}(z)$, and $\tilde{c}_{j}(z)$ are jointly analytic in $z$
and ${\bf t}$, for $z$ in a fixed sized complex
neighborhood of the interval $[z^{*},\beta]$, and for
${\bf t} \in \mathbb{T}(T_{\beta}, \gamma_{\beta})$.
\end{lem}
\begin{Proof}
From Lemma \ref{Lem:Sym}, we have an asymptotic expansion for the
function $S(z)$:
\begin{eqnarray}\label{3.010}
& & S(z) = I + \sum_{k {\mbox{\rm{ odd, }}}k \ge 1} \left(
s_{k}^{(1)}(z) \sigma_{3} + s_{k}^{(2)}(z) \sigma_{1} \right)
 N^{-k} \\
\nonumber & & \hspace{0.75in} + \sum_{k {\mbox{\rm{ even, }}} k
\ge 2} \left( s_{k}^{(1)}(z) I + s_{k}^{(2)}(z) \sigma_{2} \right)
 N^{-k},
\end{eqnarray}
in which $s_{k}^{(j)}(z)$ are functions of $z$ that are piecewise
analytic functions of $z$, which are, in particular, analytic on a
fixed size neighborhood of the interval $[0,\beta]$, and
depend analytically on ${\bf t}$ for ${\bf t} \in
\mathbb{T}(T_{\beta}, \gamma_{\beta})$.  An
asymptotic expansion for $S'(z)$ is obtained by differentiating
(\ref{3.010}) term by term.  This, together with the explicit
formula (\ref{3.002}) for $B(z)$, may be substituted into the left
hand side of (\ref{3.004}).  Now straightforward algebraic
manipulation yields an asymptotic expansion of the form displayed
in the right hand side of (\ref{3.004}).  The quantities
$\tilde{a}_{j}$ and $\tilde{b}_{j}$ may be computed via these
calculations, but all that we will require is that these functions
are analytic in a neighborhood of $[0, \beta]$, and
depend analytically on ${\bf t}$ for ${\bf t} \in
\mathbb{T}(T_{\beta}, \gamma_{\beta})$, which is manifestly true.
\end{Proof}

There is, of course, an analogous result which holds for the interval
$[\alpha - \delta, z^{*}]$, which we will only state, as the proof
follows by the same arguments as the proof of Lemma \ref{Lem:3.3}
\begin{lem}
\label{Lem:3.5} If ${\bf t} \in \mathbb{T}(T_{\alpha}, \gamma_{\alpha} )$, then there is a
fixed size neighborhood of the interval $[\alpha, z^{*}]$, on which the
following asymptotic expansion holds true:
\begin{eqnarray}\label{3.004aa}
& & \frac{i}{2N} \left[ \Big( S' \ B \ \Psi\left(\Phi_{\alpha}(z)
\right) \Big)_{11} \
\Big( S \ B \ \Psi \left(\Phi_{\alpha}(z) \right) \Big)_{21} \right. \\
& & \nonumber \hspace{1.25in} \left.
 - \
\Big( S \ B \ \Psi\left(\Phi_{\alpha}(z) \right) \Big)_{11} \
\Big( S' \ B \ \Psi\left(\Phi_{\alpha}(z)
\right) \Big)_{21}  \right] \\
\nonumber & & \hspace{1.30in} = \sum_{j {\mbox{ {\rm{even,}} }} j
\ge 2} N^{-j} \tilde{a}_{j}(z) \Psi_{11}^{2}(\Phi_{\alpha})
\gamma(z)^{2}\sqrt{\Phi_{\alpha}}\\
\nonumber & & \hspace{1.5in}+ \sum_{j {\mbox{ {\rm{even, }}}} j
\ge 2} N^{-j} \tilde{b}_{j}(z) \frac{
\Psi_{21}^{2}(\Phi_{\alpha}(z))}
{\gamma(z)^{2}\sqrt{\Phi_{\alpha}(z)} }\\
\nonumber & & \hspace{1.5in} + \sum_{j \mbox{{\rm{ odd, }}}j
\ge3}^{\infty} N^{-j} \tilde{c}_{j}(z) \Psi_{11}(\Phi_{\alpha}(z))
\Psi_{21}(\Phi_{\alpha}(z)),
\end{eqnarray}
where for each $j$, the coefficient functions $\tilde{a}_{j}(z)$,
$\tilde{b}_{j}(z)$, and $\tilde{c}_{j}(z)$ are jointly analytic in $z$
and ${\bf t}$, for $z$ in a fixed size complex neighborhood of the
interval $[\alpha, z^{*}]$, and for
${\bf t} \in \mathbb{T}(T_{\alpha},\gamma_{\alpha})$.
\end{lem}

{\bf Remark:}  Observe that if we take $T < \mbox{min}\{ T_{\beta}, T_{\alpha}\}$ and $\gamma > \mbox{max} \{ \gamma_{\beta},\gamma_{\alpha} \}$ then the conclusions of Lemmas \ref{Lem:4p2}, \ref{Lem:4p3}, \ref{Lem:3.3}, and \ref{Lem:3.5} hold true simultaneously.  From now on we will take ${\bf{t}} \in \mathbb{T}(T,\gamma)$ for such values of $T$ and $\gamma$.
\section{Proof of the Main Theorem}
\label{ProofMT}

We are now in a position to analyze the structure of the fundamental
moments (\ref{I.SMF}) which will enable us to establish the $1/N^2$
expansion of $\log Z_N$. To do this we will be using the representations
of the one-point function developed in the previous section. Precisely,
we use
\begin{equation}\label{one-point}
\rho_{N}^{(1)}(z) =
\chi_\alpha(z)\rho_N^{(1,\alpha)} + \chi_\beta(z)\rho_N^{(1,\beta)}
\end{equation}
where $\left\{ \chi_\alpha, \chi_\beta \right\}$ is a partition of unity
for $\mathbb{R}$ defined by
\begin{eqnarray}\label{partition}
\chi_\beta(z) \, &\mbox{is}& \,\, C^\infty \,\,\, \mbox{with}\\ \nonumber
0 \leq \chi_\beta(z) &\leq& 1 \\ \nonumber
\overline{supp\chi_\beta} &\subset& (z^* - \epsilon, \infty)\\
\nonumber
\chi_\beta(z) &\equiv 1&  \quad \mbox{ for } z \in
(z^* + \epsilon, \infty).
\end{eqnarray}
We then define $\chi_\alpha$ by
$$
\chi_\alpha(z) = 1 - \chi_\beta(z)
$$
which has similar properties to $\chi_\beta$ but with support in
$(-\infty, z^* + \epsilon)$

The expression (\ref{one-point}) is merely a different representation
of the one-point function $\rho_N^{(1)}$, and the partition of unity
allows us to carry out our asymptotic  analysis separately on
$\chi_\alpha(z)\rho_N^{(1,\alpha)}(z)$ in
${\bf B}_\alpha$ and on $\chi_\beta(z)\rho_N^{(1,\beta)}(z)$
in ${\bf B}_\beta$,  and then, at the end, to add the two contributions
together.  This partition of unity will prove especially useful in
avoiding the evaluation of boundary terms during repeated integrations by
parts.  Because of the complete similarity of the analysis for each of
these separate pieces it will suffice to demonstrate our claims for
$\chi_\beta(z)\rho_N^{(1,\beta)}(z)$.

\bigskip

We now begin the analysis of integrals of the form
\begin{eqnarray}
\label{4.003INT}
\hspace{0.3in}
\int_{\alpha - \delta}^{\beta+\delta} f(\lambda)
\rho_{N}^{(1)}(\lambda) d \lambda = \int_{\alpha - \delta}^{z^{*} +
\epsilon}
\chi_{\alpha}(\lambda) \rho_{N}^{(1)}(\lambda) d \lambda + \int_{z^{*} -
\epsilon}^{\beta + \delta} \chi_{\beta}(\lambda) \rho_{N}^{(1)}(\lambda)
d \lambda.
\end{eqnarray}
(Recall from Observation 2 and formula (\ref{3.002INT}) that we need only
consider integrals over the bounded set $(\alpha-\delta, \beta + \delta)$
for some sufficiently small (but independent of $N$) $\delta > 0$.)  We
will prove that if the function $f$ is $C^{\infty}$ smooth and
compactly supported within $(\alpha-\delta, \beta+ \delta)$, the second
integral in (\ref{4.003INT}) above possesses an asymptotic expansion in
even powers of $N$:
\begin{eqnarray}
\label{4.004IAS}
\int_{z^{*} -
\epsilon}^{\beta + \delta} \chi_{\beta}(\lambda) \rho_{N}^{(1)}(\lambda)
d \lambda = f_{0} + N^{-2} f_{1} + N^{-4} f_{2} + \cdots.
\end{eqnarray}
As we have explained, this will complete the proof that
\begin{eqnarray}
\label{expDerivLogASS}
\frac{\partial}{\partial t_{\ell}} \frac{1}{N^{2}} \log{ Z_{N}({\bf t})} =
\tilde{e}_{0}({\bf t}) + \frac{1}{N^{2}} \tilde{e}_{1}({\bf t}) + \cdots
\end{eqnarray}
i.e. that $\partial_{t_{\ell}} \log{Z_{N}}$ possesses a complete
asymptotic expansion in even powers of $N$.  Since this asymptotic
expansion is uniformly valid for all ${\bf t} \in
\mathbb{T}(T,\gamma)$, we may integrate (\ref{expDerivLogASS}) from ${\bf t} = {\bf
0}$, and thus we will complete the proof of Theorem \ref{I.002thm}.

From the analysis of the previous section, in particular lemma \ref{Lem:3.3},
one observes that the terms appearing in an expression of the form
$\int_{z^{*}-\epsilon}^{\beta+\delta}
\lambda^\ell\chi_\beta(\lambda)\rho_{N}^{(1,\beta)}(\lambda)$
are of one of the following four types of integrals:

\begin{eqnarray*}\label{4.001}
(0') & & \frac{1}{N}\int_{z^*-\epsilon}^{\beta + \delta} g(z)
F_0(\Phi_{\beta})\Phi_{\beta}^\prime dz, \\
(1') & & N^{-j} \int_{z^*-\epsilon}^{\beta + \delta} g(z)
Ai(\Phi_{\beta})Ai^\prime(\Phi_{\beta}) dz \ \mbox{ ($j$ odd)},\\
(2') & & N^{-j}\ \int_{z^*-\epsilon}^{\beta + \delta} g(z)
\frac{\sqrt{\Phi_{\beta}}}{\gamma(z)^{2}}Ai^{2}(\Phi_{\beta}) dz =
N^{1/3-j} \int_{z^*-\epsilon}^{\beta + \delta} \tilde{g}(z)
Ai^{2}(\Phi_{\beta}) dz \
\mbox{( $j$ even),}\\
(3') & & \hspace{-0.25in} N^{-j}\int_{z^*-\epsilon}^{\beta +
\delta} \hspace{-0.1in} g(z)
\frac{\gamma(z)^{2}}{\sqrt{\Phi_{\beta}}}[Ai^\prime]^{2}(\Phi_{\beta})
dz = N^{-1/3-j} \int_{z^*-\epsilon}^{\beta + \delta} \tilde{g}(z)
[Ai^\prime]^{2}(\Phi_{\beta}) dz, \ \mbox{ ($j$ even)},
\end{eqnarray*}
where $g(z)$ and $\tilde{g}(z)$ are general infinitely differentiable
functions of $z$, and are compactly supported within
$(z^*-\epsilon,\beta + \delta)$. The function $F_{0}$ appearing in (0') above is defined by
\begin{equation} \label{f0}
F_0(\zeta) := [Ai^\prime]^{2}(\zeta) - \zeta
Ai^{2}(\zeta).
\end{equation}

The main result would be established if we could show that all of
these quantities possess asymptotic expansions in even powers of
$N$.  We therefore must analyze the following four quantities:
\begin{eqnarray*}\label{4.001b}
(0) \hspace{0.15in} \frac{1}{N}\int_{z^*-\epsilon}^{\beta + \delta} g(z)
F_0(\Phi_{\beta})\Phi_{\beta}^\prime dz, & &
(1) \hspace{0.15in} \int_{z^*-\epsilon}^{\beta + \delta} g(z)
Ai(\Phi_{\beta})Ai^\prime(\Phi_{\beta}) dz,\\
(2) \hspace{0.15in} N^{1/3} \int_{z^*-\epsilon}^{\beta + \delta} \tilde{g}(z)
Ai^{2}(\Phi_{\beta}) dz , & &
(3) \hspace{0.15in} N^{-1/3} \int_{z^*-\epsilon}^{\beta + \delta} \tilde{g}(z)
[Ai^\prime]^{2}(\Phi_{\beta}) dz,
\end{eqnarray*}
and prove the following statements true:
\begin{eqnarray*}\label{4.002}
& & (0), \ (2), \mbox{ and } (3)
\mbox{ have asymptotic expansions in even powers of}\,\, N. \\
& & (1) \mbox{ has an asymptotic expansion in odd powers of} \,\, N. \end{eqnarray*}

In fact, once we show that integrals of the form (0) have asymptotic expansions in even powers of $N$, the following three basic observations take care of the remaining types of integrals.

{\bf Observation 1}:

(1) has an asymptotic expansion in odd powers of N if and only if
(2) has an asymptotic expansion in even powers of N:

\begin{eqnarray*}\label{4.003}
2\int_{z^*-\epsilon}^{\beta + \delta} g Ai(\Phi_{\beta})Ai^\prime(\Phi_{\beta}) dz &=&
 -\int_{z^*-\epsilon}^{\beta + \delta} \left[\frac{g}{\Phi_\beta^\prime}\right]^\prime
 Ai^2(\Phi_{\beta})dz \\
&=& -\frac{1}{N} \int_{z^*-\epsilon}^{\beta + \delta}
N^{1/3}\left[\frac{g}{\phi_\beta^\prime}\right]^\prime Ai^2(\Phi_{\beta})dz
\end{eqnarray*}
where in the first equality we've integrated by parts. (No
boundary terms are present as the integrand is compactly supported
within $(z^{*} - \epsilon, \beta + \delta)$.)  We now observe that
$\Phi_\beta^\prime(z)$ is analytic and nonvanishing throughout the
region of integration so that the integral is well-defined.
Lastly, the notation $\phi_{\beta}'(z) := N^{-2/3}
\Phi_{\beta}'(z)$ defines an analytic, nonvanishing function which
is independent of $N$.  The quantity $ \left[
{g}/{\phi_{\beta}'}\right]' $ is a new infinitely differentiable
function of $z$, compactly supported within $(z^{*} - \epsilon, \beta+\delta)$, and we have shown that integrals of the form
(2) possess asymptotic expansions in even powers of $N$ if and only if integrals of the form (1) possess expansions in odd powers of $N$.

{\bf Observation 2}:

Integrals of the form (2) have asymptotic expansions in even powers of N if and only if integrals of the form (0) have asymptotic expansions in even powers of N:

\begin{eqnarray*}\label{4.004}
N^{1/3} \int_{z^*-\epsilon}^{\beta + \delta} g Ai^{2}(\Phi_{\beta}) dz &=&
N^{1/3} \int_{z^*-\epsilon}^{\beta + \delta}
\frac{g}{\Phi_\beta^\prime} Ai^{2}(\Phi_{\beta}) d\Phi_{\beta} \\ &=&
\frac{1}{N}\int_{z^*-\epsilon}^{\beta + \delta}
\left[\frac{g}{\phi_\beta^\prime}\right]^\prime
\frac{1}{\phi_\beta^\prime}F_0(\Phi_{\beta}) \Phi_{\beta}^\prime dz
\end{eqnarray*}
where once again we have integrated by parts in the second
equality.  Now we have arrived at an integral of the form (0),
since the quantity $ \left( {1}/{\phi_\beta^\prime}
\right)\left[{g}/{\phi_\beta^\prime}\right]^\prime
 $ is infinitely differentiable on
$[z^{*}-\epsilon, \beta + \delta]$.

{\bf Observation 3}:

Finally, if integrals of the forms (0) and (1) respectively have  even and odd asymptotic
expansions, then integrals of the form (3) have even asymptotic expansions:

\begin{eqnarray*}\label{4.005}
N^{-1/3} \int_{z^*-\epsilon}^{\beta + \delta} g
[Ai^\prime]^{2} (\Phi_{\beta}) dz &=& \\ & & \hspace{-.5in}
-N^{-1} \int_{z^*-\epsilon}^{\beta + \delta}
\left[\frac{g}{\phi_\beta^\prime}\right]^\prime
\{\frac{2}{3}Ai(\Phi_{\beta})Ai^\prime(\Phi_{\beta}) + \frac{1}{3}
\Phi_\beta F_0(\Phi_\beta)\} dz
\\ &=&
-\frac{2}{3}N^{-1} \int_{z^*-\epsilon}^{\beta + \delta}
\left[\frac{g}{\phi_\beta^\prime}\right]^\prime
Ai(\Phi_{\beta})Ai^\prime(\Phi_{\beta}) dz \\ & &
-\frac{1}{3}N^{-1}\int_{z^*-\epsilon}^{\beta + \delta} \left[\frac{g}{\phi_\beta^\prime}
\right]^\prime \frac{\phi_\beta}{\phi_\beta^\prime} F_0(\Phi_\beta)
\Phi_\beta^\prime dz.
\end{eqnarray*}
There is an integration by parts in the first equation. The two
terms in  the last equation are integrals of the form (1) and (0),
respectively, and this establishes our third observation.

\smallskip

We turn now to the heart of the matter.  We will give an inductive
proof that
\begin{equation}
\label{F0Int}
 \frac{1}{N}\int_{z^*-\epsilon}^{\beta + \delta} g(z)
F_0(\Phi_{\beta})\Phi_{\beta}^\prime dz
\end{equation}
has an asymptotic expansion in even powers of $N$.  We will
integrate by parts repeatedly, peeling off contributing factors as
we go along.  The proof proceeds in four steps.  The first three steps involve
setting up an integration by parts inductive argument.

{\it Step 1}: Show that the function $F_{0}$, which is
analytic in $\zeta$, possesses the following asymptotic
expansion for $\zeta \to - \infty$:
\begin{eqnarray}
\label{4.008}
& & F_{0}(\zeta) = \sqrt{-\zeta}\left(
c_{0} + c_{1} (-\zeta)^{-3} + c_{2} (-\zeta)^{-6} \cdots \right) \\
\nonumber & & \hspace{0.2in}
+ \sqrt{-\zeta}\left(d_{0} (-\zeta)^{-3} + d_{1} (-\zeta)^{-6} + d_{2} (-\zeta)^{-9} +
\cdots \right) \sin{\left( \frac{4}{3} (-\zeta)^{3/2}\right)} \\
\nonumber & & \hspace{0.2in}
+
\left(f_{0} (-\zeta)^{-1} + f_{1} (-\zeta)^{-4} + f_{2} (-\zeta)^{-7} +
\cdots \right) \cos{\left( \frac{4}{3} (-\zeta)^{3/2}\right)}.
\end{eqnarray}

{\it Step 2}: Show that Step 1 implies that there are functions
$G_{1}(\zeta)$, $G_{2}(\zeta)$, and
$G_{3}(\zeta)$ which are analytic in $\zeta$, satisfy
\begin{eqnarray}
\label{4.009}
& &   G_{1}'(\zeta) = \left( F_{0}(\zeta) - c
(-\zeta)^{1/2}
\right), \\\label{4.010}
& & G_{2}'(\zeta) = G_{1}(\zeta), \\
\label{4.011}
& & G_{3}'(\zeta) = G_{2}(\zeta),
\end{eqnarray}
and possesses the following asymptotic expansion for $\zeta \to
-\infty$:
\begin{eqnarray}
\label{4.012}
\hspace{0.2in}G_{1}(\zeta) &=& \left( c_{0}^{(1)} (-\zeta)^{-3/2} +
c_{1}^{(1)} ( - \zeta)^{-9/2} + \cdots \right) \\
\nonumber & & + \left( d_{0}^{(1)} (-\zeta)^{-3/2} + d_{1}^{(1)}
(-\zeta)^{-9/2} + \cdots \right) \sin{ \left( \frac{4}{3}
(-\zeta)^{3/2} \right)} \\
\nonumber & & + \left(f_{0}^{(1)} (-\zeta)^{-3} + f_{1}^{(1)}
(-\zeta)^{-6} + \cdots \right) \cos{ \left( \frac{4}{3}
(-\zeta)^{3/2} \right)}\\
\label{4.013}
G_{2}(\zeta) &=& \left( c_{0}^{(2)} (-\zeta)^{-1/2} +
c_{1}^{(2)} ( - \zeta)^{-7/2} + \cdots \right) \\
\nonumber & & + \left( d_{0}^{(2)} (-\zeta)^{-7/2} + d_{1}^{(2)}
(-\zeta)^{-13/2} + \cdots \right) \sin{ \left( \frac{4}{3}
(-\zeta)^{3/2} \right)} \\
\nonumber & & + \left(f_{0}^{(2)} (-\zeta)^{-2} + f_{1}^{(2)}
(-\zeta)^{-5} + \cdots \right) \cos{ \left( \frac{4}{3}
(-\zeta)^{3/2} \right)}\\
 \label{step2}
G_{3}(\zeta) &=& c_{0}^{(3)} (-\zeta)^{1/2} + c_{1}^{(3)} (-\zeta)^{-5/2} +\cdots \\
& & \nonumber \hspace{0.2in} + \left( d_{0}^{(3)} (-\zeta)^{-5/2}
+ d_{1}^{(3)} (-\zeta)^{-11/2} + \cdots \right) \sin{\left(
\frac{4}{3}
(-\zeta)^{3/2} \right)} \\
& & \nonumber \hspace{0.2in}+ \left( f_{0}^{(3)} (-\zeta)^{-4} +
f_{1}^{(3)} (-\zeta)^{-7} + \cdots \right) \cos{ \left(
\frac{4}{3} (-\zeta)^{3/2} \right) }.
\end{eqnarray}

This defines antiderivatives of $F_{0}$ which we will also need in
what follows:
\begin{eqnarray}
\label{4.015}
F_{1} &=& G_{1} - \frac{2c_{0}}{3} (-\zeta)^{3/2},
\mbox{ ( } F_{1}' = F_{0} \mbox{ ) },\\
\label{4.016}
F_{2} &=& G_{2} + \frac{4c_{0}}{15} (-\zeta)^{5/2},
\mbox{ ( } F_{2}' = F_{1} \mbox{ ) }, \\
\label{4.017}
F_{3} &=& G_{3} - \frac{8c_{0}}{105} (-\zeta)^{7/2}, \mbox{ ( }
F_{3}' = F_{2} \mbox{ ) },
\end{eqnarray}
and we will show that these three anti-derivatives are all
exponentially decaying for $\zeta \to+\infty$:
\begin{eqnarray}
\label{4.018}
\left|F_{\ell}\right| \le C_{\ell} e^{-\frac{4}{3}\zeta^{3/2}}
\end{eqnarray}

{\it Step 3:} We express the integral (\ref{F0Int}) as
follows
\begin{eqnarray}
& & \hspace{0.3in}
\frac{1}{N}\int_{z^*-\epsilon}^{\beta + \delta}
g(z)
F_0(\Phi_{\beta})\Phi_{\beta}^\prime dz = \hat{e}_{0} + A, \\
\label{IV.020} & & \hspace{0.1in}\hat{e}_{0} =
\frac{1}{N}\int_{z^*-\epsilon}^{\beta} g(z) c_{0} \left(
-\Phi_{\beta}(z) \right)^{1/2} \Phi_{\beta}^\prime dz \\
 \label{IV.021} & & \hspace{0.05in}A =
\frac{1}{N}\int_{z^*-\epsilon}^{\beta } \hspace{-0.1in}g(z) \left(
F_0(\Phi_{\beta}) - c_{0} \left( -\Phi_{\beta}(z) \right)^{1/2}
\right)\Phi_{\beta}^\prime dz+ \frac{1}{N}\int_{\beta}^{\beta
+\delta} \hspace{-0.1in} g(z) F_0(\Phi_{\beta})
\Phi_{\beta}^\prime dz
\end{eqnarray}
Now $\hat{e}_{0}$ is in fact independent of $N$:
\begin{eqnarray}
\label{e0DEF}& & \hspace{0.01in}
\hat{e}_{0} = \frac{-c_{0}}{N} \int_{z^{*} - \epsilon}^{\beta}
\hspace{-0.1in} g(z) \frac{d}{dz} \left( \frac{2}{3} \left(
-\Phi_{\beta}(z) \right)^{3/2} \right) dz = \frac{-ic_{0}}{2}
\int_{z^{*} -\epsilon}^{\beta} \hspace{-0.1in} g(z)R_{+}(z) h(z)
dz
\end{eqnarray}
and so we must show that $A$ possesses an asymptotic expansion in
even powers of $N$.  Integrating both terms in (\ref{IV.021}) by
parts, we obtain
\begin{eqnarray}
\label{PStep3}& & \hspace{-0.2in}
A =
-\frac{1}{N}\int_{z^{*}-\epsilon}^{\beta}\hspace{-0.1in}
g'(z)  G_{1}\left(\Phi_{\beta}(z)\right)dz - \frac{1}{N}
\int_{\beta}^{\beta+\delta}\hspace{-0.1in}g'(z)
F_{1}\left(\Phi_{\beta}(z)\right) dz.
\end{eqnarray}
Observe that in (\ref{PStep3}), the boundary terms from $z^{*}-\epsilon$ and $\beta+\delta$ do not contribute because $g$ is compactly supported within $(z^{*}-\epsilon, \beta+\delta)$.  Furthermore, the boundary terms from the integrals at $z=\beta$ cancel, because $F_{1}(\Phi_{\beta}(z))$ and $G_{1}(\Phi_{\beta}(z))$ coincide at $z=\beta$.  Since $G_{1}$ and $F_{1}$ are both uniformly bounded, it follows immediately that $\left| A \right| \le CN^{-1}$.  However, it is not yet straightforward to deduce the asymptotic behavior for $A$ as $N \to \infty$, and that is our next task.

Two more integrations by parts yield
\begin{eqnarray}
\label{AExp1}
\hspace{0.3in}A &=&-\frac{1}{N}\int_{z^{*}-\epsilon}^{\beta}
\left[\frac{1}{\Phi_{\beta}'}
\left(\frac{1}{\Phi_{\beta}'}
g'(z)  \right)' \right]'
G_{3}\left(\Phi_{\beta}(z)\right)dz\\
\nonumber
& & \hspace{1.0in}
 - \frac{1}{N}
\int_{\beta}^{\beta+\delta}
\left[\frac{1}{\Phi_{\beta}'}
\left( \frac{1}{\Phi_{\beta}'}
g'(z) \right)' \right]'
F_{3}\left(\Phi_{\beta}(z)\right) dz.
\end{eqnarray}
Recalling the asymptotic expansion (\ref{step2}) for $G_{3}$ valid
for $\zeta \to -\infty$, we may rewrite (\ref{AExp1}) as follows:
\begin{eqnarray}
& & A = \hat{e}_{1} \ N^{-2} + A_{2}, \\
& & \hat{e}_{1} =-N\int_{z^{*}-\epsilon}^{\beta}
\left[\frac{1}{\Phi_{\beta}'}
\left(\frac{1}{\Phi_{\beta}'}
g'(z)  \right)' \right]'
c_{0}^{(3)} \left( - \Phi_{\beta}(z) \right)^{1/2}dz, \\
\label{A2DEF} & &A_{2} = -\frac{1}{N}\int_{z^{*}-\epsilon}^{\beta}
\left[\frac{1}{\Phi_{\beta}'} \left(\frac{1}{\Phi_{\beta}'} g'(z)
\right)' \right]' \left( G_{3}\left(\Phi_{\beta}(z)\right) -
c_{0}^{(3)}  \left( -\Phi_{\beta}(z) \right)^{1/2}
\right) dz \\
\nonumber & & \hspace{0.4in}
- \frac{1}{N}
\int_{\beta}^{\beta+\delta}
\left[\frac{1}{\Phi_{\beta}'}
\left( \frac{1}{\Phi_{\beta}'}
g'(z) \right)' \right]'
F_{3}\left(\Phi_{\beta}(z)\right) dz.
\end{eqnarray}
As with $\hat{e}_{0}$, $\hat{e}_{1}$  is independent of $N$. Indeed, using (\ref{matchbe}), we have
\begin{eqnarray}
\label{e1DEF}
& &
\hat{e}_{1}=-\int_{z^{*}-\epsilon}^{\beta}
\left[\frac{1}{\phi_{\beta}'}
\left(\frac{1}{\phi_{\beta}'}
g'(z)  \right)' \right]'
\tilde{c}_{0} \left( - \phi_{\beta}(z) \right)^{1/2}dz, \\
& &\mbox{where  for $z \in (z^{*}-\epsilon,\beta)$, }
\phi_{\beta}(z) =  -\left(\frac{-3i}{4} \int_{z}^{\beta}R_{+} (s) h(s) ds
\right)^{2/3}.
\end{eqnarray}
\vskip 0.2in

Observe that once we have established Steps 1 and 2, then Step 3, carried out above,
shows that
\begin{eqnarray}
\label{PriIndu}
\frac{1}{N} \int_{z^{*}-\epsilon}^{\beta+\delta}g(z) F_{0}(\Phi_{\beta}) \Phi_{\beta}'
dz = \hat{e}_{0} + \frac{1}{N^{2}} \hat{e}_{1} + A_{2},
\end{eqnarray}
with $\hat{e}_{0}$ and $\hat{e}_{1}$ independent of $N$ (see  (\ref{e0DEF}) and (\ref{e1DEF}), respectively), and with $A_{2}$ defined in (\ref{A2DEF}).  Furthermore, it is straightforward to prove that $\left|A_{2}\right| \le C N^{-7/3}$, and hence (\ref{PriIndu}) shows that the first two terms in the asymptotic expansion of (\ref{F0Int}) are in even powers of $N$, as desired.  The above calculations must now be automated to establish the existence of a complete asymptotic expansion in even powers of $N$.

{\it Step 4}: The procedure should now be clear to establish the induction.  We must prove that we can integrate by parts thrice, peel off the next term in the asymptotic expansion, and repeat.

We must prove that there exists $G_{m}$, $F_{m}$, $m = 4, 5,\ldots$, satisfying a number of relations, all appearing in triples.  First the differential relations:
\begin{eqnarray}
\label{GDIF}
G_{3j+1}'=G_{3j} - c_{0}^{(3j)}(-\zeta)^{1/2}, \ G_{3j+2}'=G_{3j+1}, \
G_{3j+3}' = G_{3j+2},
\end{eqnarray}
\begin{eqnarray}
\label{FDIF}
F_{3j+1}'=F_{3j}, \  F_{3j+2}'=F_{3j+1}, \ F_{3j+3}'=F_{3j+2},
\end{eqnarray}
then the asymptotic prescription
\begin{eqnarray}
\hspace{0.4in}
\label{GASS}
G_{3j+3}&=& (-\zeta)^{1/2} \left( c_{0}^{(3j+3)} + c_{1}^{(3j+3)} (-\zeta)^{-3} +\cdots  \right) \\
& & \nonumber \hspace{0.05in} + G_{S}^{(3j+3)}
\sin{\left( \frac{4}{3}
(-\zeta)^{3/2} \right)} + G_{C}^{(3j+3)} \cos{ \left(
\frac{4}{3} (-\zeta)^{3/2} \right) },
\end{eqnarray}
where $G_{S}^{(3j+3)}$ and $G_{C}^{(3j+3)}$ are asymptotic series which behave as follows:
\begin{eqnarray}& &
\mbox{ For $j$ even:} \left\{
\begin{array}{c}
G_{S}^{(3j+3)} = \left(-\zeta \right)^{-(3j+5)/2} \left( d_{0}^{(3j+3)}
+ d_{1}^{(3j+3)}(-\zeta)^{-3} + \cdots \right)\\
\hspace{-0.1in}
G_{C}^{(3j+3)} = \left(-\zeta \right)^{-(3j+8)/2} \left( f_{0}^{(3j+3)} +
f_{1}^{(3j+3)}(-\zeta)^{-3} + \cdots \right) \end{array} \right.
\\
& &
\mbox{For $j$ odd:}\left\{ \begin{array}{c}
G_{S}^{(3j+3)} = \left(-\zeta \right)^{-(3j+8)/2} \left( d_{0}^{(3j+3)}
+ d_{1}^{(3j+3)}(-\zeta)^{-3} + \cdots \right) \\
\hspace{-0.15in}G_{C}^{(3j+3)} = \left(-\zeta \right)^{-(3j+5)/2} \left( f_{0}^{(3j+3)} +
f_{1}^{(3j+3)}(-\zeta)^{-3}\cdots \right) \end{array} \right.
\end{eqnarray}
(note that this implies a similar asymptotic behavior for $G_{3j+1}$ and $G_{3j+2}$).  The quantities $F_{\ell}$ must satisfy the following estimates for $\zeta \to +\infty$:
\begin{eqnarray}
\label{FASS}
\left| F_{3j+1} \right|, \left| F_{3j+2} \right|, \left| F_{3j+3} \right| \le C_{3j} e^{-\frac{4}{3} \zeta^{3/2}},
\end{eqnarray}
and finally we require the following relation between $F_{\ell}$ and $G_{\ell}$:
\begin{eqnarray}
\label{FGEQ}
G_{m} (0) = F_{m}(0), \ \ m = 3j+1, \ 3j+2, \ 3j+3.
\end{eqnarray}

\vskip 0.2in

Once we have the existence of the functions $F_{\ell}$ and $G_{\ell}$ out of the way, then we may proceed with the inductive integration by parts.  Suppose we have arrived at the situation
\begin{eqnarray}
& &
\frac{1}{N} \int_{z^{*}-\epsilon}^{\beta+\delta}g(z) F_{0}(\Phi_{\beta}) \Phi_{\beta}'
dz = \hat{e}_{0} + N^{-2}\hat{e}_{1} + \cdots + N^{-2 \ell} \hat{e}_{\ell} + A_{\ell+1},
\end{eqnarray}
where $\hat{e}_{j}$ are all independent of $N$, and
\begin{eqnarray}
\label{4.035}
& & \hspace{0.11in}A_{\ell+1}=\frac{(-1)^{\ell}}{N} \int_{z^{*}-\epsilon}^{\beta+\delta}
\hspace{-0.05in}
\frac{d}{dz} \left( (\Phi_{\beta}')^{-1}
\frac{d}{dz} \left( (\Phi_{\beta}')^{-1}\
\cdots
\frac{d}{dz} \left( (\Phi_{\beta}')^{-1}\
g'(z)
\right)\cdots
\right)
\right)\times\\
\nonumber
& & \hspace{1.0in
}\times \left(G_{3\ell}(\Phi_{\beta}(z)) - c_{0}^{(3\ell)} (-\Phi_{\beta})^{1/2}
\right)  \ dz\\
\nonumber
& & +
\frac{(-1)^{\ell}}{N} \int_{z^{*}-\epsilon}^{\beta+\delta}
\hspace{-0.05in}
\frac{d}{dz} \left( (\Phi_{\beta}')^{-1}\
\frac{d}{dz} \left( (\Phi_{\beta}')^{-1}\
\cdots
\frac{d}{dz} \left( (\Phi_{\beta}')^{-1}\
g'(z)
\right)\cdots
\right)
\right) \times \\
\nonumber
& & \hspace{3.0in}\times F_{3\ell}(\Phi_{\beta}) dz,
\end{eqnarray}
and in each nested set of derivatives appearing in the integrals in
(\ref{4.035}), the differential operator 
\begin{eqnarray}\frac{d}{dz} \left(
\left(\Phi_{\beta}' \right)^{-1}\ \cdot \ \right)
\end{eqnarray}
appears $3 \ell-1$
times.  We further assume that in (\ref{4.035}), $G_{3\ell}$ satisfies
(\ref{GASS}), $F_{3\ell}$ satisfies (\ref{FASS}), (both with $j =
\ell-1$) and (\ref{FGEQ}) holds with $m=3\ell$.  In order to integrate
by parts, we note that from (\ref{GDIF}) and (\ref{FDIF}),
\begin{eqnarray}
& &
\left(G_{3\ell}(\Phi_{\beta}(z)) - c_{0}^{(3\ell)} (-\Phi_{\beta})^{1/2}
\right) = \frac{1}{\Phi_{\beta}'} \frac{d}{dz} G_{3\ell+1}(\Phi_{\beta}(z)), \\
& &
F_{3\ell}(\Phi_{\beta}(z)) = \frac{1}{\Phi_{\beta}'(z)}\frac{d}{dz}F_{3\ell+1}(\Phi_{\beta}(z)).
\end{eqnarray}
Using (\ref{FGEQ}) and integrating by parts, we find
\begin{eqnarray}
\label{4.038}
& & A_{\ell+1}=\\
\nonumber
& & -\frac{(-1)^{\ell}}{N} \int_{z^{*}-\epsilon}^{\beta+\delta}
\hspace{-0.05in}
\frac{d}{dz} \left( (\Phi_{\beta}')^{-1}\
\frac{d}{dz} \left( (\Phi_{\beta}')^{-1}\
\cdots
\frac{d}{dz} \left( (\Phi_{\beta}')^{-1}\
g'(z)
\right)\cdots
\right)
\right)\times\\
\nonumber
& & \hspace{3.0in
}\times \left(G_{3\ell+1}\right)  \ dz\\
\nonumber
& & -
\frac{(-1)^{\ell}}{N} \int_{z^{*}-\epsilon}^{\beta+\delta}
\hspace{-0.05in}
\frac{d}{dz} \left( (\Phi_{\beta}')^{-1}\
\frac{d}{dz} \left( (\Phi_{\beta}')^{-1}\
\cdots
\frac{d}{dz} \left( (\Phi_{\beta}')^{-1}\
g'(z)
\right)\cdots
\right)
\right) \times \\
\nonumber
& & \hspace{3.0in}\times F_{3\ell+1}(\Phi_{\beta}) dz,
\end{eqnarray}
where now the differential operator $(d/dz ( \Phi_{\beta}')^{-1} \ \cdot )$ appears
$3 \ell$ times.  At this point, using (\ref{matchbe}), it is straightforward to prove that $|A_{\ell+1}|\le C N^{ - 2\ell-2/3 }$.  However, we may integrate by parts two more times, and we then arrive at
\begin{eqnarray}
\label{4.039}
& & A_{\ell+1}=\\
\nonumber
& & -\frac{(-1)^{\ell}}{N} \int_{z^{*}-\epsilon}^{\beta+\delta}
\hspace{-0.05in}
\frac{d}{dz} \left( (\Phi_{\beta}')^{-1}\
\frac{d}{dz} \left( (\Phi_{\beta}')^{-1}\
\cdots
\frac{d}{dz} \left( (\Phi_{\beta}')^{-1}\
g'(z)
\right)\cdots
\right)
\right)\times\\
\nonumber
& & \hspace{3.0in
}\times \left(G_{3\ell+3}\right)  \ dz\\
\nonumber
& & -
\frac{(-1)^{\ell}}{N} \int_{z^{*}-\epsilon}^{\beta+\delta}
\hspace{-0.05in}
\frac{d}{dz} \left( (\Phi_{\beta}')^{-1}\
\frac{d}{dz} \left( (\Phi_{\beta}')^{-1}\
\cdots
\frac{d}{dz} \left( (\Phi_{\beta}')^{-1}\
g'(z)
\right)\cdots
\right)
\right) \times \\
\nonumber
& & \hspace{3.0in}\times F_{3\ell+3}(\Phi_{\beta}) dz,
\end{eqnarray}
where now the differential operator $(d/dz ( \Phi_{\beta}')^{-1} \ \cdot )$ appears
$3 \ell+2$ times in each integral.  Expression (\ref{4.039}) may now be rewritten as follows:
\begin{eqnarray}
& & \hspace{0.2in}A_{\ell+1}= N^{-2\ell-2}\hat{e}_{\ell+1} + A_{\ell+2}, \\
& &  \hspace{0.2in}\hat{e}_{\ell+1} = \\
\nonumber
& &
 (-1)^{\ell+1}N^{2\ell+1} \int_{z^{*}-\epsilon}^{\beta+\delta}
\hspace{-0.05in}
\frac{d}{dz} \left( (\Phi_{\beta}')^{-1}\
\frac{d}{dz} \left( (\Phi_{\beta}')^{-1}\
\cdots
\frac{d}{dz} \left( (\Phi_{\beta}')^{-1}\
g'(z)
\right)\cdots
\right)
\right)\times\\
\nonumber
& & \hspace{2.50in}
\times \left(c_{0}^{(3\ell+3)} \left( - \Phi_{\beta}\right)^{1/2}\right)  \ dz, \\
\label{Alp1}
& &\hspace{0.2in}
A_{\ell+2}  =\\
\nonumber
& & \frac{(-1)^{\ell+1}}{N} \int_{z^{*}-\epsilon}^{\beta+\delta}
\hspace{-0.05in}
\frac{d}{dz} \left( (\Phi_{\beta}')^{-1}\
\frac{d}{dz} \left( (\Phi_{\beta}')^{-1}\
\cdots
\frac{d}{dz} \left( (\Phi_{\beta}')^{-1}\
g'(z)
\right)\cdots
\right)
\right)\times\\
\nonumber
& & \hspace{1.50in}
\times \left(G_{3\ell+3} - c_{0}^{(3\ell+3)}\left( -\Phi_{\beta}\right)^{1/2}\right)  \ dz\\
\nonumber
& &
\frac{(-1)^{\ell+1}}{N} \int_{z^{*}-\epsilon}^{\beta+\delta}
\hspace{-0.05in}
\frac{d}{dz} \left( (\Phi_{\beta}')^{-1}\
\frac{d}{dz} \left( (\Phi_{\beta}')^{-1}\
\cdots
\frac{d}{dz} \left( (\Phi_{\beta}')^{-1}\
g'(z)
\right)\cdots
\right)
\right) \times \\
\nonumber
& & \hspace{3.0in}\times F_{3\ell+3}(\Phi_{\beta}) dz,
\end{eqnarray}
where the differential operator $(d/dz ( \Phi_{\beta}')^{-1} \ \cdot )$ appears
$3 \ell+2$ times in each integral.  The quantity $\hat{e}_{\ell+1}$ is easily seen to be independent of $N$ using the representation (\ref{matchbe}) of $\Phi_{\beta}$:
\begin{eqnarray}
\label{elp1}
& &\hspace{0.2in}
\hat{e}_{\ell+1}=\\
\nonumber & &
 (-1)^{\ell+1}
\int_{z^{*}-\epsilon}^{\beta+\delta}
\hspace{-0.05in} 
\frac{d}{dz} \left( (\phi_{\beta}')^{-1}\
\frac{d}{dz} \left( (\phi_{\beta}')^{-1}\
\cdots
\frac{d}{dz} \left( (\phi_{\beta}')^{-1}\
g'(z)
\right)\cdots
\right)
\right)\times\\
\nonumber
& & \hspace{2.50in}
\times \left(c_{0}^{(3\ell+3)} \left( - \phi_{\beta}\right)^{1/2}\right)  \ dz.
\end{eqnarray}
Therefore we have deduced that
\begin{eqnarray}
& &
\frac{1}{N} \int_{z^{*}-\epsilon}^{\beta+\delta}g(z) F_{0}(\Phi_{\beta}) \Phi_{\beta}'
dz = \hat{e}_{0} + N^{-2}\hat{e}_{1} + \cdots + N^{-2 \ell-2} \hat{e}_{\ell+1} + A_{\ell+2},
\end{eqnarray}
with $\hat{e}_{\ell+1}$ defined in (\ref{elp1}), and $A_{\ell+2}$ defined in (\ref{Alp1}).  Summarizing, once we establish the existence of the functions $F_{\ell}$ and $G_{\ell}$, we have established by induction that (\ref{F0Int}) has an asymptotic expansion in even powers of $N$.

\vskip 0.2in

{\bf Proof of Step 1:}
We establish Step 1 using well-known \cite{AS}
asymptotic representations of the Airy functions for large negative values
of the argument:

\begin{eqnarray} \label{Airy}\hspace{0.2in}
Ai(-z) &=& \frac{1}{\sqrt{\pi}z^{1/4}}
\left(\sin(\zeta + \frac{\pi}{4})w_1(\zeta)
- \cos(\zeta + \frac{\pi}{4})w_2(\zeta)\right),
|\mbox{arg}\,\, z| < \frac{2}{3} \pi \\
\hspace{0.25in}
\label{Airyp}
Ai^\prime(-z) &=& \frac{z^{1/4}}{\sqrt{\pi}}
\left(-\cos(\zeta + \frac{\pi}{4})w_3(\zeta)
- \sin(\zeta + \frac{\pi}{4})w_4(\zeta)\right),
|\mbox{arg}\,\, z| < \frac{2}{3} \pi,
\end{eqnarray}
where
\begin{eqnarray}
\label{AiryAS3}
\zeta = \frac{2}{3}(z)^{3/2}
\end{eqnarray}
and the $w_i(\zeta)$ are asymptotic series in large $\zeta$ defined
recursively in terms of the coefficients
\begin{eqnarray*}
u_s &=& \frac{(2s+1)(2s+3)(2s+5) \dots (6s-1)}{(216)^s s!} \\
v_s &=& -\frac{6s+1}{6s-1}u_s \\
\mbox{with} & & u_0 = 1, \,\,\, v_0 = 1.
\end{eqnarray*}

Explicitly the asymptotic series are given by
\begin{eqnarray}
\label{4.048} & & w_1 = \sum_{s=0}^{\infty} (-1)^s
\frac{u_{2s}}{\zeta^{2s}}, \hspace{0.3in}
w_2 = \sum_{s=0}^{\infty} (-1)^s \frac{u_{2s+1}}{\zeta^{2s+1}},\\
& & w_3 = \sum_{s=0}^{\infty} (-1)^s \frac{v_{2s}}{\zeta^{2s}},
\hspace{0.3in} \label{AiryAS6} w_4 = \sum_{s=0}^{\infty} (-1)^s
\frac{v_{2s+1}}{\zeta^{2s+1}}.
\end{eqnarray}
We caution the reader that in formulae (\ref{Airy})-(\ref{AiryAS6})
the variable $z>0$  is a dummy variable, as is $\zeta$ appearing in
(\ref{AiryAS3}) above.  We will replace $z$ appearing in
(\ref{Airy})-(\ref{AiryAS6})  with $-\zeta$ (where $\zeta < 0$).
Consequently, the parameter $\zeta$ appearing in
(\ref{Airy})-(\ref{AiryAS6}) becomes $(2/3)(-\zeta)^{3/2}$.

One can now substitute the expansions of the Airy functions
(\ref{Airy}) into the expression (\ref{f0}) for $F_0(\zeta)$ and
collect separately the coefficients of $\cos(4 (-\zeta)^{3/2}/3)$,
$\sin(4(-\zeta)^{3/2}/3)$ and the ``non-oscillatory'' terms. One
finds that the expansion is of the form
\begin{eqnarray}
\label{4.052}
& & F_{0}(\zeta) =(-\zeta)^{1/2}\
S_{0}^{(0)}  + \\
\nonumber
& & \hspace{0.4in} + (-\zeta)^{1/2} \ S_{1}^{(0)}  \sin{\left( \frac{4}{3} (-\zeta)^{3/2}\right)} +  \ S_{2}^{(\ell)}  \cos{\left( \frac{4}{3} (-\zeta)^{3/2}\right)},
\end{eqnarray}
where $S_{0}^{(0)}$, $S_{1}^{(0)}$ and $S_{2}^{(0)}$ are
asymptotic series valid for $\zeta \to - \infty$:
\begin{eqnarray}
& &S_{0}^{(0)} = \frac{1}{2\pi}(w_1^2 + w_2^2 + w_3^2 + w_4^2), \\
& & S_{1}^{(0)} = \frac{1}{2 \pi} \left(w_{1}^{2} - w_{2}^{2} - w_{3}^{2} + w_{4}^{2}\right), \\
& & S_{2}^{(0)} =  \frac{(-\zeta)^{1/2} }{\pi} \left( w_{3} w_{4} - w_{1} w_{2} \right).
\end{eqnarray}
Now using the definitions (\ref{4.048})-(\ref{AiryAS6}) of the asymptotic series $\{w_{j}\}_{j=1}^{4}$ (with $\zeta$ replaced by $(-\zeta)^{3/2}$), we see that $S_{0}^{(0)}$ is an asymptotic expansion in powers of $(-\zeta)^{-3}$, starting with the constant term.  Similarly, $S_{1}^{(0)}$ is an asymptotic expansion in powers of $(-\zeta)^{-3}$ starting with $(-\zeta)^{-3}$, and $S_{2}^{(0)}$ is an asymptotic expansion in powers of the form $(-\zeta)^{-3p -1}$, $p=0, 1, \ldots$.  This establishes the form of the expansion
specified in Step 1, formula (\ref{4.008})

{\bf Proof of Steps 2 and 4:}
Now we must establish the existence of $F_{\ell}$ and $G_{\ell}$, $\ell\ge 1$, satisfying (\ref{4.009})-(\ref{4.018}) (for $\ell = 1, 2, 3$), and
(\ref{GDIF})-(\ref{FGEQ}) (for $\ell \ge 4$).

We observe that if $\{F_{\ell}\}_{\ell \ge 1}$ satisfies (\ref{4.015})-(\ref{4.017}) and (\ref{FDIF}), then clearly $F_{\ell}$ is the $\ell$-th antiderivative of the function $F_{0}$.  Now $F_{0}(\zeta) = \left(Ai'(\zeta) \right)^{2} - \zeta \left(Ai(\zeta)\right)^{2}$, and it straightforward to prove that the $\ell$-th antiderivative of $F_{0}$ is given by
\begin{eqnarray}
\label{4.056}
F_{\ell}(\zeta) = q_{0}^{(\ell)}(\zeta) \left(Ai'(\zeta)\right)^{2} + q_{1}^{(\ell)}(\zeta) \left(  Ai(\zeta) \right)^{2} + q_{3}^{(\ell)}(\zeta) Ai(\zeta) Ai'(\zeta)
\end{eqnarray}
where $\{q_{\mu}^{(\ell)}\}$ are polynomials in $\zeta$.  Now (\ref{4.056}), together with the asymptotics for $\zeta \to \infty$ of the Airy function $Ai(\zeta)$ imply that
for each $\ell$, $F_{\ell}$ satisfies (\ref{4.018}) and (\ref{FASS}).  Furthermore, using the asymptotics (\ref{Airy}) and (\ref{Airyp}) (again replacing $z$ with $-\zeta$, and $\zeta$ with $(2/3)(-\zeta)^{3/2}$), and letting $\zeta \to - \infty$, we find that for each $\ell$, $F_{\ell}$ possesses an expansion of the following form
\begin{eqnarray}
\label{4.057}
& & F_{\ell}(\zeta) = (-\zeta)^{1/2} \
S_{0}^{(\ell)}  + \\
\nonumber
& & \hspace{0.4in} + (-\zeta)^{1/2} \ S_{1}^{(\ell)}  \sin{\left( \frac{4}{3} (-\zeta)^{3/2}\right)} + S_{2}^{(\ell)} \cos{\left( \frac{4}{3} (-\zeta)^{3/2}\right)},
\end{eqnarray}
where $S_{0}$, $S_{1}$, and $S_{2}$ are asymptotic series valid for $\zeta \to -\infty$.  We will refine our knowledge of these series in what follows, but for now we observe the following important basic property of these asymptotic series:

{\bf Property \refstepcounter{equation}\theequation\label{BasProp}}:  The asymptotic series $S_{0}^{(\ell)}$, $S_{1}^{(\ell)}$ and $S_{2}^{(\ell)}$ are all in integer powers of $(-\zeta)$.

\vskip 0.2in

In order to obtain more detailed information about the asymptotic series $S_{0}^{(\ell)}$, $S_{1}^{(\ell)}$ and $S_{2}^{(\ell)}$, we will make use of the following Lemma.

\begin{lem}\label{sublem}
If a function $G(\zeta)$ is $C^{\infty}$ on $(-\infty,
a)$ for some $a > 0$, and possesses the following
asymptotic expansion (which may be differentiated
repeatedly to obtain asymptotic expansions for
derivatives),
\begin{eqnarray}
G(\zeta) = \left(   a (-\zeta)^\frac{-j}{2} + b
(-\zeta)^\frac{-(j+3)}{2}  + \dots \right) trig\left(
\frac{4}{3}(-\zeta)^{3/2}\right),
\end{eqnarray}
in which the asymptotic expansion in parentheses descends in
powers of $(-\zeta)^{3/2}$, $j \ge 1$, and $trig(\cdot)$ denotes either
$\sin{(\cdot)}$ or $\cos{(\cdot)}$, then the function
$\int_{-\infty}^{s} G(\zeta) d \zeta$ possesses the following
asymptotic expansion for $s \to -\infty$:
\begin{eqnarray}
\nonumber
& & \hspace{-1.0in}
\int_{-\infty}^{s}G(\zeta)d \zeta = \left( \tilde{a}(-s)^\frac{-(j+1)}{2} + \tilde{b}
(-s)^\frac{-(j+4)}{2}  + \dots \right) trig\left( \frac{4}{3}(-s)^{3/2}\right)
\end{eqnarray}
where the expression in parentheses descends in powers of $(-s)^{3/2}$.
\end{lem}

\begin{Proof}
\begin{eqnarray*}
& & \int_{-\infty}^s \left( a (-\zeta)^\frac{-j}{2} + b (-\zeta)^\frac{-(j+3)}{2}
+ \dots \right) trig\left( \frac{4}{3}(-\zeta)^{3/2}\right)d\zeta \\
&=&
\int_{-\infty}^s \left( a (-\zeta)^\frac{-j}{2} + b (-\zeta)^\frac{-(j+3)}{2}
+ \dots \right)
\frac{-1}{2(-\zeta)^{1/2}}\frac{d}{d\zeta}
trig\left( \frac{4}{3}(-\zeta)^{3/2}\right)d\zeta  \\
&=& -\left( a (-s)^\frac{-j}{2} + b (-s)^\frac{-(j+3)}{2} + \dots \right)
\frac{1}{2(-s)^{1/2}} trig\left( \frac{4}{3}(-s)^{3/2}\right)\\
& & +\frac{1}{2} \int_{-\infty}^s \left( a (-\zeta)^\frac{-(j+1)}{2}
+ b (-\zeta)^\frac{-(j+4)}{2} + \dots \right)^\prime
trig\left( \frac{4}{3}(-\zeta)^{3/2}\right)d\zeta \\
&=& -\frac{1}{2}\left( a (-s)^\frac{-(j+1)}{2} + b (-s)^\frac{-(j+4)}{2} + \dots \right)
trig\left( \frac{4}{3}(-s)^{3/2}\right)\\
& & + \frac{1}{2}\int_{-\infty}^s \left( a \frac{j+1}{2}(-\zeta)^\frac{-(j+3)}{2} +
b \frac{j+4}{2}(-\zeta)^\frac{-(j+6)}{2} + \dots \right)\times\\
& & \hspace{2.0in}\times
\frac{-1}{2(-\zeta)^{1/2}}\frac{d}{d\zeta}
trig\left( \frac{4}{3}(-\zeta)^{3/2}\right)d\zeta,
\end{eqnarray*}
and continuing in this way establishes the lemma.
\end{Proof}

Let us write $F_{1}$ in the following convoluted way:
\begin{eqnarray}
\label{4.060}
F_{1}(\zeta) = \int_{-\infty}^{\zeta} \left( F_{0}(s) - c_{0} (-s)^{1/2} \right) ds - \frac{2}{3}c_{0} (-\zeta)^{3/2}.
\end{eqnarray}
One sees that $F_{1}$ so defined is an antiderivative of $F_{0}$ immediately.  To verify that $F_{1}$ so defined is exponentially decaying for $\zeta \to +\infty$, one uses the asymptotic expansion of $F_{0}$ obtained in Step 1 together with Lemma \ref{sublem} to compute the asymptotic expansion of the integral in (\ref{4.060}), and compares to the asymptotic expansion (\ref{4.057}).  Since there is no constant term in (\ref{4.057}), neither in the asymptotic expansion of (\ref{4.060}), $F_{1}$ defined by (\ref{4.060}) must be expressible in the form (\ref{4.056}), and hence it is exponentially decaying for $\zeta \to +\infty$.

So we have learned that (1) $F_{1}$ exists satisfying $F_{1}'=F_{0}$, (2) $F_{1}$ satisfies (\ref{4.018}) for $\zeta \to +\infty$, and (3) $F_{1}$ possesses an asymptotic expansion of the form (\ref{4.057}) for $\zeta \to -\infty$.  Furthermore, we learn that the asymptotic series $ (-\zeta)^{1/2}S_{0}^{(1)}$ appearing in (\ref{4.057}) with $\ell=1$ is obtained by computing the term-by-term anti-derivative of the asymptotic series $(-\zeta)^{1/2}S_{0}^{(0)}$ appearing in (\ref{4.052}).  This provides us with more detailed information about the asymptotic series $S_{0}^{(1)}$:
\begin{eqnarray}
(-\zeta)^{1/2} S_{0}^{(1)} = -\frac{2}{3}c_{0}(-\zeta)^{3/2} + \frac{2}{3}c_{1}(-\zeta)^{-3/2} +\frac{2}{9}c_{2}
(-\zeta)^{-9/2} + \cdots.
\end{eqnarray}
If, rather than using Lemma \ref{sublem}, one is more careful integrating by parts repeatedly, it is straightforward to prove the following two properties:

\vskip 0.1in
\noindent
{\bf Property \refstepcounter{equation}\theequation\label{S1Prop}}:  The expansion $S_{1}^{(1)}$  starts at $(-\zeta)^{-2}$ and descends in powers of $(-\zeta)^{3}$.

\vskip 0.1in
\noindent{\bf Property \refstepcounter{equation}\theequation\label{S2Prop}}:  The expansion $S_{2}^{(1)}$  starts at $(-\zeta)^{-3}$ and descends in powers of $(-\zeta)^{3}$.

Setting $G_{1} = F_{1} + (2 c_{0}/3) (-\zeta)^{3/2}$, we have proven the existence of $G_{1}$ satisfying (\ref{4.009}) and (\ref{4.012}), with $c_{0}^{(1)} = -2c_{1}/3$ (the constant $c_{1}$ appears in (\ref{4.008})).  We may now define
\begin{eqnarray}
& &G_{2} = \int_{-\infty}^{\zeta}G_{1}(s) ds,
\end{eqnarray}
The function $G_{2}$ clearly satisfies (\ref{4.010}), and using the asymptotic expansion (\ref{4.012}) satisfied by $G_{1}$ together with careful integration by parts shows that $G_{2}$ satisfies the asymptotic expansion (\ref{4.013}) for $\zeta \to -\infty$.
We next define $F_{2}$ via
\begin{eqnarray}
F_{2} = G_{2} + \frac{4 c_{0}}{15} (-\zeta)^{5/2}
\end{eqnarray}
which, it turns out, is equivalent to the definition $F_{2} = \int_{\infty}^{\zeta} F_{1}(s) ds$.  The function $F_{2}$ then clearly satisfies (\ref{4.016}) and (\ref{4.018}).

The functions $G_{3}$ and $F_{3}$ are defined in similar fashion.  First we define
\begin{eqnarray}
G_{3} = \int_{-\infty}^{\zeta} \left( G_{2}(s) - c_{0}^{(2)}(-s)^{-1/2} \right) ds - 2 c_{0}^{(2)} (-\zeta)^{1/2}.
\end{eqnarray}
Second, using the asymptotic expansion (\ref{4.013}) satisfied by $G_{2}$ and careful integration by parts in the manner used to prove Lemma \ref{sublem}, we deduce that $G_{3}$ possesses the asymptotic expansion (\ref{step2}).  Third, we define $F_{3}$ in the obvious way:
\begin{eqnarray}
F_{3} = G_{3} - \frac{8c_{0}}{105} (-\zeta)^{7/2}.
\end{eqnarray}
Uniqueness of asymptotic expansions then implies that $F_{3} = \int_{\infty}^{\zeta} F_{2}(s) ds $, and hence $F_{3}$ satisfies (\ref{4.017}) and (\ref{4.018}).

\vskip 0.15in

The inductive argument goes as follows:  suppose we have the existence of  $F_{\ell}$ and $G_{\ell}$ for $\ell = 1,\ldots, 3k$, satisfying (\ref{4.009})-(\ref{4.018}), and possibly (\ref{GDIF}) - (\ref{FGEQ}) with $j = k-1$ (if $k \ge 2$).   We define $G_{3k+1}$ and $G_{3k+2}$ as follows:
\begin{eqnarray}
& &
G_{3k+1}(\zeta) = \int_{-\infty}^{\zeta} \left(G_{3k}(s) - c_{0}^{(3k)}(-s)^{1/2} \right) ds \\
& & G_{3j+2} = \int_{-\infty}^{\zeta}G_{3k+1}(s) ds.
\end{eqnarray}
The asymptotic expansion satisfied by $G_{3k}$ (either (\ref{step2}) if $k=1$, or (\ref{GASS}) with $j=k-1$ if $k\ge 2$), together with a careful integration by parts argument as outlined in the proof of Lemma \ref{sublem} shows that $G_{3k+1}$ satisfies the following asymptotic expansion:
\begin{eqnarray}
& &G_{3k+1} = (-\zeta)^{1/2} \left(
c_{0}^{(3k+1)} (-\zeta)^{-2}+c_{1}^{(3k+1)}\zz^{-5} + \cdots \right) \\
\nonumber
& & +
G_{S}^{(3k+1)} \sin{\left(\frac{4}{3} \zz^{3/2}\right)} + G_{C}^{(3k+1)} \cos{\left(  \frac{4}{3}\zz^{3/2}  \right)},
\end{eqnarray}
where $G_{S}^{(3k+1)}$ and $G_{C}^{(3k+1)}$ are asymptotic expansions:
\begin{eqnarray}& &\mbox{ For $3k+1$ even:} \\ \nonumber
& & \hspace{0.3in}\left\{
\begin{array}{c}
G_{S}^{(3k+1)} = \left(-\zeta \right)^{-(3k+6)/2} \left( d_{0}^{(3k+1)}
+ d_{1}^{(3k+1)}(-\zeta)^{-3} + \cdots \right)\\
\hspace{-0.1in}
G_{C}^{(3k+1)} = \left(-\zeta \right)^{-(3k+3)/2} \left( f_{0}^{(3k+1)} +
f_{1}^{(3k+1)}(-\zeta)^{-3} + \cdots \right) \end{array} \right.
\\
& & \mbox{For $3k+1$ odd:}\\ \nonumber
& & \hspace{0.3in}\left\{ \begin{array}{c}
G_{S}^{(3k+1)} = \left(-\zeta \right)^{-(3k+3)/2} \left( d_{0}^{(3k+1)}
+ d_{1}^{(3k+1)}(-\zeta)^{-3} + \cdots \right) \\
\hspace{-0.15in}G_{C}^{(3k+1)} = \left(-\zeta \right)^{-(3k+6)/2} \left( f_{0}^{(3k+1)} +
f_{1}^{(3k+1)}(-\zeta)^{-3}\cdots \right) \end{array} \right. .
\end{eqnarray}

Similarly, $G_{3k+2}$ satisfies the following asymptotic description for $\zeta \to -\infty$:
\begin{eqnarray}
& &G_{3k+2} = (-\zeta)^{1/2} \left(
c_{0}^{(3k+2)} (-\zeta)^{-1}+c_{1}^{(3k+2)}\zz^{-4} + \cdots \right) \\
\nonumber
& & +
G_{S}^{(3k+2)} \sin{\left(\frac{4}{3} \zz^{3/2}\right)} + G_{C}^{(3k+2)} \cos{\left(  \frac{4}{3}\zz^{3/2}  \right)},
\end{eqnarray}
where $G_{S}^{(3k+2)}$ and $G_{C}^{(3k+2)}$ are asymptotic expansions:
\begin{eqnarray}& & \mbox{ For $3k+2$ even:} \\
 \nonumber & & \hspace{0.3in} \left\{
\begin{array}{c}
G_{S}^{(3k+2)} = \left(-\zeta \right)^{-(3k+7)/2} \left( d_{0}^{(3k+2)}
+ d_{1}^{(3k+2)}(-\zeta)^{-3} + \cdots \right)\\
\hspace{-0.1in}
G_{C}^{(3k+2)} = \left(-\zeta \right)^{-(3k+4)/2} \left( f_{0}^{(3k+4)} +
f_{1}^{(3k+4)}(-\zeta)^{-3} + \cdots \right) \end{array} \right.
\\
& &\mbox{For $3k+2$ odd:}\\
\nonumber & & \hspace{0.3in} \left\{ \begin{array}{c}
G_{S}^{(3k+2)} = \left(-\zeta \right)^{-(3k+4)/2} \left( d_{0}^{(3k+2)}
+ d_{1}^{(3k+2)}(-\zeta)^{-3} + \cdots \right) \\
\hspace{-0.15in}G_{C}^{(3k+2)} = \left(-\zeta \right)^{-(3k+7)/2} \left( f_{0}^{(3k+2)} +
f_{1}^{(3k+2)}(-\zeta)^{-3}\cdots \right) \end{array} \right. .
\end{eqnarray}

We may now define $G_{3k+3}$:
\begin{eqnarray}
& & G_{3k+3}(\zeta) = \int_{-\infty}^{\zeta}\left(
G_{3k+2}(s) - c_{0}^{(3k+2)}(-s)^{-1/2} \right) ds - 2 c_{0}^{(3k+2)} \zz^{1/2}.
\end{eqnarray}
It is by now a straightforward exercise to establish that $G_{3k+3}$ so defined satisfies (\ref{GDIF}) as well as the asymptotic description (\ref{GASS}) for $\zeta \to -\infty$.

Turning now to the existence of $F_{3k+1}$, $F_{3k+2}$ and $F_{3k+3}$, we define them through
\begin{eqnarray}& &
F_{3k+1} = \int_{\infty}^{\zeta} F_{3k}(s) ds, \ \ \
F_{3k+2} = \int_{\infty}^{\zeta} F_{3k+1}(s) ds, \ \ \
F_{3k+3} = \int_{\infty}^{\zeta} F_{3k+2}(s)ds.
\end{eqnarray}
It is immediately clear from these definitions that $\{ F_{3k+\mu} \}_{\mu=1}^{3}$ satisfy  (\ref{4.018}) for $\zeta \to \infty$.

Because of the recursive definitions of $F_{\ell}$ and $G_{\ell}$, it follows that
\begin{eqnarray}\label{4.080}
& &
F_{3k+1} =G_{3k+1} -\frac{2 c_{0}^{(3k)}}{3} \zz^{3/2}  \\
\nonumber
& & \hspace{0.75in}
+ \frac{2^{4}c_{0}^{(3k-3)}}{945} \zz^{9/2} + \cdots +
 \frac{(-1)^{3k+1}2^{3k+1}c_{0}}{\prod_{j=1}^{3k+1}(2j+1)} \zz^{(6k+3)/2}, \\
 & & F_{3k+2} = G_{3k+2}  \label{4.081}  +
 \frac{4 c_{0}^{(3k)}}{15} \zz^{5/2} \\
 \nonumber
 & & \hspace{0.75in}+ \frac{2^{5}c_{0}^{(3k-3)}}
 {10395} \zz^{11/2} + \cdots +
 \frac{(-1)^{3k+2}2^{3k+2}c_{0}}{\prod_{j=1}^{3k+2}(2j+1)} \zz^{(6k+5)/2} ,\\
 & & F_{3k+3} = G_{3k+3}
 - \frac{8 c_{0}^{(3k)}}{105} \zz^{7/2} \label{4.082} \\
 \nonumber
 & & \hspace{0.75in}+ \frac{2^{6}c_{0}^{(3k-3)}}
 {135135} \zz^{13/2} + \cdots +
 \frac{(-1)^{3k+3}2^{3k+3}c_{0}}{\prod_{j=1}^{3k+3}(2j+1)} \zz^{(6k+7)/2}.
\end{eqnarray}
we observe that there are no constant terms in the relations (\ref{4.080})-(\ref{4.082}).  Indeed, this is so because (1) for each $\ell$, $F_{\ell}$ has  the representation (\ref{4.056}) for some polynomials $\{q_{\mu}^{(\ell)}\}_{\mu=0}^{2}$, and consequently the asymptotic expansion for $F_{\ell}$ valid for $\zeta \to -\infty$ is of the form (\ref{4.057}) with no constant term and (2)  $G_{\ell}$ as defined possesses an asymptotic expansion with no constant term for $\zeta \to -\infty$.  Therefore the arbitrary constants of integration which may appear in (\ref{4.080})-(\ref{4.082}) are all $0$.

Finally, (\ref{4.080})-(\ref{4.082}) imply that $\{F_{3k+\mu} \}_{\mu=1}^{3}$ and $\{G_{2k + \mu} \}_{\mu=1}^{3}$ satisfy (\ref{FGEQ}) with $j=k$.

This completes the proof of Step 2 and Step 4, and we have established
the first claim of Theorem \ref{1PTASS}, that asymptotic expansions of the form
(\ref{I.Weakasym}) hold true.  The explanation for why the
coefficients in such an asymptotic expansion are analytic is as
follows.

We return to the integral (\ref{4.004IAS}), and show that each term
in {\it its} asymptotic expansion depends analytically on ${\bf t}$.
Since the same is true of the other integral appearing in (\ref{4.003INT}),
this will complete the proof of Theorem \ref{1PTASS}.
We begin by observing that each term in the asymptotic
expansion of (\ref{4.004IAS}) is of the form 
\begin{eqnarray}
\nonumber
\int_{z^{*}-\epsilon}^{\beta} \chi_{\beta}^{(j)}(\lambda) Q(\lambda,
{\bf t})  \left( - \phi_{\beta} \right)^{1/2} d \lambda,
\end{eqnarray}
where 
\begin{itemize}
\item{$\chi_{\beta}^{(j)}(\lambda)$ represents some finite number of
derivatives of $\chi_{\beta}(\lambda)$;}
\item{$Q(\lambda, {\bf t})$
represents terms obtained from the integration by parts procedure but
do not depend on the partition of unity $\chi_{\beta}$ (these terms
depend analytically on $\lambda$ and on ${\bf t}$);}
\item{$(-\phi_{\beta}(\lambda))^{1/2}$ may be taken to be analytic
with branch cut emanating from $\lambda = \beta$, down the real axis,
and passing in particular through 
$z^{*}+\epsilon$.  For $\lambda$ away from the cut, this 
function also depends analytically on ${\bf t}$.}
\end{itemize}
We now decompose the integral into a sum:
\begin{eqnarray}
\label{AnaLISSUE}
& & 
\int_{z^{*}-\epsilon}^{z^{*}+\epsilon} \chi_{\beta}^{(j)}(\lambda) Q(\lambda,
{\bf t})  \left( - \phi_{\beta} \right)^{1/2} d \lambda = 
\int_{z^{*}-\epsilon}^{\beta} \chi_{\beta}^{(j)}(\lambda) Q(\lambda,
{\bf t})  \left( - \phi_{\beta} \right)^{1/2} d \lambda + 
\int_{z^{*}+\epsilon}^{\beta} Q(\lambda,
{\bf t})  \left( - \phi_{\beta} \right)^{1/2} d \lambda.
\end{eqnarray}
The first integral on the
right hand side of (\ref{AnaLISSUE}) is clearly analytic in ${\bf t}$,
because the integrand is analytic in ${\bf t}$.  For the second
integral, note that $\chi_{\beta}$ does not appear.
This is because the partition of unity is such that $\chi_{\beta}
\equiv 1$ on $(z^{*}+\epsilon, \beta)$.  Now this second integral
is seen to be analytic by 
expressing it as $1/2$ the value of a contour integral encircling the interval
$(z^{*}+\epsilon, \beta)$, passing through $z^{*}+\epsilon$.  This
can be done because the integrand is analytic, with a square-root
branch point at $\lambda = \beta$.  This completes the proof of
Theorem \ref{1PTASS}. 

\newpage

\section{Appendix}
Here we present a set of explicit formulae for $S_{1}(z)$, the
solution of Riemann--Hilbert problem \ref{RHRec}.
\begin{enumerate}
\item[1.]
for $z \in \cb \setminus ( B_{\delta}^{\al} \cup
B_{\delta}^{\beta})$
\begin{eqnarray}
& & S_{1}(z) = \\
\nonumber
& & \frac{5}{144 h(\alpha)} \left\{
3 (z - \alpha)^{-2} + \frac{3}{5} \left[ \frac{1}{\alpha-\beta} - 3
\frac{h'(\alpha) }{h(\alpha)} \right] (z-\alpha)^{-1} \right\}
\pmtwo
{-1}{-i}{-i}{1} \\
\nonumber
& &
+ \frac{7}{48 (\alpha - \beta) h(\alpha)} (z-\alpha)^{-1}
\pmtwo
{-1}{i}
{i} {1}
\\
\nonumber
& &
+
\frac{5}{144 h(\beta)} \left\{
3 (z - \beta)^{-2} + \frac{3}{5} \left[ \frac{1}{\beta-\alpha} - 3
\frac{h'(\beta) }{h(\beta)} \right] (z-\beta)^{-1} \right\}
\pmtwo
{-1}{i}{i}{1} \\
\nonumber
& & +
\frac{7}{48 (\beta - \alpha) h(\beta)} (z-\beta)^{-1}
\pmtwo
{-1}{i}
{i} {1}.
\end{eqnarray}

\item[2.]  For $z \in B_{\delta}^{\beta}$,
\begin{eqnarray}
& & S_{1}(z) = \\
\nonumber
& &
\frac{5}{144 h(\alpha)} \left\{
3 (z - \alpha)^{-2} + \frac{3}{5} \left[ \frac{1}{\alpha-\beta} - 3
\frac{h'(\alpha) }{h(\alpha)} \right] (z-\alpha)^{-1} \right\}
\pmtwo
{-1}{-i}{-i}{1} \\
\nonumber
& &
+ \frac{7}{48 (\alpha - \beta) h(\alpha)} (z-\alpha)^{-1}
\pmtwo
{-1}{i}
{i} {1}
\\
\nonumber
& & +
\frac{5}{72}
\left\{
\frac{3}{2h(\beta)} (z-\beta)^{-2}
+ \frac{3}{10 h(\beta)} \left[ \frac{1}{\beta-\alpha} - \frac{
3h'(\beta)}{h(\beta)} \right] (z-\beta)^{-1} \right. \\
\nonumber
& & \hspace{1.75in}\left. - \frac{ (z -
\alpha)^{1/2}}
{(z-\beta)^{1/2} \int_{\beta}^{z} R(s) h(s) ds } \right\}
\pmtwo
{-1}{i}
{i}{1} \\
\nonumber
& & + \frac{7}{72} \left\{
\frac{3}{2 ( \beta-\alpha) h(\beta) } ( z - \beta)^{-1} - \frac{ ( z -
\beta)^{1/2} }{( z - \alpha)^{1/2} \int_{\beta}^{z} R(s) h(s) ds }
\right\}
\pmtwo
{1}{i}
{i}{-1}.
\end{eqnarray}

\item[3.] For $z \in B_{\delta}^{\alpha}$,
\begin{eqnarray}
& & S_{1}(z) = \\
\nonumber
& &
\frac{5}{144 h(\beta)} \left\{
3 (z - \beta)^{-2} + \frac{3}{5} \left[ \frac{1}{\beta-\alpha} - 3
\frac{h'(\beta) }{h(\beta)} \right] (z-\beta)^{-1} \right\}
\pmtwo
{-1}{i}{i}{1} \\
\nonumber
& &
+ \frac{7}{48 (\beta - \alpha) h(\beta)} (z-\beta)^{-1}
\pmtwo
{1}{i}
{i} {-1}
\\
\nonumber
& & +
\frac{5}{72}
\left\{
\frac{3}{2h(\alpha)} (z-\alpha)^{-2}
+ \frac{3}{10 h(\alpha)} \left[ \frac{1}{\alpha-\beta} - \frac{
3h'(\alpha)}{h(\alpha)} \right] (z-\alpha)^{-1} \right. \\
\nonumber
& & \hspace{1.75in}\left. - \frac{ (z -\beta)^{1/2}}
{(z-\alpha)^{1/2} \int_{\alpha}^{z} R(s) h(s) ds } \right\}
\pmtwo
{-1}{-i}
{-i}{1} \\
\nonumber
& & + \frac{7}{72} \left\{
\frac{3}{2 ( \alpha-\beta) h(\alpha) } ( z - \alpha)^{-1} - \frac{ ( z -
\alpha)^{1/2} }{( z - \beta)^{1/2} \int_{\alpha}^{z} R(s) h(s) ds }
\right\}
\pmtwo
{1}{-i}
{-i}{-1}.
\end{eqnarray}

\end{enumerate}
For comparative purposes, if $t_{j}=0, j \neq 4$, and if we set $t_{4} =
t$, then we have
\begin{enumerate}
\item[1.]
for $z \in \cb \setminus ( B_{\delta}^{\al} \cup
B_{\delta}^{\beta})$
\begin{eqnarray}
& &  \ \ S_{1}(z) = \\
& & \ \ \ \ \left( \frac{ - 7 \beta}{ 96 ( 8 - \beta^2) ( \beta -
z ) } + \frac{\beta ( - 56 z - 136 \beta + 15 \beta^2 z + 25
\beta^{3} )}
{96 ( 8 - \beta^{2})^{2} ( z + \beta)^{2}} \right) \pmtwo{1}{i}{i}{-1}\\
& & \ \ \ \ + \left( \frac{7 \beta}{96 ( 8 - \beta^{2})(z +
\beta)} - \frac{ \beta( - 136 \beta + 56 z + 25 \beta^3 - 15
\beta^{2} z)}{96 ( 8 - \beta^{2})^{2} ( \beta - z )^{2}} \right)
\pmtwo{-1}{i}{i}{1}
\end{eqnarray}
\item[2.] For $z \in B_{\delta}^{\beta}$,
\begin{eqnarray}
& & \ \ \ S_{1}(z) = \left( \frac{ - 7 \beta}{ 96 ( 8 - \beta^2) (
z - \beta) } + \right. \\
& & \  \ \ \left. \frac{\beta ( - 56 z - 136 \beta + 15 \beta^2 z
+ 25 \beta^{3} )} {96 ( 8 - \beta^{2})^{2} ( z + \beta)^{2}} -
\frac{ 7 \gamma^{2}}{72 \int_{\beta}^{z} R(s) h(s) ds}
\right) \pmtwo{1}{i}{i}{-1} + \\
& & \  \ \ + \left( \frac{7 \beta}{96 ( 8 - \beta^{2})(z +
\beta)}   + \right. \\
& & \  \ \ \left.  - \frac{ \beta( - 136 \beta + 56 \beta z + 25
\beta^3 - 15 \beta^{2} z)}{96 ( 8 - \beta^{2})^{2} ( \beta - z
)^{2}} - \frac{5}{72 \gamma^{2} \int_{\beta}^{z} R(s) h(s) ds}
\right) \pmtwo{-1}{i}{i}{1}
\end{eqnarray}
\item[3.] For $z \in B_{\delta}^{(-\beta)}$, we have
\begin{eqnarray}
& & \hspace{0.15in}
S_{1}(z) = \left(
\frac{ -\beta( - 136 \beta + 56 z + 25 \beta^3 - 15
\beta^{2} z)}{96 ( 8 - \beta^{2})^{2} ( \beta - z )^{2}} \right. \\
\nonumber
& & \hspace{1.0in}
\left. - \frac{7}{96} \left[ \frac{4\gamma^{-2}}{3 \int_{-\beta}^{z} R(s) h(s)
ds} + \frac{\beta (z+\beta)^{-1}}{(8 - \beta^{2}) } \right] \right)
\pmtwo
{-1}{i}
{i} {1} \\
\nonumber
& & + \left(
\,{\frac {\beta\,\left (25\,{\beta}^{3}-136\,\beta+15\,{\beta}^{2}z
-56\,z\right )}{96 \left ({\beta}^{2}-8\right )^{2}\left (z+\beta\right )
^{2}}} + \frac{  \gamma^{2}}{72 \int_{-\beta}^{z}R(s) h(s) ds}
 \right. \\
\nonumber
& & \hspace{2.6in}
\left. - \frac{7 \beta }{96} \left[ \frac{(z - \beta)^{-1}}{8 -
\beta^{2}} \right] \right)
\pmtwo
{-1} {-i}
{-i} {1}
\end{eqnarray}
\end{enumerate}

\section{Conclusions}

This paper provides the first mathematical derivation of the form of the
asymptotic expansion of the Hermitian random matrix partition function, 
conjectured in \cite{BIZ}.  This uses the connection to non-classical 
orthogonal polynomials, together with the Deift-Zhou steepest descent and 
stationary phase method for the asymptotic analysis of Riemann-Hilbert 
problems. In particular, the asymptotic analysis of (\ref{I.001}) presented 
here demonstrates that

\noindent $\bullet \log{ \left( \hat{Z}_{N} \right)}$ has an asymptotic 
expansion of the form of (\ref{I.003});

\noindent $\bullet$ For arbitrary integers 
$g \geq 0$, $e_{g}(t_{1},\ldots, t_{\nu})$ is an analytic function of the 
(complex) vector $\mbox{ {\bf t}} := (t_{1},\ldots, t_{\nu})$, 
in a neighborhood of $(0,\ldots, 0)$;

\noindent $\bullet$ For times, {\bf t}, within the domains of holomorphy for 
the $e_{g}$, the limiting mean density $\rho_{N}^{(1)}$ of eigenvalues for the 
Hermitian random matrix ensemble is supported on a single interval 
$(\alpha, \beta)$. The computed asymptotic expansion of $\rho_{N}^{(1)}$ 
is valid for all $\lambda \in (\alpha, \beta)$ and in fact depends only
on the equilibrium measure; 

\noindent $\bullet$ the coefficients of the $e_g$ are generating functions for 
graphical enumeration in the sense that 
$$
(-1)^n \frac{\partial^n}{\partial t_k^n} e_g(\bf{0})
= \#  \{\mbox{k-valent, n-vertex, g-maps}\}. 
$$

As a consequence of these results one is now  situated to be able to combine 
asymptotic analytical techniques with the results of many ingenious ideas, 
contributions, and calculations which have arisen over the past 20 years from 
the theory of 2D quantum gravity \cite{Difrancesco95}, to investigate the 
connection between the asymptotic expansion of the partition function and 
solutions of integrable systems such as the KdV hierarchy, the Toda lattice, 
and singular limits of these equations described by modulation equations
\cite{Witten, Tian, McL}.

\bibliographystyle{amsplain}

\end{document}